\documentclass[twocolumn, twocolappendix, dvipsnames]{aastex63}
\submitjournal{ApJ}

\usepackage{amsmath,natbib, float, color, soul, xspace, tabularx}
\usepackage[normalem]{ulem}
\usepackage{listings}

\usepackage{array}
\newcolumntype{R}[1]{>{\raggedright\arraybackslash\hspace{0pt}}p{#1}}

\long\def\***#1{\textsf{\textbf{***#1***}}}

\newcommand*\Msun[0]{\ensuremath{M_{\odot}}}
\newcommand*\Msolarh[0]{h^{-1} \, M_{\odot}}
\newcommand*\msolarh[0]{h^{-1} \, M_{\odot}}
\newcommand*\Mpch[0]{h^{-1}\,\mathrm{Mpc}}

\newcommand*\sig{\textsc{$\sigma_8$}\xspace}
\newcommand*\Sig{\ensuremath{\sigma_{8}}\xspace}

\newcommand*\Mgas{\textsc{$M_\mathrm{gas}$}\xspace}

\newcommand*\new[1]{\textbf{\color{RoyalBlue}#1}}
\renewcommand*\new[1]{{#1}}

\shorttitle{Interpretable Cluster Encoder}
\shortauthors{Ntampaka \& Vikhlinin}

\begin{document}

\correspondingauthor{Michelle Ntampaka}
\email{mntampaka@stsci.edu}

\author{Michelle Ntampaka}
\affiliation{Space Telescope Science Institute, Baltimore, MD 21218, USA}
\affiliation{Department of Physics \& Astronomy, Johns Hopkins University, Baltimore, MD 21218, USA}
\affiliation{Center for Astrophysics $|$ Harvard \& Smithsonian, Cambridge, MA 02138, USA}
\affiliation{Harvard Data Science Initiative, Harvard University, Cambridge, MA 02138, USA}

\author{Alexey Vikhlinin}
\affiliation{Center for Astrophysics $|$ Harvard \& Smithsonian, Cambridge, MA 02138, USA}
\affiliation{Space Research Institute (IKI), Profsoyuznaya 84/32, Moscow, Russia}

\title{The Importance of Being Interpretable: \\ Toward An Understandable Machine Learning Encoder for Galaxy Cluster Cosmology} 

\begin{abstract}
	We present a deep machine learning (ML) approach to constraining cosmological parameters with multi-wavelength observations of galaxy clusters.  The ML approach has two  components: an \new{encoder} that builds a compressed representation of each galaxy cluster and a flexible CNN to estimate the cosmological model from a cluster sample.  It is trained and tested on simulated cluster catalogs built from the \texttt{Magneticum} simulations.  From the simulated catalogs, the ML method estimates the amplitude of matter fluctuations, \sig, at approximately the expected theoretical limit.  More importantly, the deep ML approach can be interpreted.  We lay out three schemes for interpreting the ML technique: a leave-one-out method for assessing cluster importance, an average saliency for evaluating feature importance, and correlations in the terse layer for understanding whether an ML technique can be safely applied to observational data.  These interpretation schemes led to the discovery of a previously unknown self-calibration mode for flux- and volume-limited cluster surveys. We describe this new  mode, which uses the amplitude and peak of the cluster mass PDF as anchors for mass calibration. We introduce the term ``overspecialized'' to describe a common pitfall in astronomical applications of machine learning in which the ML method learns simulation-specific details, and we show how a carefully constructed architecture can be used to check for this source of systematic error.\\
\end{abstract}


\section{Introduction}
\label{sec:intro}

Machine learning can be both a blessing and a curse \textemdash{} the
flexibility of data-driven methods often lead to \mbox{improved}
performance, which can come at the \mbox{expense} of interpretability.  But without
understanding, how can the results be trusted?  In this work, we address
the issue of understanding a deep machine learning method that is
trained to estimate cosmological parameters from simulated catalogs of
galaxy clusters.

Galaxy clusters are systems of hundreds of galaxies and hot gas
embedded in a massive dark matter halo.  Clusters have halo masses
$\gtrsim 10^{14}\, M_\odot$ and they populate the high mass tail of
the distribution of gravitationally bound cosmic structures.  The
number density of galaxy clusters as a function of mass is sensitive to the
underlying cosmological model, particularly to the level of matter
density perturbations, usually expressed as \sig, the amplitude of
linear perturbations at the length scale $8\,h^{-1}\,$Mpc,
approximately corresponding to the cluster mass scale \citep{1990ApJ...351...10F}.

Cosmological constraints with cluster abundance often
proceed with two distinct steps: first, by deriving mass estimates for
each cluster and second, by comparing the population of cluster masses
to theoretically or computationally derived halo mass functions, such
as those of \cite{2008ApJ...688..709T} and \cite{2016MNRAS.456.2361B}. 
The first step of this process, estimating a
cluster mass, can be performed across the electromagnetic spectrum.
In visible wavelengths, masses can be estimated through strong and
weak gravitational lensing
\citep[e.g.,][]{2015MNRAS.449..685H,2014MNRAS.439...48A} or through
galaxy dynamics \citep[e.g.,][]{2016ApJ...819...63R,
  2018A&A...620A...8F}. Observations of the hot intracluster gas in
the X-rays \citep[e.g.][]{2006ApJ...640..691V, 2010MNRAS.406.1759M}, 
as well as in microwave via the Sunyaev-Zeldovich effect
\citep[e.g.][]{2011ApJ...737...61M, 2013ApJ...763..127R}, 
provide cluster mass estimates via the hydrostatic
equilibrium technique.
Although the hydrostatic mass estimates may have systematic biases
\citep{2007ApJ...655...98N}, X-ray properties of the hot gas can be used to
construct a number of low-scatter proxies for the total cluster
mass. Even the simplest X-ray parameter --- the total luminosity ---
correlates with mass with a 25\% scatter 
\citep{2019ApJ...871...50B}. The scatter is expected to be as low as
$5\%-7\%$ for higher-quality proxies such as $Y_{X}$, the product of
the X-ray derived gas mass and temperature
\citep{2006ApJ...650..128K}. The availability of both the low-scatter
mass proxies for individual objects and unbiased \emph{on average}
mass measurements (e.g., via weak lensing) opens a possibility of both
precise and accurate determinations of the cluster mass function and
hence cosmological constraints \citep{2009astro2010S.304V, 2011ARA&A..49..409A}.
This approach is a foundation for a
number of modern cluster cosmology studies 
\citep[e.g.][]{2009ApJ...692.1060V, 2010MNRAS.406.1759M, 
2014A&A...571A..20P, 2019ApJ...878...55B}.

Approaches involving ML are potentially very useful for derivation of
the low-scatter mass proxies from the data and the subsequent
calibration of their absolute mass scale, because this procedure
typically involves complex manipulations of the input data. Indeed,
ML techniques have already been developed to reduce scatter in cluster
mass estimates from optical, X-ray, or multi-wavelength observations
\citep[e.g.,][]{2015ApJ...803...50N, 2016ApJ...831..135N,
  2018arXiv181008430A, 2019MNRAS.tmp.2686C, 2019ApJ...887...25H,
2019ApJ...884...33G, 2019ApJ...876...82N, 2020arXiv200613231H,
  2020arXiv200305951K}.  These methods typically
take advantage of complicated
and subtle signals in the data to reduce scatter and produce tighter
mass estimates. Furthermore, methods have been developed recently to
bypass the cluster mass estimation step and instead obtain
cosmological constraints directly from cluster observables
\citep[e.g.][]{2012PhRvD..86l2005W, 2014arXiv1411.8004H,
  2016MNRAS.462.4117C, 2017A&A...607A.123P, 2017ApJ...835..106N,
  2019ApJ...880..154N}. To our knowledge, such methodology have not
yet been developed to X-ray observables, even though they offer the
highest-quality and most accurate characterization of galaxy clusters
\citep{2011ARA&A..49..409A}. 

Because of their ability to extract complex and subtle signals from
data, deep learning methods are an enticing tool for cluster cosmology.  These
methods are often used for their remarkable improvements over
traditional methods, though the improvement may come at the cost of
lost understanding.  We do not, however, have to resign ourselves to
treating them as black boxes.  Driven in part by the desire to derive
physical intuition from deep methods, interrogation and interpretation
methods for deep learning are an active and important area of research
\citep[e.g.,][]{doshi2017roadmap,2017arXiv170303717S,2018arXiv180607366C,
rudin2019stop,winkler2019association,wu2020regional,
  yu2019classifying}.

We present the first paper in a series to explore how ML interpretability can drive astrophysical understanding. In this paper, we present a deep learning technique that follows a physically
motivated approach to cluster cosmology: first, the X-ray data of each
cluster are compressed into a single number, and second, these
parameters together with weak lensing measurements for the population
of clusters are used to constrain the cosmological model.  The ML
method deviates subtly from a human approach, however, because 
the data compression is flexible and is not explicitly
required to map to cluster mass. Because our machine learning
architecture is physically motivated, it is inherently more
interpretable: we are able to trace how the model depends on input
data, identify trends, and understand how the data are being used. 
We lay out three schemes for learning from the
machine learning model to identify informative trends in the data and present a previously unknown self-calibration technique for cluster surveys that resulted from the interpretation of the ML model.

The paper is organized as follows. We describe the process of building
mock cluster observations in \S\ref{sec:mocks} and the deep learning
model in \S\ref{sec:cnn}.  We present the constraints on \Sig\ in
\S\ref{sec:results}.  We lay out interpretation schemes in
\S\ref{sec:interpretation} and present a discussion and conclusion in
\S\ref{sec:conclusion}.  In Appendix~\ref{sec:cheat}, we present a
pedagogical description of how interrogation schemes can be used to
identify when it would be unsafe to apply a ML technique to
observational data.

\begin{figure*}[t]
	\begin{centering}
		\includegraphics[width=0.8\textwidth]{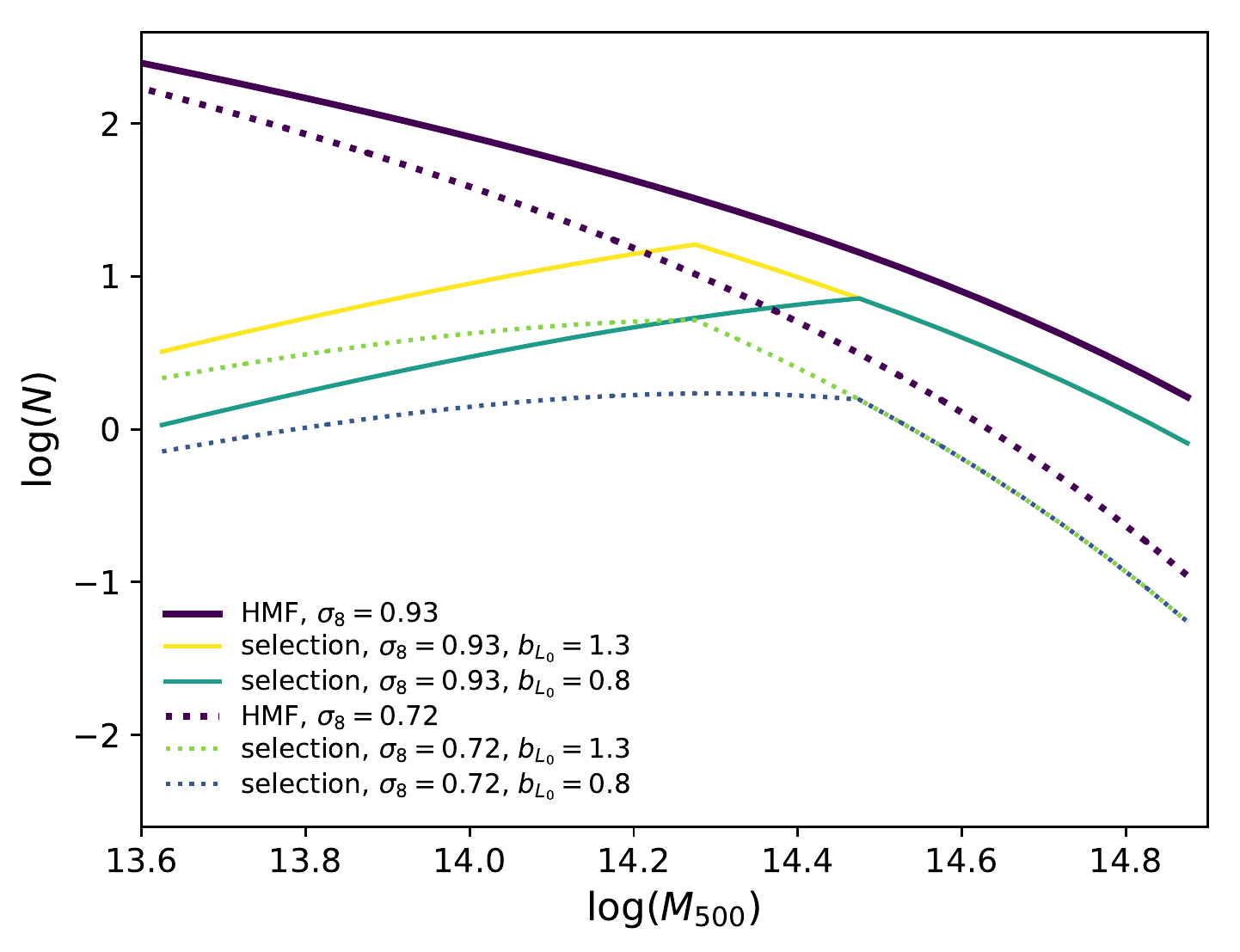}\\
	\end{centering}
	\caption[]{The number of clusters per mass bin, $N$, for $\Sig=0.93$ (purple solid) and $\Sig=0.72$ (purple dotted), for a $(352h^{-1}\,Mpc)^3$ volume, $z=0.07$, and $0.01dex$ logarithmic mass bins.  Realistic cluster abundances for half this volume (yellow, teal, lime, and blue) are also shown.  \new{Cluster abundance depends on \Sig in the volume-limited regime at the high mass end, and also on the details of the mass-luminosity relation (such as the power law bias for the mass-luminosity scaling relation, $b_{L_0}$) in the flux-limited regime at lower masses; the sample \sig and $b_{L_0}$ values shown in this figure are near the extremes of the catalogs created for this study and most catalogs lie in the range of these four selections.}  Note that the effects of \Sig are most pronounced above the transition between these two regimes.  At high masses, increasing \sig also increases the number of massive clusters in the sample.  The details of the mass-luminosity scaling relation determine the mass  at which the flux-limited sample at the lower mass end transitions to a volume-limited sample. \\}
       	\label{fig:selfunc}	
\end{figure*}

\section{Methods:  Cosmological Mock Observations of Galaxy Clusters}
\label{sec:mocks}

In this section, we describe the process of building mock cluster
observations that emulate a realistic observing scenario, span a range
of cosmological parameters, allow for scatter in the observed cluster
parameters, and retain correlated errors among cluster observables. In
\S\ref{sec:clusters}, we describe the process for producing mock
observations of a simulated cluster sample and in
\S\ref{sec:cosmomocks}, we describe the process for building mock
catalogs of clusters. The key data preparation
steps of this section are summarized in Table \ref{table:dataprep}.

\subsection{Simulated Cluster Sample}
\label{sec:clusters}

Mock observations of individual clusters are built from the \texttt{Magneticum}\footnote{\href{www.magneticum.org/}{www.magneticum.org/}} \texttt{Box2} high resolution cosmological hydrodynamical simulation \citep{2015ApJ...812...29T,2016MNRAS.463.1797D,2016MNRAS.456.2361B,2017MNRAS.469..787P,2017Galax...5...49R}.  X-ray mock observations of the simulated clusters are used to infer cluster gas density profiles; the process of producing these profiles from simulated clusters is described below.

\texttt{Magneticum} \texttt{Box2} 
 is a high resolution and cosmological volume
hydrodynamical simulation that models 
cooling and star formation \citep{2003MNRAS.339..289S}, 
black holes and AGN feedback \citep{2005MNRAS.361..776S, 2010MNRAS.401.1670F},
and magnetic fields \citep{2009MNRAS.398.1678D} among others, as described
in \cite{2014MNRAS.444.2938H} and \cite{2017A&C....20...52R}.  
The simulation assumes the WMAP7 cosmological parameters from
\cite{2011ApJS..192...18K}: $\Omega_m=0.226$, $\Omega_b=0.046$,
$\Lambda_0=0.728$, $h=0.704$, $n=0.963$, and $\sigma_8=0.809$
and it has a box side length of $352\Mpch$.

The lowest-redshift output, $z=0.07$ has 1165 unique clusters with
masses $M_{500}\gtrsim10^{13.58}\,\Msolarh$\footnote{Throughout,
  cluster masses are defined at the enclosed sperical overdensity
  threshold of 500 relative to the critical density.} The cluster
catalogs constructed from the simulation tabulate total cluster masses
($M_{500}$), X-ray luminosities ($L_X$), mass-weighted temperatures
($T$), and masses of the hot gas ($M_\mathrm{gas}$).  
\cite{2016MNRAS.456.2361B} models the cluster mass 
function of the simulation. 

\subsubsection{Survey Selection Function}
\label{sec:selection1}

Here we describe the pure observational selection.
Typical X-ray surveys do not result in volume- and mass-limited
cluster samples. Instead, the observational selection is typically in
the form of X-ray flux or luminosity cuts 
\citep{2004A&A...425..367B, 2007A&A...469..363B}. We use a simplified approach
to model an X-ray survey selection applied to the \texttt{Magneticum}
clusters. The approach emulates the selection function for a survey that
is both X-ray flux-limited and implements an upper and lower redshift
boundaries, such as the lower-$z$ part of the
\cite{2009ApJ...692.1033V} sample (c.f.\ their Fig.~14). The selection
function discussed below is defined as a ratio of the search volume for a
cluster with mass $M$ to the total geometric volume between
$z_{\text{min}}$ and $z_{\text{max}}$:
\begin{equation}
  \label{eq:sel:obs}
  S_{\text{obs}}(M) = V_{\text{obs}}(M) / V_{\text{tot}}.
\end{equation}

For the highest-mass clusters, $V_{\text{obs}}(M) =1$, meaning that
such clusters are well above the flux threshold at
$z=z_{\text{max}}$. For $M$ below some threshold value, the clusters
are no longer detectable at the redshift $z_{\text{max}}$. The
selection function in this regime takes a power-law form,
$V_{\text{obs}}(M) \propto M^{A}$ where the slope $A$ reflects a
Eucledian search volume, $V_{\text{obs}}(L_{X})\propto L_{X}^{3/2}$,
and an observed power-law relation between the cluster mass and
luminosity. We use $A=2.4$, appropriate for the observed slope of the
$L_{X}-M$ relation from \cite{2009ApJ...692.1033V}. 

The transition between the $V_{\text{obs}}(M) =1$ and
$V_{\text{obs}}(M) \propto M^{A}$ regimes should be smooth, reflecting
the substantial scatter in the $L_{X}-M$ scaling relation \citep[c.f.\ Fig.~14
in][]{2009ApJ...692.1033V}. We neglect this and assume that the
selection function changes from the power law to 1 abruptly at
$M=M_{\text{break}}=10^{14.4}\,\Msun$. This value of $M_{\text{break}}$
corresponds to the same X-ray flux limit as in the
\cite{2009ApJ...692.1033V} sample, but a lower value of the maximum
redshift, $z_{\text{max}}=0.07$, chosen such that the total
survey volume corresponds to 1/2 of the \texttt{Magneticum} simulation box.

At the low-mass end, we introduce a cutoff at
$M_{\text{min}}=10^{13.6}\,M_{\odot}$. Such a cutoff approximates
an effect from combination of an X-ray flux threshold and a minimal
redshift boundary in realistic surveys \citep[see, e.g., Fig.~14
in][]{2009ApJ...692.1033V}. The threshold we use would correspond
approximately to $z_{\text{min}}=0.01$ for the X-ray flux
threshold used in the selection of the \cite{2009ApJ...692.1033V}
sample.


When later in the course of this work we simulate samples with a
different $L_{X}-M$ relation (Section \ref{sec:YLTscatter} below), 
we appropriately
scale the value of $M_{\text{break}}$, but we keep the slope of the
$V_{\text{obs}}(M) \propto M^{A}$ branch at $A=2.4$\footnote{This choice, as
well as using the piece-wise power law approximation to
$V_{\text{obs}}(M)$, does not improve or significantly change the quality of the cosmological constraints presented below in \S\ref{sec:results}.}.

With the selection function described above, the application of the
observational selection to \texttt{Magneticum} clusters is straightforward. For
each cluster with mass $M$, we simulate a random number $0<x<1$, and
if $x<S_{\text{obs}}(M)$, the cluster is added to the input catalog.

\subsubsection{Mock X-ray Data}
\label{sec:selection2}{
For the \texttt{Magneticum} clusters, we generated mock X-ray
observations via \texttt{Magneticum}'s public implementation of the
\texttt{PHOX} algorithm \citep{2012MNRAS.420.3545B,
  2013MNRAS.428.1395B}.  This algorithm models ICM thermal emission
from simulated gas particles, simulating a large number of photons.
These are redshifted, projected into the plane of the sky, and
subselected for the instrument field of view (FoV) and observing time ($t_\mathrm{obs}$).  Further
details of the \texttt{Magneticum} implementation of \texttt{PHOX} can
be found in the publicly available \texttt{Magneticum} Cosmological Web Portal\footnote{\href{https://c2papcosmosim.uc.lrz.de/}{https://c2papcosmosim.uc.lrz.de/}}
and in \cite{2017A&C....20...52R}. 

Mock
data were generated for a generic instrument with a flat effective area,
$A=600\,\mathrm{cm}^2$, in the $0.5-2.0\,$keV energy band, a square
$42'\times42'$ field of view, $6''$ pixels, and in the idealized
regime of no background and infinitely sharp angular resolution. The
simulated exposure time is $t_\mathrm{obs}=100\,\mathrm{ks}$.  X-ray
emission from correlated structure within $\pm50\,\mathrm{Mpc}$ along
the line of sight are included in the mock observations, but active
galactic nuclei (AGN) are not, assuming that they can be easily
removed in preprocessing, as is indeed routinely possible with
high-resolution X-ray telescopes such as \emph{Chandra}.

\begin{figure}[]
	\begin{centering}
		\includegraphics[width=1.0\linewidth]{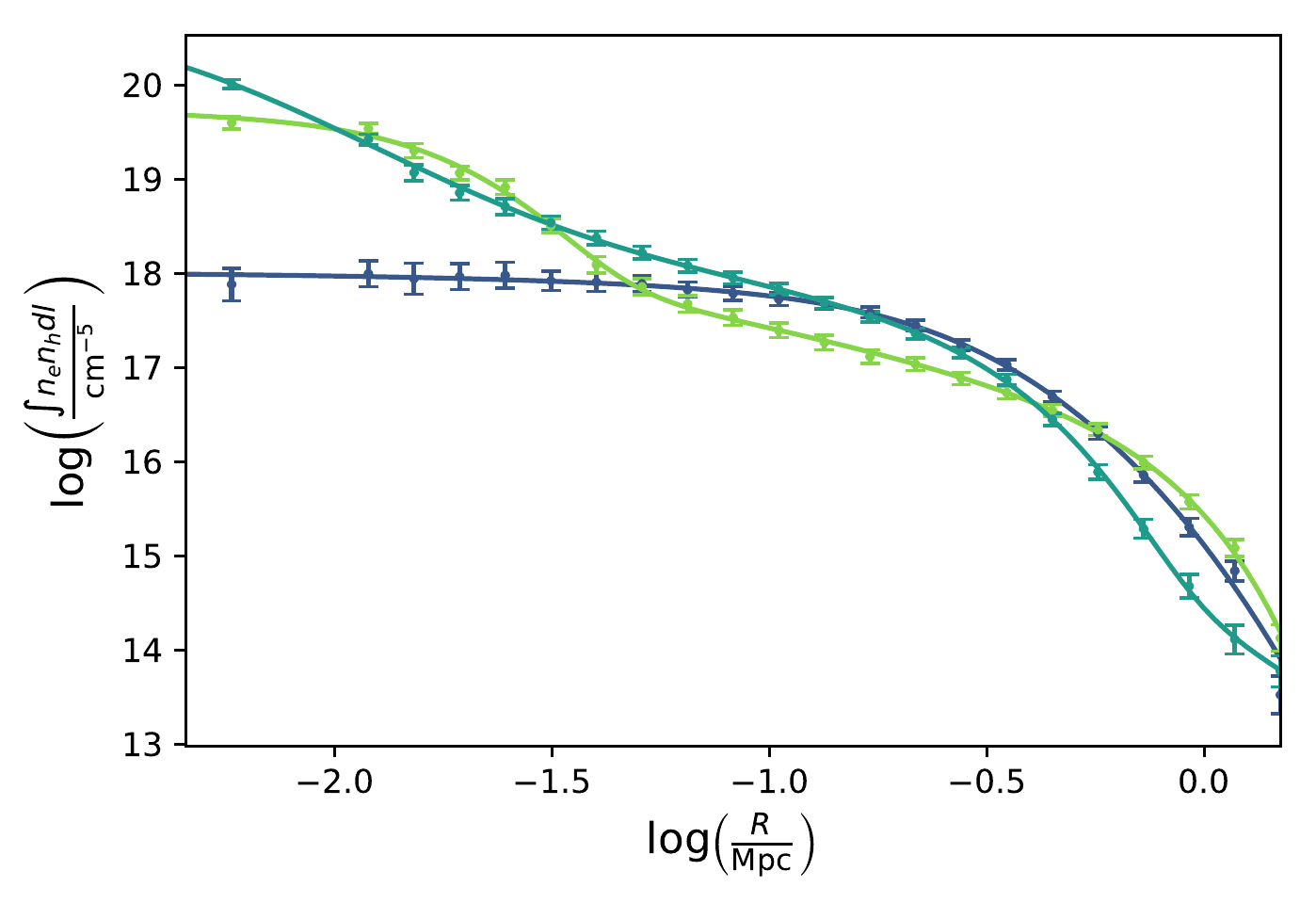}\\
	\end{centering}
	\caption[]{{Density-squared profiles of three representative clusters (points with Poisson error bars).  These mock observations are fit to a model that has flexibility to capture a range of shapes, including a cluster with a flat core (dark blue), a cluster with a peaked core (teal blue), and a cluster with a cool core (lime green).}}
       	\label{fig:sampleprof}	
\end{figure}

Mock data were reduced following typical analysis procedure employed
in the analysis of real observations (see \cite{2009ApJ...692.1033V} and 
 \cite{2007ApJ...655...98N} for reference).  
 Substructures within each cluster were masked using
the wavelet decomposition algorithm \citep{1998ApJ...502..558V}.
Under the assumption of spherical symmetry, the projected X-ray
brightness profiles are fit to a flexible model
\citep[c.f.][]{2006ApJ...640..691V}, as shown in Figure
\ref{fig:sampleprof}. The fit provides a 3D run of the gas density
which in turn can be used to obtain the integrated gas mass. Our
mock data include the Poisson noise, but the statistical errors on the
derived $M_\mathrm{gas,\,total}$ values are $\lesssim 2.5\%$ across
the cluster mass range.
  
A typical cosmological analysis employs one or more cluster parameters (e.g., cluster temperature, X-ray luminosity,  inferred gas mass, or weak lensing mass) to estimate the masses for each cluster in a sample.  Because deep learning is an inherently flexible approach, it can learn from noisy and complex data sets and this simplification is not necessary.  This allows us to omit the simplifying step of inferring masses from these observables and, instead, the full data set can be included as input.  Our mock observations will be reduced to several parameters \textemdash{} temperature, luminosity, gas mass in spherical radial shells, gas density slope, and weak lensing mass \textemdash{} and the deep learning model will be allowed to identify the informative and meaningful parameters in this list.  

\subsubsection{Mock Weak Lensing Data}
\label{sec:selection3}

Weak lensing (WL) measurements are essential for obtaining an unbiased
normalization of the scaling relations between observables and cluster\
mass. In our model, we assume that all selected clusters will have WL
measurements. The public interface to the \texttt{Magneticum} output does not provide
an option of raytracing via full light cones to obtain the lensing
sheer maps, as done, e.g., in \cite{2011ApJ...740...25B}. Instead,
we implement a much more simplified algorithm based on the results
from the Becker \& Kravtsov work.

We assume that the weak lensing mass measurements of galaxy clusters
are almost unbiased \citep[see,
e.g.,][]{2011ApJ...740...25B,2014MNRAS.439...48A,2015MNRAS.449..685H},
but have a significant scatter for individual objects.   The scatter
cannot be calibrated with a sufficient accuracy a priori because it
depends on many factors, such as the unavoidable line-of-sight
projection effects \citep{2011ApJ...740...25B} and observational
uncertainties including statistics of background galaxies, their
estimated redshifts,
etc.~\citep{2014MNRAS.439...48A,2015MNRAS.449..685H}. We can, however,
expect that it is in the neighborhood of $\sim
20-30\%$~\citep{2011ApJ...740...25B} and we make the simplifying assumption 
that the weak lensing mass measurements are 
 uncorrelated with cluster X-ray observables.

To reflect these expected properties of the weak lensing mass
estimates, we use the following procedure. For each cluster, the weak
lensing mass $M_\mathrm{WL}$ is selected according to
\begin{equation}
	\log(M_\mathrm{WL}) = \log\left(M_{500} \right) \pm \delta_\mathrm{WL}, 
	\label{eq:WLscatter}
\end{equation}
where $M_{500}$ is the true cluster mass.  
For each realization of the mock catalog (see \S\ref{sec:cosmomocks}
below), the weak lensing scatter $\sigma_\mathrm{WL}$ is selected
randomly between $10$ and $30\%$ with flat priors.  For each cluster
within a mock catalog, $\delta_\mathrm{WL}$ is selected randomly
from a Gaussian distribution with variance $\sigma_\mathrm{WL}$.

\begin{table*}[t] 
\caption{Data Preparation and Galaxy Cluster Properties}
\label{table:dataprep}
\begin{center}
\begin{tabular}{| R{1.5cm} | R{3cm} | R{3cm} | R{2.75cm} | R{3cm} | R{2.75cm} |}
\hline
\hline
Property & Definition & Property is derived from$\dots$ & Scatter is introduced by$\dots$ & Property is used$\dots$ & Comments \\
\hline
\hline
$N_\mathrm{cluster}$		
	& number of clusters in a simulated catalog 
	& counting clusters under the selection function. 
	& implementing a random draw to populate each cluster mass bin according to the desired selection function.
	& implicitly.  This parameter is not explicitly included in the input data, but it sets the size of the input data.
	& The selection function varies with \Sig and the details of the $M-L$ scaling relation; see Figure  \ref{fig:selfunc} and \S\ref{sec:cosmomocks}. \\
\hline
$M_\mathrm{500}$		
	& cluster mass, \linebreak $\Delta=500\rho_\mathrm{crit}$
	& simulation.
	& Not applicable.
	& This property is not used in the input data.
	& \\
\hline
$M_\mathrm{WL}$		
	& weak lensing mass, $\Delta=500\rho_\mathrm{crit}$
	& true mass with 20\%-40\% lognormal scatter.
	& selecting a lognormal scatter size for each mock cluster sample.
	& in the supplementary input (inserted at the terse layer).
	& See Equation \ref{eq:WLscatter}. \\
\hline
$L_X$		
	& $X$-ray luminosity
	& simulation.
	& adopting $\delta_L^\star$ from a nearby neighbor.
	& in the supplementary input (inserted at the terse layer) and also for sorting clusters.
	& See table note\tablenotemark{a}, ($X=L_X$). \\
\hline
$T$		
	& mass-weighted temperature
	& simulation.
	& adopting $\delta_T^\star$ from the same nearby neighbor.
	& in the 2D input.
	& See table note\tablenotemark{a}, ($X=T$). \\
\hline
$M_\mathrm{gas}$		
	& total simulated gas mass
	& simulation.
	& adopting $\delta_{M_\mathrm{gas}}^\star$ from the same nearby neighbor.
	&  implicitly.  This property is not used in the input data, but is used to scale inferred gas mass profiles.
	& See \S\ref{sec:cosmomocks} and table note\tablenotemark{a}, ($X=M_\mathrm{gas}$). \\
\hline
$M_\mathrm{gas}(r)$		
	& inferred gas mass profile
	& mock observation by fitting luminosity profile, inferring a gas density profile, and integrating in logarithmically spaced radius bins to $R_{500}$.
	& adopting gas mass profile from the same nearby neighbor and scaling according the the ratio of total gas masses.
	&  in the 2D input.
	& See \S\ref{sec:cosmomocks}. \\
\hline
$\frac{d\log\left[\rho_\mathrm{gas}(r)\right]}{d\log r}$		
	& inferred gas density slope profile, $\gamma$
	& mock observation by fitting luminosity profile and inferring a gas density profile.
	& adopting gas density profile from the same nearby neighbor.
	& in the 2D input.
	& See \S\ref{sec:cosmomocks}. \\
\hline
\end{tabular}
\end{center}
\tablenotetext{a}
{Uncertainties in the luminosity ($L_X$) temperature ($T$), and gas mass ($M_\mathrm{gas}$) power law scaling relations are modeled by varying the normalization, slope, and intrinsic scatter of these power law relationships.  The baseline scaling relation is derived from the simulation; a power law is fit to the cluster mass-observable relation of the simulated clusters, according to
\begin{equation}
	\log(X_\mathrm{fit}) = \alpha_x \log\left(\frac{M^\star_\mathrm{500}}{10^{14.4}\,h^{-1}\,M_\odot}\right) + \log(X_0), \nonumber
	\label{eq:PL}
\end{equation}
where $X$ is a placeholder variable for any parameter (e.g., $L_X$, $T$, or $M_\mathrm{gas}$) that scales with cluster mass as a power law.  This baseline fit is for the full cluster sample at the \texttt{Magneticum} WMAP7 fiducial cosmology.  See \S\ref{sec:YLTscatter} for further details.}
\end{table*}

\subsection{Mock Catalogs}
\label{sec:cosmomocks}

The \texttt{Magneticum} simulation was run at a single (WMAP-7) cosmology. The
simulated clusters follow a specific set of scaling relations between the
cluster mass and observables representing a set of physical
assumptions built into the numerical model. We need a mechanism to
replicate this catalog for different cosmologies (\Sig) and also different
scaling relations representing a plausible range of uncertainties in
the physical model of the intracluster gas. Changes in cosmology are implemented 
as changes in the cluster mass function and selection function, while changes 
in the scaling relations are achieved by varying the assigned parameters for fixed cluster mass.

\subsubsection{Varying Cosmology}
\label{sec:selfunc}

For our purposes, the main effect of the \Sig\ variation on the cluster
population is the corresponding change in the cluster mass function,
the number density of objects at a given mass. We implement this by
modifying the observational selection function (eq.~\ref{eq:sel:obs})
by the ratio of the model mass function computed for the given
cosmology and for the ``reference'' combination of parameters used in
the \texttt{Magneticum} simulation,
\begin{equation}
  \label{eq:sel:cosm}
    S = S_{\text{obs}}(M) \times n(M) / n_{\text{ref}}(M),
\end{equation}
where $S_{\text{obs}}(M)$ is given by eq.~(\ref{eq:sel:obs}) and
$n(M)$ is the differential cluster mass function model \citep[we use
the \texttt{COLOSSUS} implementation, described in][]{2018ApJS..239...35D}. 

In reality, changes in \Sig\ affect the cluster merger trees, and thus
affect the clustering properties of clusters, and possibly the scaling
mass-observable scaling relations. Our procedure ignores these
changes. This is justified because we operate with
sample sizes much smaller than those useful for the cluster
clustering measurements  and
thus ignore this information in the \Sig\ constraints. The explicit
effect of \Sig\ on the cluster scaling relations is expected to be
small \citep[e.g.,][]{Evrard:2008aa, 2019arXiv191105751S}, and can also
be ignored.

\subsubsection{Varying the Luminosity, Temperature, and Gas Mass Scaling Relations}
\label{sec:YLTscatter}

The detailed state of the clusters' baryonic component are not
uniquely predicted from first principles.  And while the training
simulation adopts a realistic choice of subgrid physics, a single
choice does not capture the range of subgrid physics models and the
variety of clusters that might result from them.  Numerical models are
not sufficiently robust, particularly for $L_X$ predictions, and so we
need to include a plausible range of uncertainties for modeling
cluster baryons.

This uncertainty is added by varying the $L_{X}$, $M_\mathrm{gas}$,
and $T$ power law normalizations, slopes, and scatters for each
realization.  Specifically, we randomly vary the normalizations and
scatter by $\pm30\%$ relative to the baseline values established by
the power-law fit to the full Magneticum sample. \new{The 30\%
  level is not based on any observational or numerical
  uncertainties. We simply chose a range that is sufficiently wide
  such that the accuracy of the \Sig\ reconstruction in most cases is
  not driven by the priors on the scaling relations for the Magneticum
  clusters (c.f.\ the results reported in \S\ref{sec:Ncl}, where the \Sig\
  constraints are almost entirely dependent on the scaling relation
  priors).}
The nominal values
for these power law fits are given in  Table~\ref{table:scaling}, and
are in good agreement with observations
\citep[e.g.,][]{2019arXiv191105751S}.  In addition, we vary the power
law slope in the $M-L_{X}$ relation by 30\% because this relation is
the one most affected by uncertainties in the numerical models.

This level of additional variations ensures that the accuracy of cosmological constraints does not depend on a particular set of scaling relations produced by the Magneticum simulations. For example, because of this additional added uncertainty in the power law scaling relations, the cluster masses cannot be inferred sufficiently accurately from $L_{X}$ or $T$. The absolute cluster mass scale in our analysis is established entirely using weak lensing observables, or via self-calibration (see below, \S\ref{sec:nowlresults}) --- just as it is done in the analysis of real cluster samples.

\begin{deluxetable}{l l r r r r r}
\tablecaption{Simulation power law best fit parameters.}
\label{table:scaling}
\tablehead{
\colhead{Item} & \colhead{Units} &\colhead{$\alpha_x$} & \colhead{$\log(X_0)$}  & \colhead{$\sigma_{X_0}$} 
}
\startdata
$\log(L_x)$				& $10^{44}$\textit{erg}/s 	& $1.74$ 		& $-24.81$ 	& $0.13$ \\
$\log(T)$					& $keV$ 				& $0.61 $		& $-8.25$ 		& $0.03$ \\
$\log(M_\mathrm{gas})$		& $\Msolarh$ 			& $1.16$ 		& $-3.32$ 		& $0.10$ \\
\enddata
\end{deluxetable}

While we ultimately do not use Magneticum
   $L_{X}$, $M_{\text{gas}}$, and $T$ outputs directly, we
  preserve the simulation's guidance on the details of the scatter of
  these parameters from the nominal relations and covariances and correlations among
  them. 
This is done with the goal of moving the analysis beyond two common simplifying assumptions:  1.~the assumption that scatter in mass-observable relationships is lognormal\footnote{Some observable properties are well-described by lognormal scatter \citep[e.g.,][]{2018MNRAS.478.2618F} and standard statistical approaches \citep[e.g.,][]{2014MNRAS.441.3562E} build on this premise to provide flexible, interpretable analytic models.  Recent work expands on this canonical approach to account for deviations from log-normality \citep[e.g.,][]{2011ARA&A..49..409A,2019ApJ...880..154N,2020MNRAS.495..686A}, showing the benefit of a more nuanced treatment of scatter in cosmological analyses of cluster samples.} and 2.~the assumption that errors among cluster observables are uncorrelated.
 Note that we do not
  anticipate that the Magneticum simulation fully realistically reproduces all
  details of the scatter in the cluster scaling relations. This
  uncertainty is approximated by variation of the mean levels of
  scatter, implemented as described
  below.

    \begin{figure}[]
  \centering
   \includegraphics[width=0.45\textwidth]{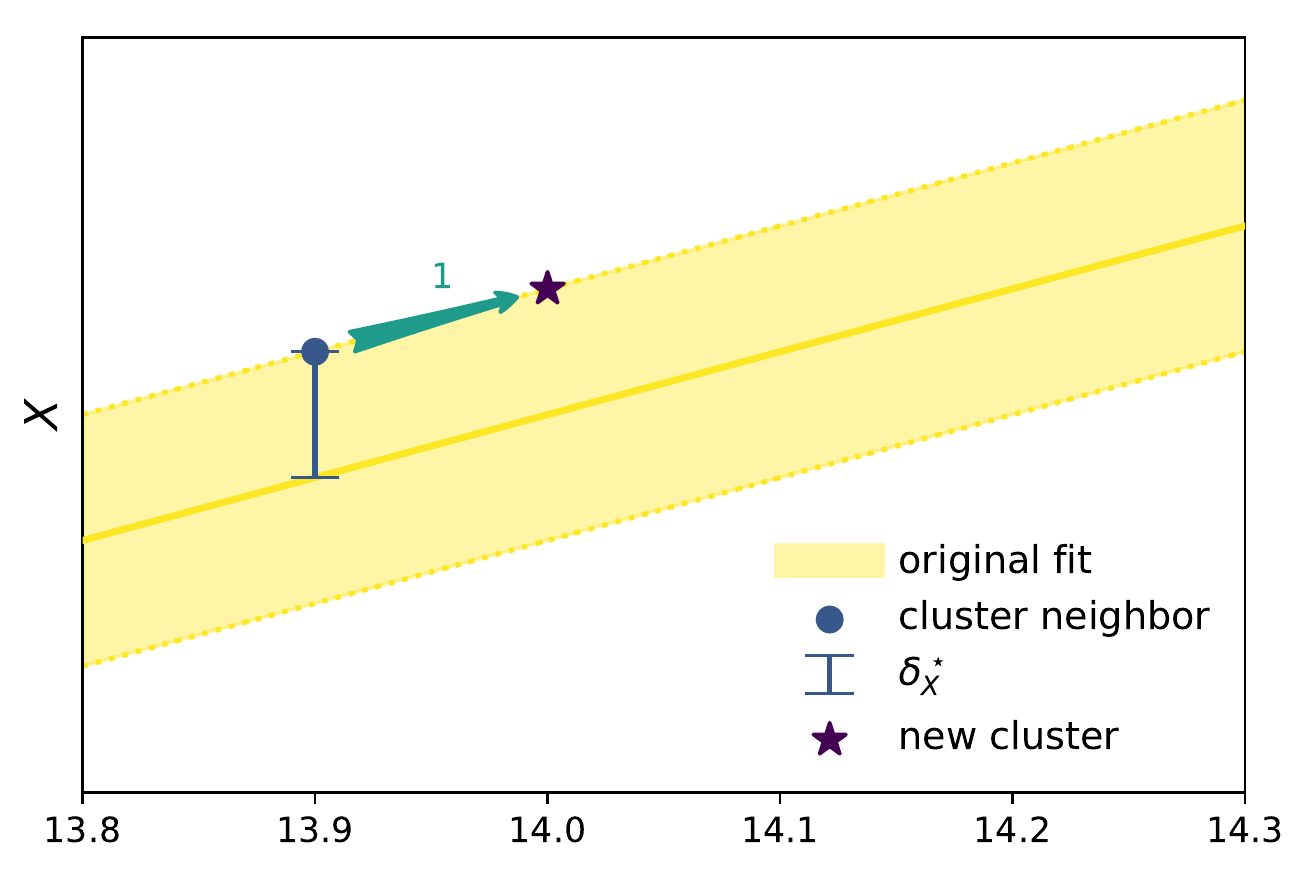}
   \includegraphics[width=0.45\textwidth]{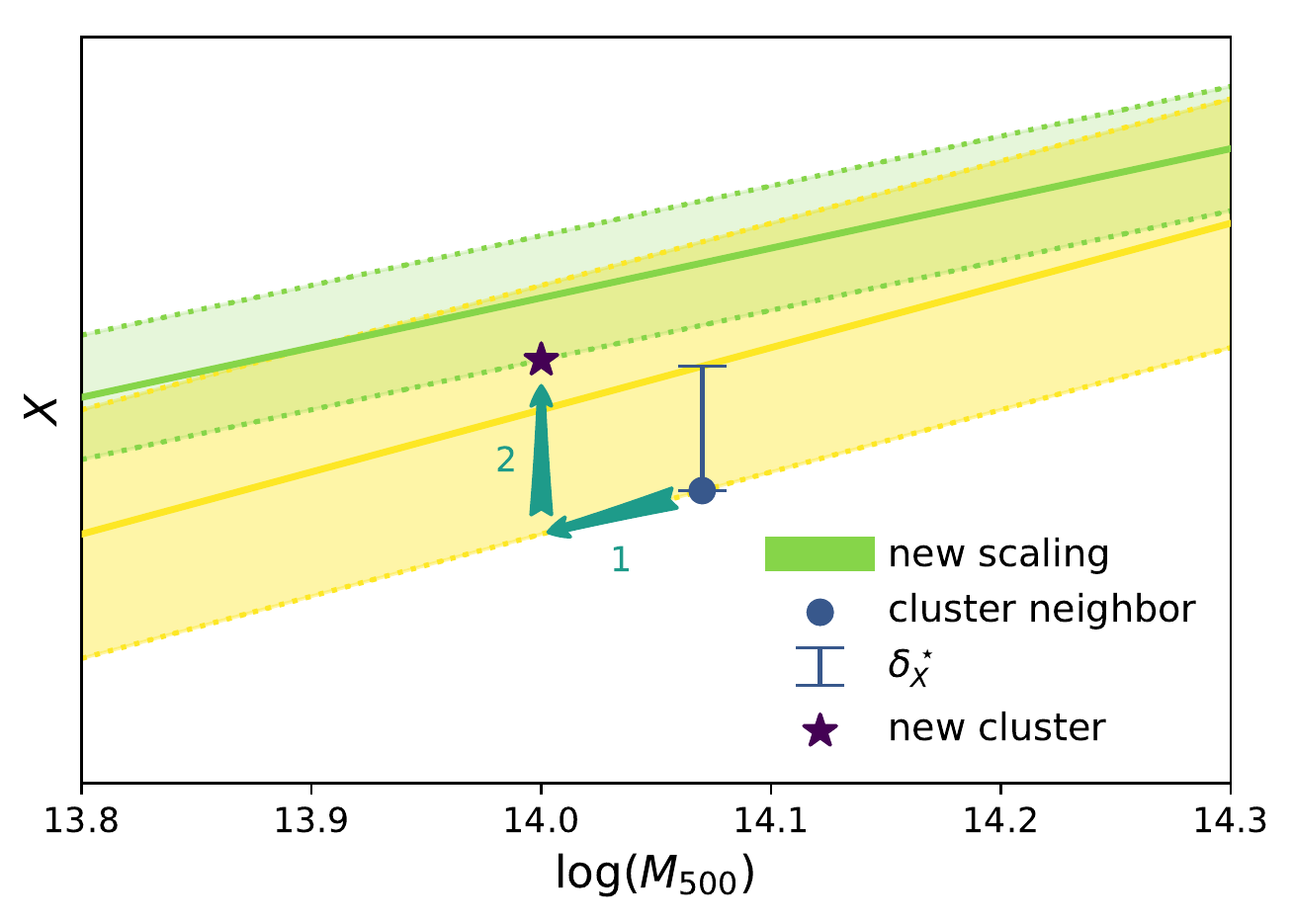}
    \caption[]{{Example of introducing scatter to a $M_{500}=10^{14}\msolarh$ cluster.  Top:  a power law with scatter (yellow band) describes the $M-X$ power law for the simulated galaxy clusters.  $X$ is a placeholder for $L_X$, $T$, or $\Mgas$.  A \new{simulated cluster with a mass close to the target mass} (blue dot) is identified \new{in the simulation} and the offset from the best fit power law ($\delta_X^\star$, blue bar) is \new{measured from the simulation data}.  This offset is translated (teal arrow 1) to the target mass, resulting in a new\new{, realistic} cluster (purple star).  \new{This new cluster is effectively a mass-shifted version of a cluster in the \texttt{Magneticum} simulation.} Bottom:  a more complex example of translating from the nearby neighbor to the new simulated cluster.  Here, the scaling relation has changed in slope, amplitude, and scatter. The nearby neighbor's offset from the best fit power law ($\delta_X^\star$, blue bar) is translated to the target mass (teal arrow 1) and to the target scaling relation (teal arrow 2).} }
\label{fig:introscatter}
\end{figure}
  
The following process for introducing scatter in these power law scaling relations is summarized  in Figure \ref{fig:introscatter} and Table \ref{table:dataprep} 
\new{and is motivated by two considerations:  1. We want to preserve realistic correlations among cluster observables from simulations.  For example, it is known that clusters with higher-than-expected temperatures will tend to have lower-than-expected gas masses \citep{2006ApJ...650..128K}.  However, gas mass profiles are not well-described by theoretical models and because of this,  theoretical guidelines for making realistic mock profiles are severely limited.  2. We want to build many realistic mock cluster samples without re-using identical cluster data.}  New values for the seven parameters are selected:  $L_X$ normalization, $T$ normalization, $\Mgas$ normalization, $L_X$ scatter, $T$ scatter, $\Mgas$ scatter, and $L_X$ slope.  These new values are $\pm30\%$ of the fiducial simulated values and a new choice of each parameter is made for each of the simulated cluster samples.  The normalization and slope parameters adjust the underlying power law scaling relation and the scatter parameters adjust the intrinsic scatter of clusters around the baseline power law scaling relation.

In implementing the scatter directly from Magneticum outputs, we need
to ensure that the generated mock catalogs are suitable for training
and testing a deep learning model. If identical combinations of
cluster parameters are repeated many times in the mock catalogs, this
can drive the machine towards an unfair and unphysical handling of the
data set by memorizing clusters and exploiting this information.  To
address this potential concern, we select a set of $L_X$, $T$, and
$M_{\mathrm{gas}}$ offsets for each simulated cluster randomly from a
neighbor with
$|\log(M_\mathrm{500})-\log(M_\mathrm{neighbor})|\leq0.1$ (as
illustrated in Figure \ref{fig:introscatter}).  The gas mass profiles are scaled by the ratio of the target cluster gas mass to the simulated cluster gas mass.

\new{In all, we vary 8 parameters.  Four are sampled on a grid (\sig between 0.7 and 0.92 with $\Delta\sig=0.01$, and also power law biases for $L_X$, $T$, and
$M_{\mathrm{gas}}$ each independently between 0.8 and 1.3, with $\Delta=0.1$), and four parameters are selected randomly (weak lensing scatter and also power law biases for $L_X$, $T$, and
$M_{\mathrm{gas}}$, each of which is sleected randomly between 0.7 and 1.3 with flat priors). The process produces 4,968 simulated cluster samples.} The process of introducing scatter in our mock catalog is borne from necessity; 
large suites of cosmological volume hydrodynamical simulations for building 
appropriate mock X-ray observations of galaxy clusters at many cosmologies and for many subgrid physics models
are not yet available.  However, in Section \ref{sec:tersecorr}, we will 
show that this approach does, indeed, make it possible to train a model that generalizes.

\section{Methods:  Machine Learning Model}
\label{sec:cnn}

\begin{figure}[]
  \centering
   \includegraphics[width=0.45\textwidth]{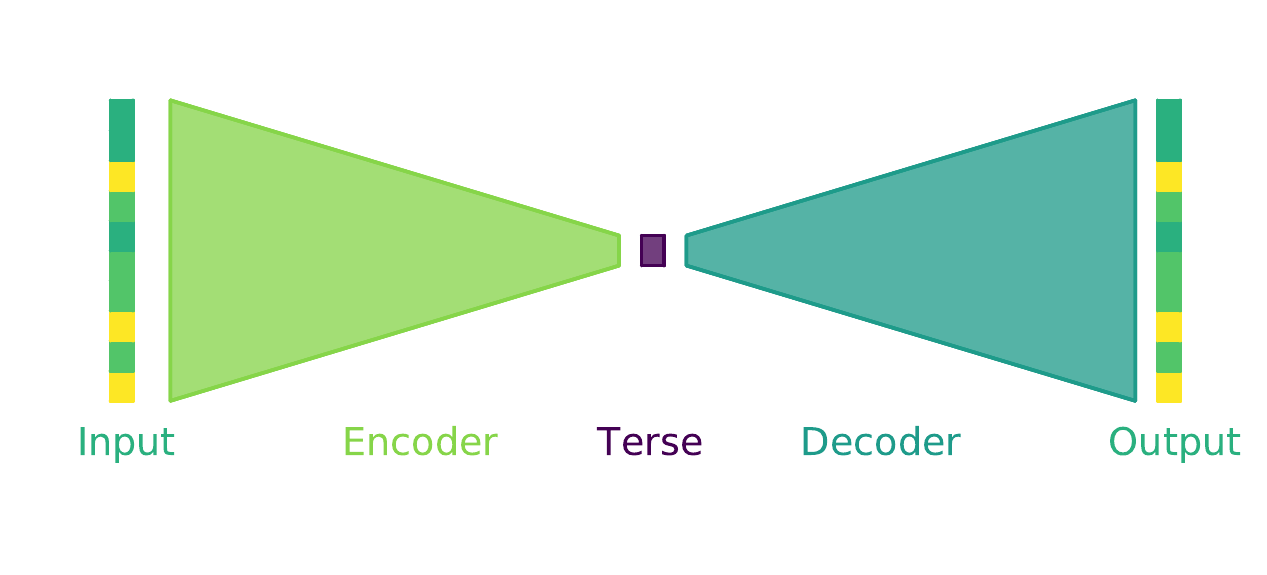}
    \caption[]{An autoencoder is a machine learning technique that summarizes a signal.  In this pedagogical example, the input signal (left) passes through the encoder (lime green) to a narrow bottleneck  or \textit{terse} space (purple).  The terse values are used by the decoder (teal) to reconstruct some approximation of the original signal (right).  Because labeled training data are unnecessary,  this is an example of an \textit{unsupervised} ML technique.  Traditionally, the bottleneck layer (purple) is referred to as the \textit{latent} layer to emphasize that it unseen.  In this work, we will explicitly view, analyze, and interpret this space.  A change in terminology is needed to illuminate the use of the bottleneck layer, and so we refer to this layer as the ``terse'' space.}
\label{fig:AE}
\end{figure}
  
Our aim is to estimate \sig from simulated observations of a population of galaxy clusters.  The function mapping a list of cluster observables (such as luminosity, gas mass, and temperature) to the underlying cosmology (\sig) is a complex and nonlinear function, and because of this, a flexible machine learning model is a good choice for the task at hand.  One subset of machine learning approaches that are particularly adept at extracting complicated patterns from data are deep learning models. 

{Deep learning is a class of machine learning (ML) algorithms that
  find patterns in data by processing it through multiple unseen
  layers.  The unseen (or \textit{latent}) layers encode and exploit
  increasingly complex patterns in the data.  Convolutional Neural
  Networks \citep[CNNs, ][]{fukushima1982neocognitron,
    lecun1999object, NIPS2012_4824} are deep learning class of
  algorithms that learn a series of filters, weights, and biases to
  extract meaningful patterns from an input image.  For astronomical
  applications, CNNs can be applied to natural images \citep{2015MNRAS.450.1441D} or can
  be applied to data cast into \new{a two- or three-dimensional array} by, for example, applying a
  smoothing filter to a point cloud of data
  \citep[e.g.,][]{2019ApJ...887...25H, 2020arXiv200613231H} or by
  exploiting symmetries in the data
  \citep[e.g.,][]{2018arXiv181011030P, 2020arXiv200705144S}.}

{Autoencoders \citep{10.5555/104279.104293} are a separate class
  of machine learning algorithms that are sometimes built within the
  CNN framework.  The fundamental component that differentiates deep
  autoencoders from other deep methods is an explicit encoding or
  compression of input data.  Autoencoders have been applied in
  astronomy for a range of tasks, including generating realistic
  galaxy images \citep{2016arXiv160905796R}, denoising gravitational
  waves \citep{2017arXiv171109919S}, classifying supernovae
 \citep{2020ApJ...905...94V}, and generating realistic SZ mock images of 
 galaxy clusters \citep{2021arXiv211002232R}.  A traditional application of an autoencoder
  can be thought of as a flexible version of principle component
  analysis, summarizing a complicated signal in a few essential values
  from which an approximation of the original signal can be
  reconstructed.  Autoencoders typically include both encoding layers
  to summarize the signal, and also decoding layers to recreate (an
  approximation of) the original signal from the encoding.  Because
  the input signal is also the output of an autoencoder, this is an
  example of \textit{unsupervised} learning, where training labels are
  not necessary.  A pedagogical autoencoder architecture is shown in
  Figure \ref{fig:AE}.}

\subsection{Architecture}
\label{sec:architecture}

\begin{figure*}[]
  	\centering
		\includegraphics[width=0.9\textwidth]{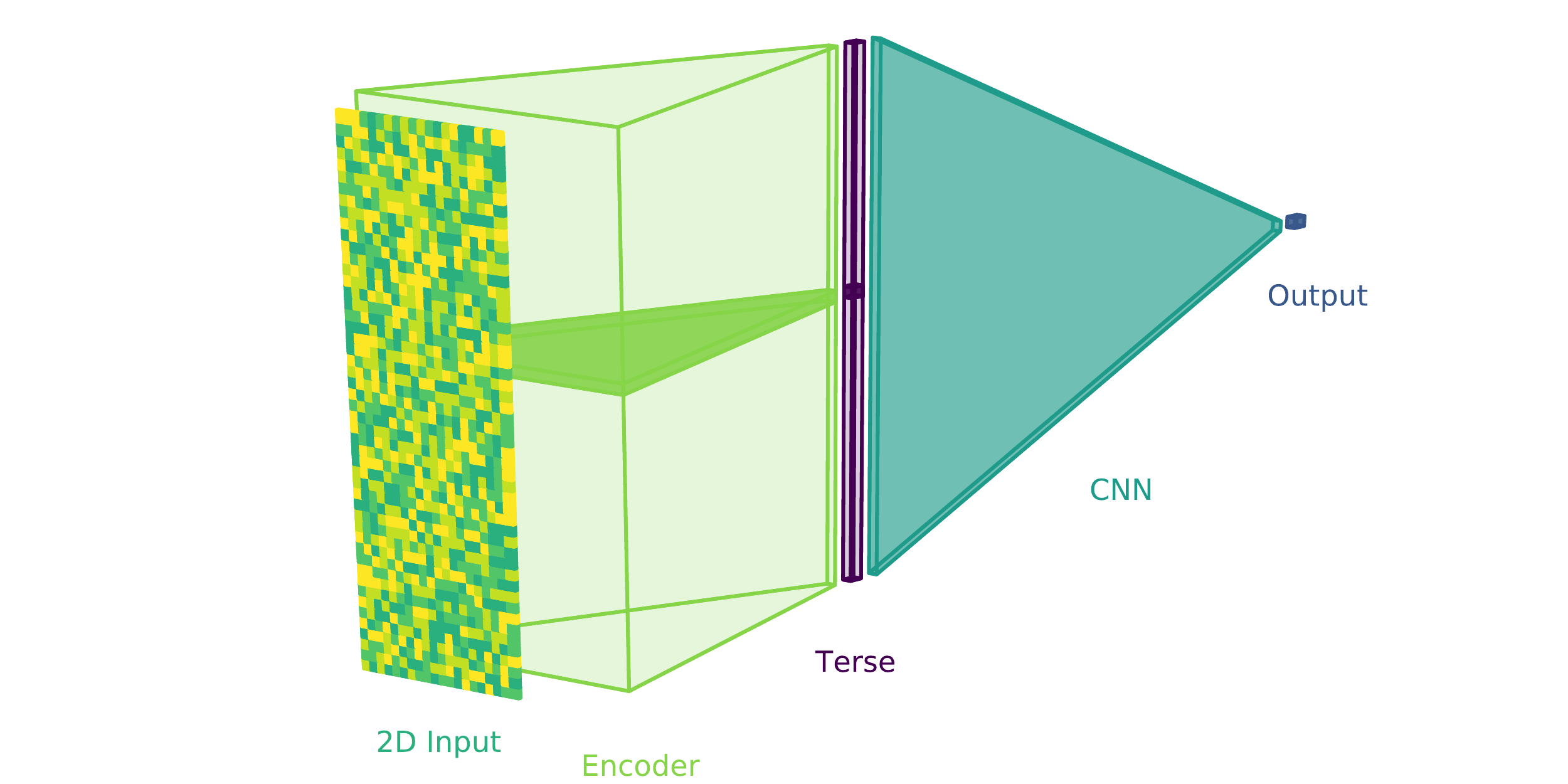}\\[5ex]
		\begin{tabular}{c c}
			\includegraphics[width=0.45\textwidth]{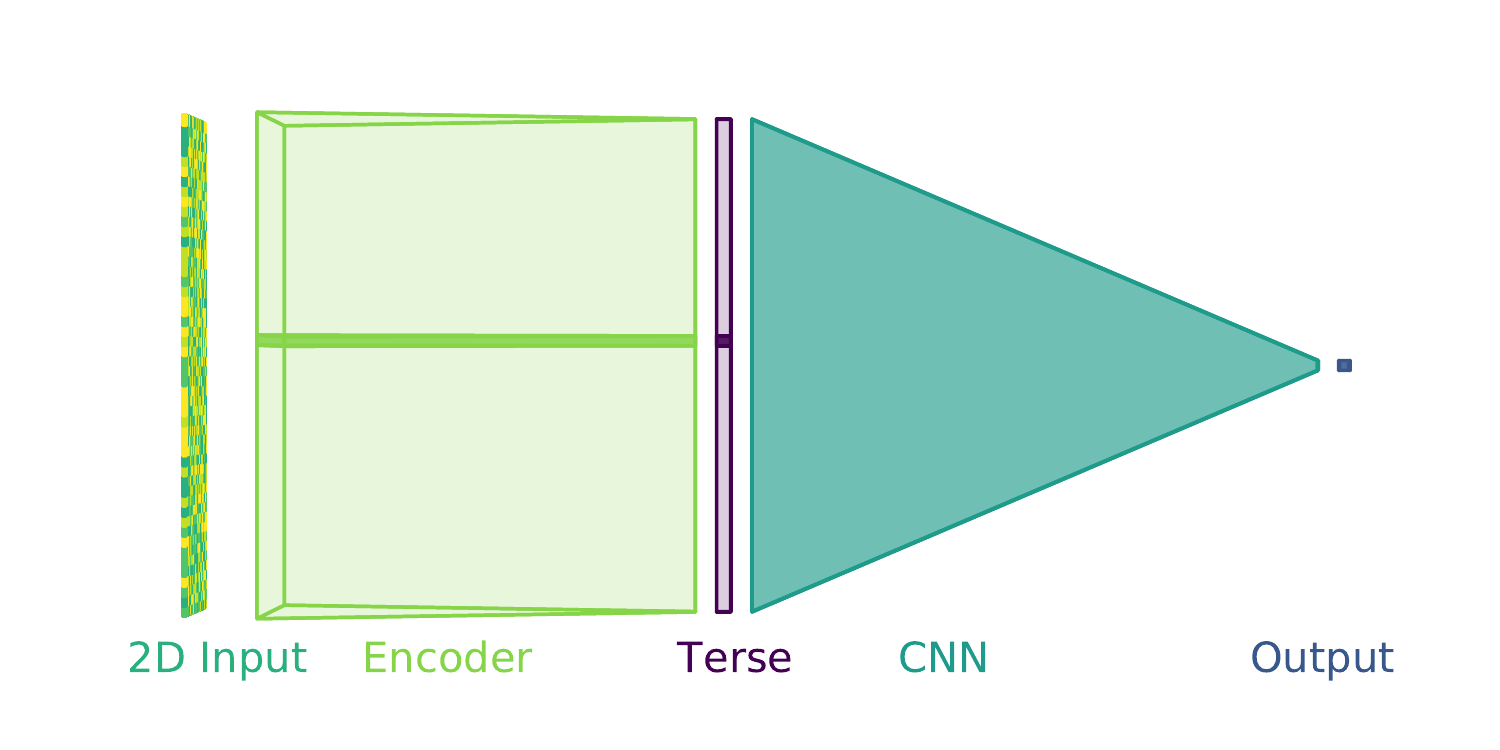} & \includegraphics[width=0.45\textwidth]{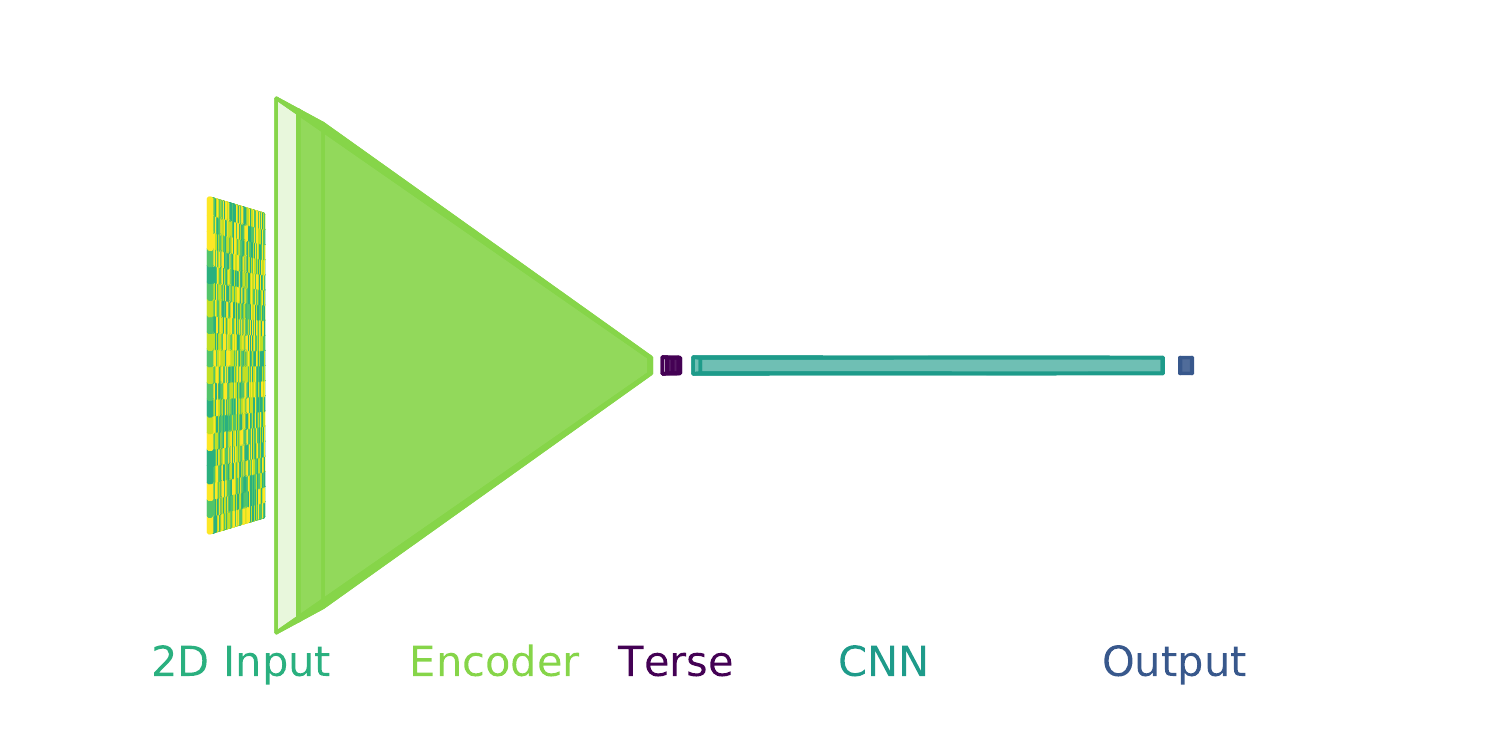} \\
		\end{tabular}
    \caption[]{A schematic view of the neural network architecture.  Top:  A nearly side-on view of the architecture, which is more complex than the pedagogical autoencoder shown in Figure \ref{fig:AE}. The cluster encoder (lime green) summarizes the 2D input  from each cluster as a single, meaningful parameter in the bottleneck terse layer (dark purple).  The supplementary data vector is inserted at the terse layer so that each cluster is summarized by three numbers:  the terse value, the cluster luminosity, and the cluster weak lensing mass.  The CNN (teal) uses the values in the terse layer and also the values in the supplementary data to estimate an output value of \Sig (dark blue).  Because this encoder requires  the input signal to be properly labeled with a cosmological parameter \Sig, this encoder is a \textit{supervised} machine learning model.  Bottom Left:  A completely side-on view of the same architecture.  The opaque slice through the encoder and terse layer highlights the information path for a single, randomly selected cluster.  This shows that the architecture does not allow information from neighboring clusters to inform the terse parameter. Bottom Right:   A top view of the same architecture.  From this vantage point, the role of the encoder is more clear \textemdash{} it summarizes the complex, multi-wavelength information in each cluster into a single parameter at the bottleneck.  The 2D and Supplementary Inputs are discussed next in \S \ref{sec:input} and are illustrated in Figure \ref{fig:sample_images}}
\label{fig:architecture}
\end{figure*}

The deep architecture is built to map a cluster sample to the underlying cosmological parameter \sig.  Rather than building a black-box architecture that performs this task, our architecture is built in a physically motivated way.  First, it summarizes each cluster with a mass proxy.  Next, it uses the full cluster sample to infer \sig.  By sculpting the architecture in this way, the model design is inherently more interpretable than a black box
network, giving us an opportunity to check and
understand what the model has learned. 

The network has two primary components.  The first component is an encoder to produce a terse representation of individual galaxy clusters, e.g., it compresses the cluster observations into a cluster mass proxy.  The second component is a more standard CNN that uses these terse values to estimate the cosmological parameter \Sig, e.g., it uses the list of cluster masses to infer the underlying cosmology.  The network is implemented in Keras \citep{chollet2015} with a Tensorflow \citep{45381} backend, and a schematic of the architecture is shown in Figure \ref{fig:architecture}.

The model takes two inputs, and the details of the observational parameters in these inputs is described in the following section.  The first input is a single-channel, $N_\mathrm{cluster}\times w \times1$ matrix.  The second input is a $N_\mathrm{cluster}\times 1 \times 2$ supplementary input.  This first dimension, $N_\mathrm{cluster}$, is the number of clusters; this parameter varies per \new{2D data array}.  The second dimension denotes the width of the inputs and the number of observational parameters per cluster.  For the Standard Catalog, described in the next section, $w=15$.  The final dimension denotes the number of channels (or colors).  While it is not standard to tabulate the number of channels, we opt here to include this as the final dimension to emphasize that the encoder network is creating complex, nonlinear combinations of the 2D input data.  

When training, the network learns how to summarize individual clusters optimally for regressing estimates of the cosmological parameter \Sig.  The full architecture is given in Table \ref{table:architecture}.  This model has 10,472,834 trainable parameters; $\sim2.8$ million in the encoder and $\sim7.6$ million in the cosmological CNN.

\begin{deluxetable*}{l r l r r l}
\tablecaption{Network Architecture}
\label{table:architecture}
\tablehead{
\colhead{Layer} &\colhead{Filters} &\colhead{Shape} & \colhead{Activation}  & \colhead{Dropout} & \colhead{Comments}
}
\startdata
2D input$\dagger$						&\nodata			& $N_\mathrm{cluster}\times w \times 1$		& \nodata		& \nodata		& Input 1 of 2\\
repeated network  						&  2048 			& $N_\mathrm{cluster}\times 1 \times 2048$	& leaky ReLU	& 25\%		& \nodata\\
repeated network 						&  1024 			& $N_\mathrm{cluster}\times 1 \times 1024$	& leaky ReLU	& 0\%		& \nodata\\
repeated network 						&  512 			& $N_\mathrm{cluster}\times 1 \times 512$	& leaky ReLU	& 0\%		& \nodata\\
repeated network 						&  256 			& $N_\mathrm{cluster}\times 1 \times 256$	& leaky ReLU	& 0\%		& \nodata\\
repeated network 						&  128 			& $N_\mathrm{cluster}\times 1 \times 128$	& leaky ReLU	& 0\%		& \nodata\\
repeated network 						&  64 			& $N_\mathrm{cluster}\times 1 \times 64$		& leaky ReLU	& 0\%		& \nodata\\
repeated network 						&  32 			& $N_\mathrm{cluster}\times 1 \times 32$		& linear		& 0\%		& \nodata\\
repeated network 						&  1 				& $N_\mathrm{cluster}\times 1 \times 1$		& \nodata	& 0\%		& layer output = ``terse value''\\
\tableline
supplementary data vector input$\dagger$	&\nodata			& $N_\mathrm{cluster}\times 1 \times 2$		& \nodata		& 0\%		& Input 2 of 2\\
concatenation$\dagger$					&\nodata			& $N_\mathrm{cluster}\times 1 \times 3$		& \nodata		& 0\%		& Joins terse and suppl.~input\\
$4\times1$ convolution 					&$128$			& varies 								& leaky ReLU	& 0\%		& zero padded\\
$2\times1$ mean pooling					&\nodata			& varies								& \nodata		& 0\%		& \nodata \\
$4\times1$ convolution 					&$256$			& varies								& leaky ReLU	& 0\%		& zero padded\\
$2\times1$ mean pooling					&\nodata			& varies								& \nodata		& 0\%		& \nodata \\
$4\times1$ convolution					&$512$			& varies								& leaky ReLU	& 0\%		& zero padded\\
$2\times1$ mean pooling					&\nodata			& varies								& \nodata		& 0\%		& \nodata \\
$4\times1$ convolution 					&$1024$			& varies								& leaky ReLU	& 0\%		& zero padded\\
$2\times1$ mean pooling					&\nodata			& varies								& \nodata		& 0\%		& \nodata \\
global max pooling						&\nodata			& $1024$								& \nodata		& 0\%		& \nodata\\
fully connected							&\nodata			& $2048$								& leaky ReLU	& 30\%		& \nodata \\
fully connected							&\nodata			& $1024$								& leaky ReLU	& 30\%		& \nodata \\
fully connected							&\nodata			& $512$								& leaky ReLU	& 30\%		& \nodata \\
fully connected							&\nodata			& $256$								& leaky ReLU	& 30\%		& \nodata \\
fully connected							&\nodata			& $128$								& linear		& 0\%		& \nodata \\
output								&\nodata			& $1$								& \nodata		& \nodata		& linearly scaled \sig\\
\enddata
\tablecomments{All layers follow directly from the line above, except for the three layers that are marked with $\dagger$.  Of these, two are input layers and the third is a concatenation.}
\end{deluxetable*}

In this model, cluster feature extraction is performed by the encoder.
In Table \ref{table:architecture}, we refer to each layer of the encoder as a ``repeated network.'' 
The repeated networks can be thought of as a fully connected network that inputs a cluster $L_X$, $T$, and $M_\mathrm{gas}$, and outputs a single value.
In practice, the repeated network is coded as a convolutional
layer with kernel size $1\times w$ (where $w$ is the layer
width) and with no padding.  This careful filter and padding choice ensures that the convolutional filter evaluates just one cluster at a time and that it treats each input independently.  It acts as a fully connected network, with one subtle difference:  each connection has a multiplicative, but not an additive, free parameter.
Via the encoder, X-ray observations of individual clusters are summarized as a single
terse value\footnote{The final results that are discussed in \S\ref{sec:results} showed no improvement when the architecture included more than one terse value per for a cluster, an unsurprising result because cosmological changes in \sig were modeled entirely by changing the selection function, which in turn only depended on a single cluster parameter, the cluster mass. }, and the
simulated catalog is interpreted as an ensemble in the final CNN
portion of the network, as shown in Figure \ref{fig:architecture}.

For our training data, the width of the \new{2D data array} is fixed, but the height is
set by the number of clusters,
$N_\mathrm{cluster}$, which is a function of \Sig and the $M-L$ scaling relation parameters. 
Most convolutional neural networks require fixed-size \new{inputs}, and the fact that the input \new{array} varies in size required special attention.
Two architecture implementation choices allow us to train the model on
variable-size arrays: 1) the global pooling layer sets the
number of model free parameters so that it does not vary as a function
of \new{2D data array} size and 2) the model is trained with batches of identically
sized \new{arrays}.  Regarding the latter, the order of batches is shuffled
randomly before each epoch.

\subsection{2D and Supplementary Input}
\label{sec:input}

\begin{figure*}[]
	\centering
	\includegraphics[width=0.65\textwidth]{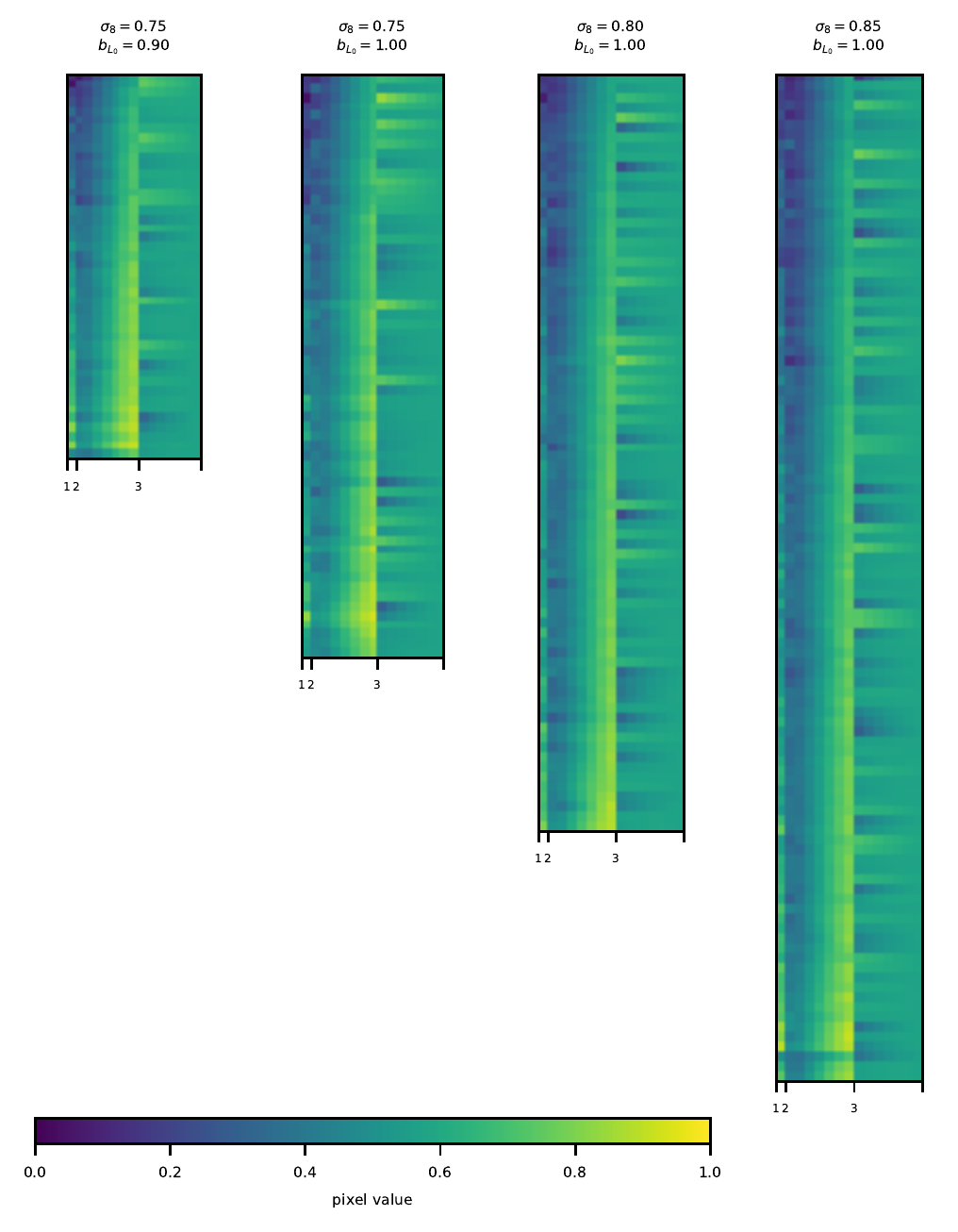} 
		\caption[]{A sample of 2D input matrices.  These are not sky-projected X-ray images of the cluster; instead, they are ordered cluster observables.  Each cluster is summarized as a data vector of the cluster temperature, gas mass profile, and gas density slope profile.  A row of the 2D input comprises summary statistics from one cluster. The clusters in the sample are sorted by $\log(L_X)$, with the highest luminosity clusters at the bottom of the 2D input matrix.
		  The number of clusters, which is a function of both \Sig and the $M-L_X$ scaling parameters (see Figure \ref{fig:selfunc}), sets the height of each 2D input matrix, while the number of features sets the width.  The technical challenge of training a neural network on input ``images'' of varying sizes is addressed in two ways:  first, the global pooling layer in the architecture fixes the number of free parameters regardless of 2D input height, and second, the training proceeds on image batches with identical dimensions.  The features included in each image are noted along the bottom edge of each 2D input and are 1: $\log(T)$, 2: the profile of $\log(M_\mathrm{gas}(r))$, and 3: the profile of $d\log(\rho)/d\log(r)$. }
       	\label{fig:sample_images}	
\end{figure*}

In Section \ref{sec:architecture}, we described a supervised \new{encoder architecture}
 that uses two input data sets,  1) a two-dimensional input \new{array} that includes a number of $X$-ray-inferred cluster parameters, and 2) a supplementary data vector, which includes weak lensing masses and cluster luminosities.  Here, we describe the process for casting each mock cluster catalog into an appropriate format for the supervised \new{encoder}.

Convolutional neural networks typically learn from \new{2D data arrays.  For many applications, they learn from natural images, but they are not restricted to this \textemdash{} CNNS can be a powerful tool for analyzing any ordered 2- or 3D data.}  In our work, the 2D inputs here are not images in a traditional sense (they are not, for example, X-ray pictures of galaxy clusters showing an extended and nearly spherically symmetric ICM).  Rather, they are an ordered sequence of data vectors that are cast into a 2D array so that each row describes a single cluster and each column describes a cluster observable.  We opt for learning from \new{radial} profiles (rather than, for example, X-ray images of clusters) because, though X-ray substructure information would give marginal improvements \citep[e.g.][]{2011ApJ...731L..10N, 2019ApJ...884...33G}, but it would likely lead to overfitting.  Our data vector includes the following observables:  cluster temperature $\log(T)$, binned gas mass in spherical radial shells,
$\log[M_\mathrm{gas}(r)]$, and binned gas density slope,
$\gamma \equiv d\log[\rho(r)]/d\log(r)$. 
The data vectors are rank-ordered by cluster luminosity,
$L_X$. A sample of these 2D inputs is shown in Figure \ref{fig:sample_images}.

While the
2D inputs tally X-ray information about each cluster, additional
cluster parameters are useful and necessary for deriving useful
cosmological constraints from X-ray observations.  Weak lensing
estimates, for example, are necessary to constrain the absolute mass
scale.  X-ray luminosity is used to infer both survey volume and the
mass scale where the observation transitions from flux- to
volume-limited.  \new{Looking ahead to the interpretation schemes in \S \ref{sec:interpretation}, it will be interesting to isolate the X-ray inputs (temperature, gas mass profile, and gas density profile) and investigate their role in the ML encoder.  For these reasons,} two parameters \textemdash{} weak lensing mass
estimates and X-ray luminosity \textemdash{} comprise a
separate supplementary data vector.

The machine learning method described previously (Section~\ref{sec:architecture})
converges faster if the input (or ``pixel'') values do not vary wildly, but are
approximately constrained between $0$ and $1$.  Therefore, the 2D inputs, supplementary inputs, and \sig output labels are
re-scaled linearly to arbitrary units so that they vary between $0$
and $1$.  In the case of the 2D inputs, this is done for groups of pixels with the same physical interpretation (e.g., all seven bins of gas masses are scaled with the same linear parameters; they are not scaled individually).  \new{No information is lost; the same rescaling parameters are used to rescale every simulated cluster catalog identically. }

In addition to the Standard Catalog, two other simulated catalogs are built and tested.  These catalogs are:  1) the Shuffled Catalog: to understand whether the sorting is informative, the sort is removed and the clusters are shuffled randomly and 2) the No WL Catalog:  to assess the information content in weak lensing mass estimates, this parameter is excluded from the supplementary data vector. The machine learning model is trained independently for each of these catalogs.

For each realization of the mock catalog (including selection of
clusters as described in Section \ref{sec:selection1} and generation of their mock
observations  in Sections \ref{sec:selection2}-\ref{sec:selection3}), 
the mock cluster sample is randomly
assigned to one of three data sets.  The training set comprises 80\% of the
simulated catalogs and is used to train the model.  The validation set
comprises 10\% of the simulated catalogs and is used to assess when the
model has optimally trained.  The testing set comprises the remaining
10\% of the simulated catalogs; predictions for the testing set are
presented in Section \ref{sec:results}.

\subsection{Training and Model Selection}
\label{sec:train}

\begin{figure}[]
	\includegraphics[width=0.45\textwidth]{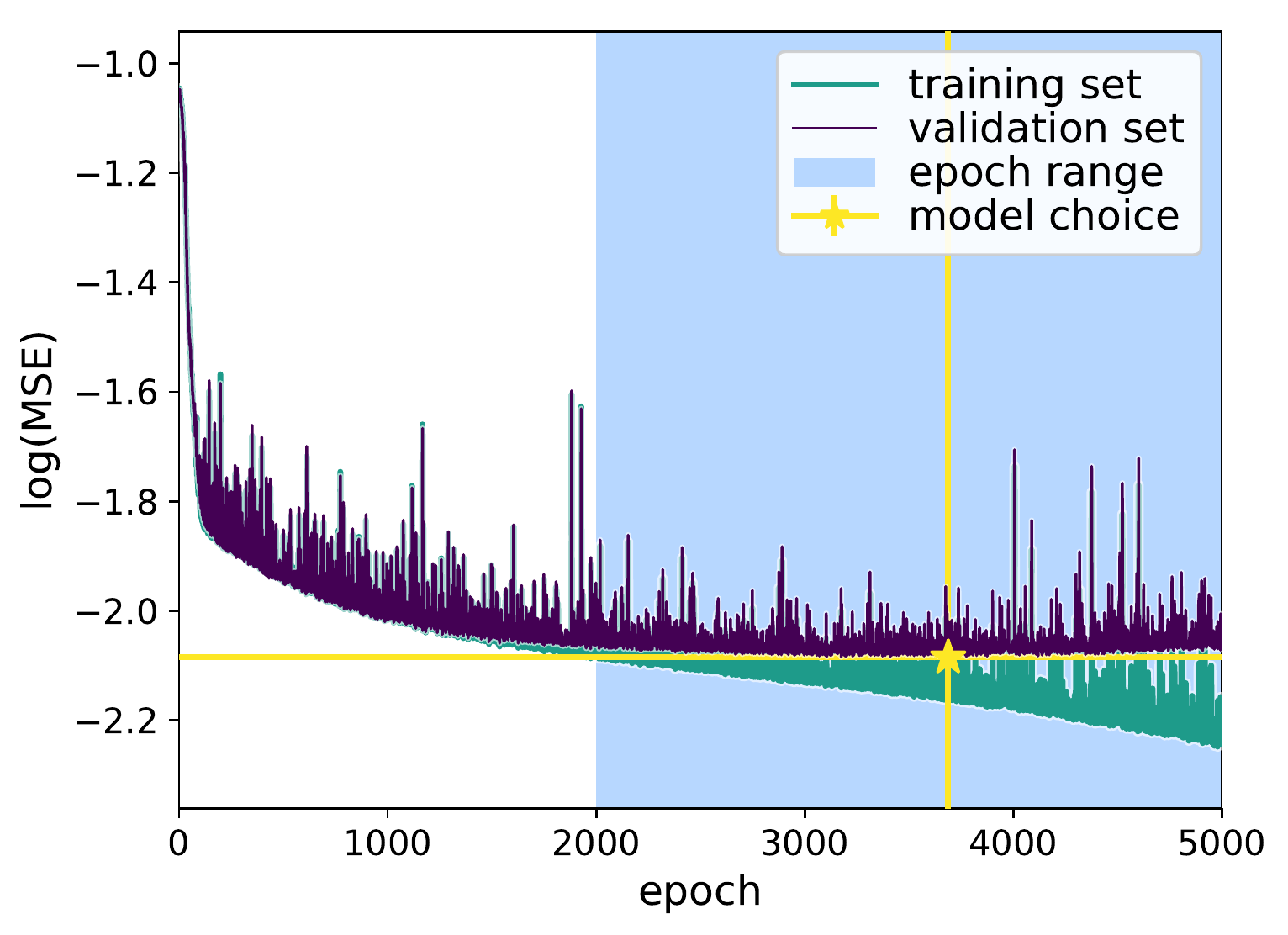} 
	\caption[]{Model training.  The $\log(\mathrm{MSE})$ on the training set (teal) decreases steadily for the 5000 epochs of training, though with noise due to the stochasticity of the process.  The MSE for the validation set (dark purple), however, has a minimum at epoch 3688; this is the model selected for training (yellow star).  Beyond this epoch, the model overfits, steadily decreasing MSE for the training set, but increasing MSE for the validation set.  Notice that the validation set MSE sits above the yellow horizontal line; this line is the validation set MSE of the best fit model.  Overfit models (such as this one at epoch $\sim5000$) will produce pessimistic results due to the model memorizing details of the training set but failing to generalize to the validation and testing sets.}
       	\label{fig:train}	
\end{figure}

The model is trained with a mean absolute error loss function and the Stochastic Gradient Descent (SGD) optimizer with the default learning rate ($\mathrm{lr}=0.001$), and the step size is updated after each batch of identically-sized \new{2D data arrays}.  See \cite{ruder2016overview} for a review of SGD and other gradient descent optimization algorithms.  

The training proceeds for 5000 epochs and the final model is selected
afterward.  Figure \ref{fig:train} shows the mean squared error (MSE)
of both the training and validation sets as the training progresses.
ML models often converge more quickly if the output is scaled to be between -1 and 1; the MSE shown is relative to this linearly scaled output \sig.  Models in the
$2000-5000$ epoch range are assessed according to MSE in the
validation set, and the model with the lowest validation MSE is
selected as the final model.  The selected final model is toward the middle of
this range (epoch 3688).   At the
final epochs, the model is clearly overfit:  it produces very small, decreasing errors on the training set and at the same time, the error on the validation set begins to rise.  The model has ``memorized'' details in the training data that do not generalize.

It is worth noting here that, in the course of training the model, we performed many routine numerical checks, including training with 10 different random initializations of the model.  The trends reported in upcoming Section 
\ref{sec:interpretation} are qualitatively similar across the ten randomly seeded and independently trained models.
The test set was properly set aside during this process; it was used neither for
training the model nor for assessing the best fit model.  

\section{{Results}}
\label{sec:results}
  
In this section, we verify that the ML method indeed works by assessing the model predictions for \sig.  We first present the results of the Standard Catalog, then we check that the model produces sensible results for alternative sets of input data, in which information has been selectively removed from the training data.

\subsection{Standard Catalog \Sig Constraints}
\label{sec:stdcat}

\begin{figure}[]
	\includegraphics[width=0.5\textwidth]{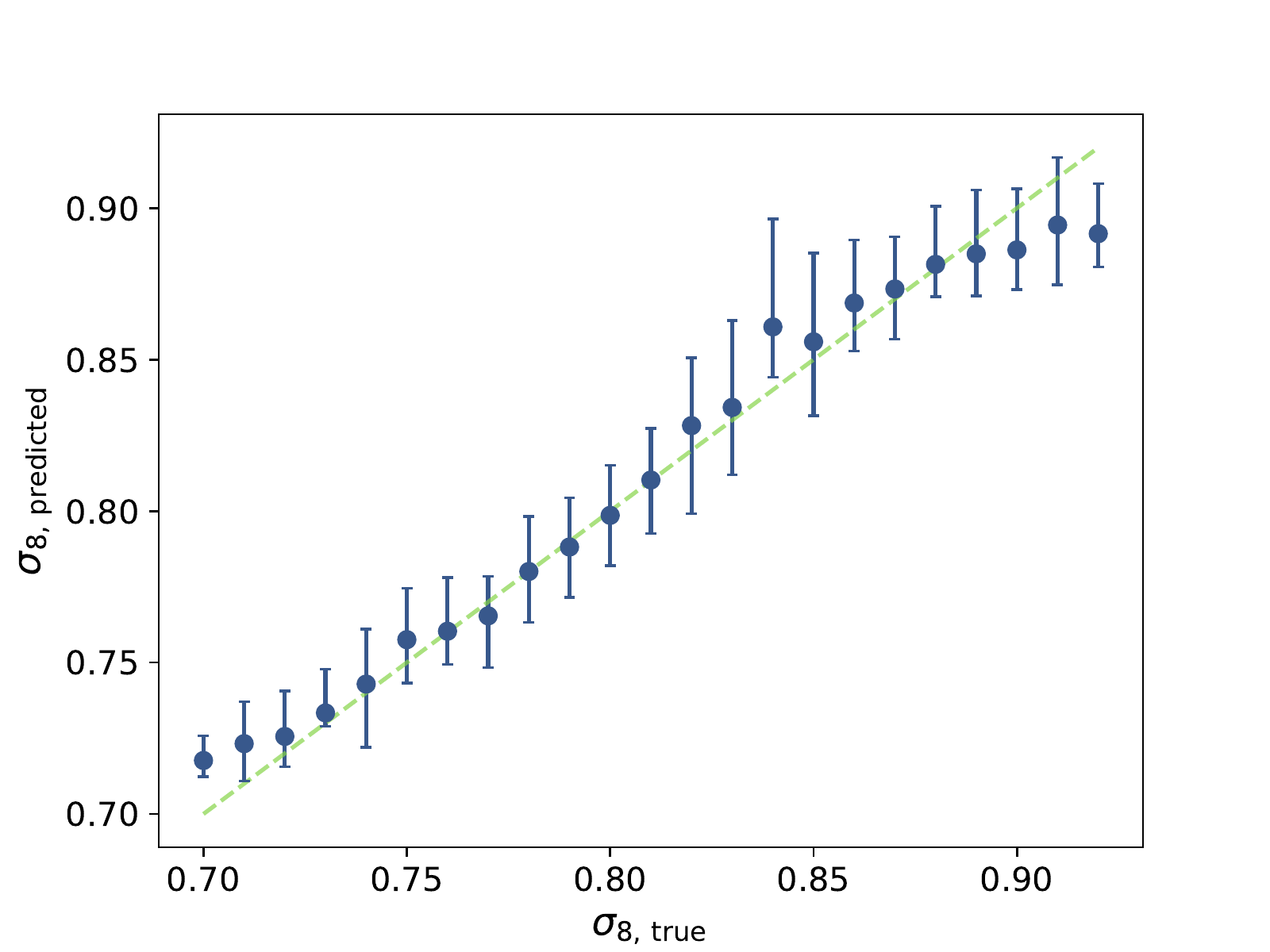} 
	\caption[]{True and predicted \Sig values for the
          testing catalog.  Errors $\delta \Sig \equiv
          \sigma_\mathrm{8,\,predicted}
          -\sigma_\mathrm{8,\,true} $ have intrinsic scatter of $0.0212\pm0.0009$ (error bars show the middle 68\% in each \sig bin).  This intrinsic scatter is
          near the theoretical limit of cosmological constraints for
          an observation of 100 clusters.  Predictions toward the center of the
          training \Sig range ($0.73 \lesssim \Sig \lesssim 0.87$) are
          consistent with the one-to-one line (lime green dashed) and
          are not noticeably biased high or low.  However, outside of
          this range, the model shows a slight biasing toward the
          mean; this tendency is common in regression tasks with deep
          ML.\\ \\}       
	\label{fig:results}	
\end{figure}

\begin{deluxetable*}{p{2in}lc}
\tablecaption{Scatter in \Sig Prediction Errors}
\label{table:results}
\tablehead{
\colhead{Test Case} & \colhead{$\Delta \Sig$} & \colhead{Discussion}}
\startdata
Standard\dotfill & $0.0212\pm0.0009 $  &  \S \ref{sec:stdcat} \\
Shuffled\dotfill &$0.0262\pm0.0011 $  &  \S \ref{sec:shuffled} \\
No WL\dotfill    &$0.0297\pm0.0013 $  &  \S \ref{sec:nowlresults} \\
$N_\mathrm{cluster}$ mapping\dotfill &$0.0650\pm0.0030 $  &  \S\ref{sec:Ncl} \\
Array of Ones\dotfill &$0.0620\pm0.0027 $  &  \S\ref{sec:Ncl} \\
\enddata
\end{deluxetable*}

The machine learning model described in the previous section uses a
collection of cluster observations to infer the underlying
cosmological parameter \sig, and Figure \ref{fig:results} shows how
well the model performs at this task. Measurement errors on \sig are
calculated by $\delta\Sig \equiv \sigma_{8,\,\mathrm{predicted}}
-\sigma_{8,\,\mathrm{true}}$.  For the Standard Catalog, the standard
deviation of errors is $\Delta \Sig=0.0212\pm0.0009$; this is also
reported in Table \ref{table:results} for reference.  Note that our
Standard Catalog model --- which uses all available information ---
produces uncertainties in \Sig\ at approximately the theoretical limit
for a cluster sample of this size and median cluster mass \citep[c.f.,
e.g., purely statistical uncertainties on \Sig\
in][]{2009ApJ...692.1060V}.

In Figure \ref{fig:results}, we also see that toward the middle of the \Sig range, the estimates are
consistent with the one-to-one line, but estimates at the edges of the
\Sig range bias toward the mean.  This is a common outcome with ML
regression techniques, which tend to interpolate well, but are less adept at extrapolation.  
When applying such models to observational data, then, it is
important to be aware of this trend and to train the model on labeled
data that extend well beyond the expected output range.  
Encouragingly, when it has access to the full data set, the machine learning model estimates \sig at about the theoretical limit.  Next, we will explore how well the model performs when the full input data are not available.

\subsection{Alternative Sets of Input Data \sig Constraints}

\label{sec:nonstdresults}

In this section, we explore how well the ML model can recover \sig when information is removed.  This serves as an important check of the model, which we expect to underperform when it does not have access to vital data.  
The Shuffled Catalog assesses the information in the ordering of the clusters and
the No Weak Lensing Catalog explores the measure of information in the
supplementary input.  Models that are trained only on the number of clusters in the sample ($N_\mathrm{cluster}$ Mapping and
Array of Ones)  assess the constraints that are possible from severely limited cluster data.  These scenarios are described in subsequent subsections and the results are tabulated in Table \ref{table:results}.  As expected, these tests show that removing information increases uncertainties on \sig.

\subsubsection{Shuffled Catalog}
\label{sec:shuffled}

  Here, we explore how well the ML model can recover information that is, in principle, contained in the simulated catalog but not given explicitly.  Rather than removing inputs relative to the Standard Catalog, we shuffle the cluster order\new{; the 2D Inputs and Supplementary Inputs for each simulated cluster catalog are each shuffled identically so that the clusters are no longer ranked by luminosity, but the weak lensing masses can still be used for calibration.  This catalog} tests how much information is
implicitly encoded in the ordering of the clusters according to
luminosity.  As expected, removing this
ordering increases the scatter, $\Delta \sig = 0.0262\pm0.0011 $ (compared to the Standard Catalog's $0.0212\pm0.0009 $).

The information for sorting, $L_X$,  is still included in the supplementary data, so why is the scatter increase expected?  This is due to the nature of the convolutional filters, which are tools for finding
\textit{spatially correlated} trends in data.  Our shuffled data and architecture choice are an ill-suited pair.  
This highlights an often-ignored nuance: despite their flexibility, deep machine
learning methods are not universally suited for all data sets.  Method and data should be carefully matched to address the task at hand.\\

\subsubsection{Cluster Count Catalog}
\label{sec:Ncl}
The number of clusters in the sample, $N_\mathrm{cluster}$, is weakly correlated with the underlying cosmology, but it is also affected by luminosity scaling relation parameters (which sets the cluster selection function and cluster abundance at low mass).  Here, we explore how much information is contained in this single value.  

First, we test a simple statistical mapping of
  $N_\mathrm{cluster}$ to $\Sig$.  This approach does not use ML.  Instead, \new{it is a simple binned regression on $N_\mathrm{cluster}$;} the data are binned (with width $\Delta
  N_\mathrm{cluster}=10$) and the median \Sig is the naive estimate of the cosmological model as a function of cluster count.  \new{Near $N_\mathrm{cluster}\sim100$ (e.g., near the mean  of $N_\mathrm{cluster}$ values in our catalogs)}, this approach gives $\Delta \sig = 0.0650\pm0.0030$ (compared to the Standard Catalog's $0.0212\pm0.0009 $), which  is well
  outside of the $\Delta \Sig$ results of the machine learning models.  It is unsurprising that this simplistic approach performs much worse than our Standard Catalog approach.

Next, we assess the CNN's ability to count the number of clusters and infer $\Sig$.  To implement this test, we create input arrays of 1's (rather than using more informative parameters such as weak lensing masses or temperatures).  The resulting constraint is \new{consistent with the   $N_\mathrm{cluster}$ regression:} $\Delta \sig = 0.0620\pm0.0027$.  

From these tests, we see that the number of clusters is not sufficiently informative to explain the narrow \sig constraints from the Standard Catalog.  $N_\mathrm{clusters}$ alone is simply not a good probe of the underlying cosmology; more information is needed to calibrate the volume and mass of the sample.




\subsubsection{No WL Catalog}

The final test case is a catalog for which weak lensing masses are removed from the
supplementary inputs. This test case produced unexpected results which
led to a discovery of a new mode of self-calibration in the analysis
of cluster catalogs. The discussion of this test case relies on the
techniques we developed for interpretation of the ML models, presented
next. Therefore the discussion of the no-WL case is deferred to
\S~\ref{sec:nowlresults}. \\

\section{ML Model Interpretation}
\label{sec:interpretation}

Here, we lay out three interpretation schemes for understanding the signals that are being used by the supervised encoder.  This is an important step because, unfortunately, deep ML models are often used as un-interpretable black boxes.  Though it is difficult to interrogate and interpret deep models, this is an essential step toward building trustworthy machine learning methods that can be applied to astronomical observations.  Our physically motivated architecture offers several opportunities for interpretability that are discussed here:  a leave-one-out method that probes the importance of a single cluster, average encoder saliency that explores the model's sensitivity to the gas mass profile, and correlations in the terse layer that allows us to understand how the model is inferring cosmology from mock cluster observations.

\subsection{Leave One Out}
\label{sec:oneout}

\begin{figure*}[]
	\centering
	\includegraphics[width=\textwidth]{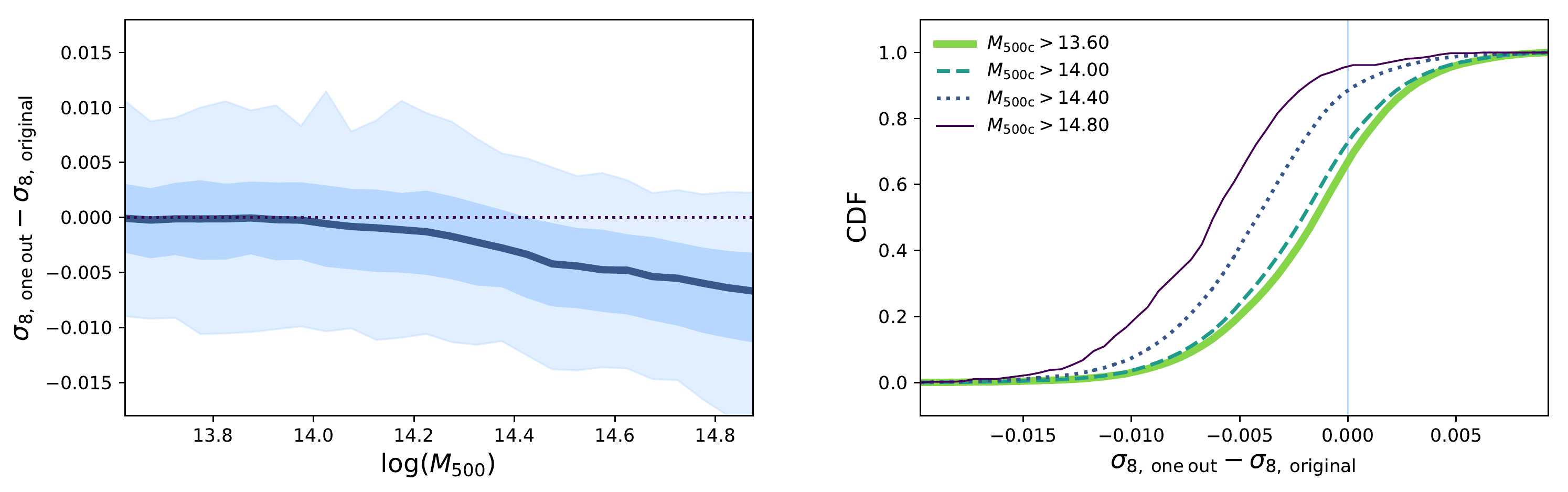}
	\caption[]{Left: removing a single cluster from a simulated catalog affects $\Delta \Sig$, the difference between the new \Sig prediction and the original.  For low mass clusters ($M_\mathrm{500}\lesssim 10^{14.2}\msolarh$), removing a single cluster has no systematic effect on the new prediction ($\sigma_\mathrm{8,\,one\,out}$) relative to the original \Sig prediction ($\sigma_\mathrm{8,\,original}$).  The median (dark blue curve) shows that the average effect at low masses tends to neither increase nor decrease the prediction, while the 68\% and 95\% regions (shaded light blue regions) give a sense of the stability of the model under the removal of a single cluster.  Removing a higher-mass cluster, however, systematically reduces the \Sig prediction.  This is unsurprising; removing a high mass cluster lowers the observed HMF in the volume-limited regime, mimicking a lower-\Sig cosmology.
			Right: the cumulative distribution function (CDF) of $\Delta \Sig$ for clusters above several benchmark masses.  For all clusters in the sample (lime green solid), removing a single cluster has a slightly more-than-even probability of lowering \Sig.  The highest-mass sample (purple solid), however, almost always lowers the \Sig prediction.  
			This leave-one-out method is useful for assessing information content in a cluster as a first step of understanding how the neural network builds cosmological predictions; if the model was simply counting the clusters in the sample, $\sigma_\mathrm{8,\,one\,out}-\sigma_\mathrm{8,\,original}$ would always be negative and would not show trends with cluster mass. \\}
       	\label{fig:oneout}	
\end{figure*}

\begin{figure*}[]
	\centering
	\includegraphics[width=0.65\textwidth]{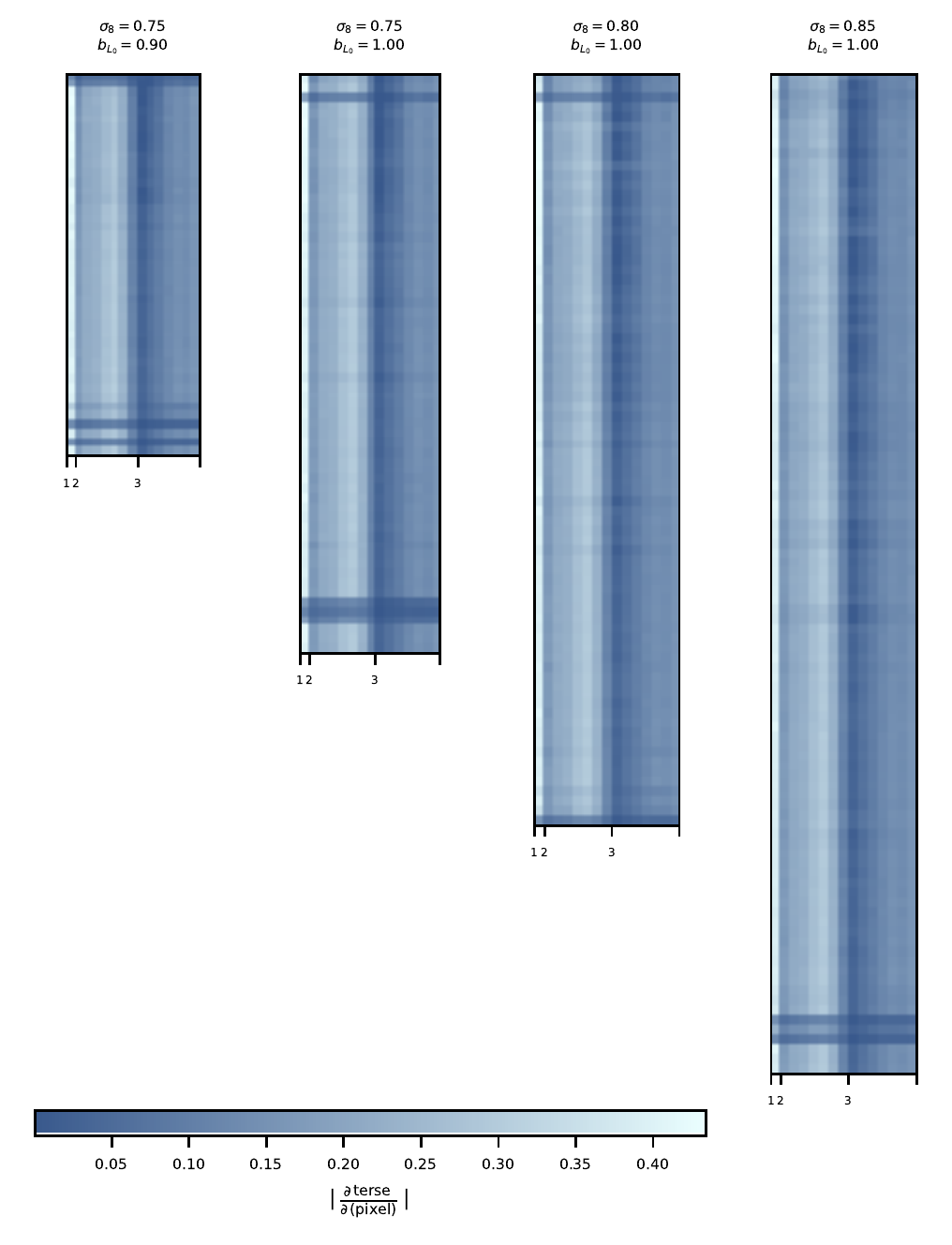} 
	
		\caption[]{Saliency maps, showing $| \partial \, \mathrm{terse}/ \partial \, \mathrm{pixel}| $, for the same sample of simulated catalogs that are shown in Figure \ref{fig:sample_images}.  
		As before, the features are 1: $\log(T)$, 2: the profile of $\log(M_\mathrm{gas}(r))$, and 3: the profile of $d\log(\rho)/d\log(r)$.  Clusters are sorted by $\log(L_X)$.  Light blue pixels of large saliency highlight values that are important to the terse value; changes in these high-saliency values will change the terse value dramatically.  Dark blue pixels, on the other hand, indicate that the terse value is relatively insensitive to changes in the input pixel value.  It is particularly interesting to look at trends in the features (by, for example, averaging the columns) to identify which of the input features are the highest saliency, as the terse value is sensitive to changes to these features. \\}
       	\label{fig:sample_sal}	
\end{figure*}

The leave-one-out technique is used for assessing the information content in a single cluster.  Physically, we are testing how the inferred \sig would change if a single cluster were removed from the sample (from both the 2D input matrix and also from the supplementary input).  After the cluster is removed, the simulated catalog is labeled with an estimated \sig by the \textit{trained} network.  The \sig estimates from before and after the cluster is removed are compared to one another.

Recall that, in Section \ref{sec:architecture}, we addressed the technical challenge of varying-sized \new{2D data arrays}.  The network has been carefully engineered to work on a variety of \new{input} sizes, and so smoothing, padding, and other \new{image-processing} acrobatics are not needed.  

The results are shown in Figure \ref{fig:oneout}.  Low mass clusters have roughly equal probability of increasing or decreasing the predicted \Sig.  The width of the distribution in $\Delta \Sig$ gives us a sense of the stability of the trained model for small perturbations; the middle 68\% of values have $|\Delta \Sig| < 0.003$.  We find that low mass clusters are less important for cosmological constraints.  This is unsurprising because, as we saw in Figure \ref{fig:selfunc}, changing \sig does not appreciably affect the halo mass distribution at low masses.

The story changes at higher cluster masses.  Omitting a single high-mass ($M_\mathrm{500}\gtrsim 10^{14.2}\msolarh$) cluster from a simulated catalog will systematically lower \Sig, an effect that grows  with cluster mass.  Unsurprisingly, the rare, massive clusters are much more important to the determination of \Sig compared to their lower-mass counterparts.  The reason can again be inferred from the halo mass function (Figure \ref{fig:selfunc}); we saw there that the effects of changing \sig were most pronounced at the high mass end.  Removing a massive cluster effectively pulls the observed HMF down, imitating a lower-\Sig cosmology.\\

\subsection{Average Encoder Saliency}
\label{sec:sal}

\begin{figure*}[]
	\centering
	\includegraphics[width=\textwidth]{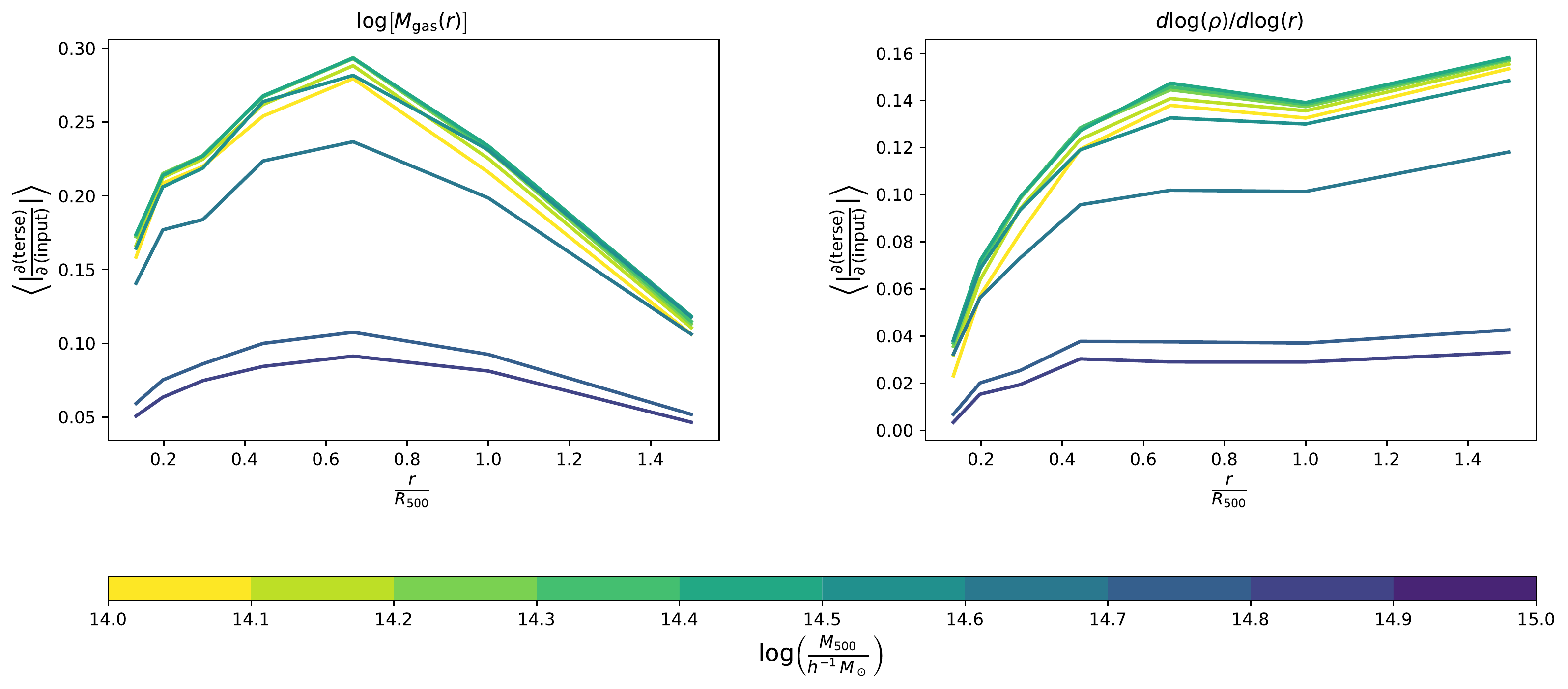} 
		\caption[]{Median saliency for all clusters in the testing set, plotted for each $0.1\,dex$ cluster mass bin (color bar).
		Left:  the average saliency of $\log\left[ M_\mathrm{gas}(r)\right]$ shows the relative importance of cluster gas mass as a function of radius.  Regardless of cluster mass, there is a clear trend in saliency, with a dip at the cluster cores, a gentle rise toward $R\sim0.6\,R_{500}$, and a slow decline beyond.
		Right:  average saliency of gas density slope profile, $d\log(\rho)/d\log(r)$ shows the relative importance of cluster gas density.  Again, we see a low sensitivity to cluster cores \textemdash{} a surprising result due to the fact that core densities are informative about the accretion history and dynamical state of the cluster.  
		Both average saliency maps show the same trend with cluster mass.  Clusters with $\log(M_\mathrm{500})\lesssim 14.7$ have similar mean saliency curves, while higher-mass clusters have terse values that are less sensitive to changes in \Mgas and gas density slopes. }
       	\label{fig:av_sal}	
\end{figure*}

Next, we use saliency maps to assess the terse value's sensitivity to small changes in the gas mass profile.  Saliency maps \citep{2013arXiv1312.6034S} are a deep learning interpretation technique that allows the user to visualize and understand which the portions of the input  that are most important in determining the label.  These maps show the gradient of the output label with respect to the input \new{2D data array}.  Visualizations such as saliency maps can be informative for illuminating the important features of the input and can be used for assessing the validity of the results\footnote{\cite{winkler2019association}, for example, used saliency maps to show that an image classification model with superb accuracy for classifying skin cancer was learning from the presence of skin markings.  These markings  are sometimes drawn on the patient's skin by medical professionals preparing to photograph worrisome moles.
In another example, \cite{2019NatCo..10.1096L} used a different visualization technique to show that an image classification model had identified an equine photographer's copyright stamp as excellent evidence for the photo containing a horse. }.  In astronomy, the method has been used to evaluate and assess trained CNNs \citep[e.g.,][]{2019ApJ...882L..12P,  2020arXiv200614305V, 2020arXiv200511066Z, 2020arXiv200706529Z}.

Because we want to understand how changes to the input affect changes in the terse value, we will consider the saliency of the encoder portion of the architecture.  This saliency, $\mathcal{S}$, is given by
\begin{equation}
	\mathcal{S} \equiv \left| \frac{\delta\,\mathrm{terse}}{\delta\,\mathrm{pixel}} \right|,
	\label{eq:saliency}
\end{equation}
the gradient of the terse value with respect to pixel values, where a ``pixel value'' here is a single number in the $N_\mathrm{cluster}\times w \times 1$, 2D input (that is, a single pixel from a \new{2D data array} in Figure \ref{fig:sample_images}, such as the temperature of the $13^{th}$ cluster or the core gas mass of the $42^{nd}$ cluster).  Figure \ref{fig:sample_sal} shows saliency maps for a representative sample of simulated catalogs.  Several clusters appear as ``unimportant'' dark horizontal stripes in the sample saliency maps.  These are discussed in further detail in Section \ref{sec:tersecorr}.

From the saliency maps, we can glean information beyond the relative importance of each cluster; we can understand the relative importance of each \textit{feature} (e.g., temperature, gas mass at a given radius, or gas mass profile at a given radius).  Vertical stripes in Figure \ref{fig:sample_sal} show the sensitivity of the terse value to input features.  While it might be difficult to summarize this information for a natural image, our input is simply arranged and can be collapsed into feature averages (shown in Figure \ref{fig:av_sal}).

We can infer from the dip in saliency near the cluster cores that the neural network has learned that cores are high scatter with cluster mass  \textemdash{} a result that was found by traditional means in \cite{2007ApJ...668..772M} and has also been shown with machine learning in \cite{2019ApJ...876...82N}.

The importance of the gas mass measurements peaks near $R=0.6 R_{500}$ and falls off at higher radii.  The trend is not driven by clusters in a particular mass range; while the overall magnitude of the effect is a function of mass, the qualitative trends do not depend on cluster mass.  This function of gas mass importance is often treated as a step function \textemdash{} gas masses within $R_{500}$ are equally important, while gas beyond some set cluster ``edge'' (for example, $R_{500}$) are not used.  Some treatments of cluster gas mass exclude the inner regions as well (this is because gas mass within $\sim0.15R_\mathrm{500}$ are high-scatter with cluster mass).  This mean saliency result suggests that a simple step function may not be sufficient to fully capture the relationship between gas masses and total mass; a more nuanced approach could take advantage of gas masses outside of the traditional $R_\mathrm{500}$ cluster boundary.

The mean saliencies of the gas density profiles show a similar sensitivity to gas masses near $\sim0.6R_\mathrm{500}$, with a gently rising sensitivity beyond.  These $\partial\,\mathrm{terse}/\partial\,\mathrm{pixel}$ values are negative, indicating that a rise in gas density corresponds to a decrease in terse value; the absolute value of these negative values is plotted.  

The average saliencies, $\bar{\mathcal{S}} \equiv \left< \frac{\partial \left( \mathrm{terse} \right)}{\partial \left( \mathrm{pixel} \right)} \right>$, for each input are:
the temperature saliency \mbox{$\bar{\mathcal{S}}_T=0.391^{+0.014}_{-0.008}$,} 
the gas mass saliency \mbox{$\bar{\mathcal{S}}_{M_\mathrm{gas}}=0.217^{+0.034}_{-0.066}$,} and
the gas density slope saliency
\mbox{$\bar{\mathcal{S}}_\gamma=0.119^{+0.019}_{-0.053}$}. Note that
the gas mass saliency here is somewhat low artificially because we
compute it at a single radius ($0.6\,R_{500}$) while in reality the gas mass
information is combined from several radial apertures. Taking this
into account, we estimate that the gas mass and temperature are of roughly equal importance to the terse value.

The low sensitivity to gas density slope in the cluster cores is surprising.  Cluster cores encode information about the cluster accretion history and if the encoder had learned methods for differentiating the cluster population according to dynamical state, one would expect a high saliency to $d\log(\rho)/d\log(r)$ at small radii.  We postulate that the lack of attention to cluster dynamical state is a result of limitations of the data set.  Cluster dynamical state is a weak function of \Sig, with higher \Sig cosmologies resulting in more merger events and fewer relaxed clusters.  The clusters for this work selected from a simulation at a single fiducial \Sig, with no attention paid to the cluster dynamical state.  Therefore, every simulated catalog has roughly the same ratio of disturbed to relaxed clusters.  We expect that, if the dynamical state ratio was modeled to capture this detail, the saliency in gas density slope profile would show more attention to this cosmological dependence.

Naively, Figure \ref{fig:oneout} suggests that the massive clusters are the most important for cosmological constraints, so one might expect these clusters to also have high saliency.  In fact,  the mass trends in Figure \ref{fig:av_sal} shows that this is not the case.  To understand why, we need to emphasize a detail of Equation \ref{eq:saliency}:  our saliency is measured and reported for the encoder portion of the network only and saliency measures sensitivity of the terse value, not of the final network output of \Sig.  Figure \ref{fig:av_sal} shows that, compared to their lower mass counterparts, the high mass clusters' terse values are less sensitive to small perturbations in the inputs.  

\begin{figure*}[t]
	\begin{centering}
	\includegraphics[width=\textwidth]{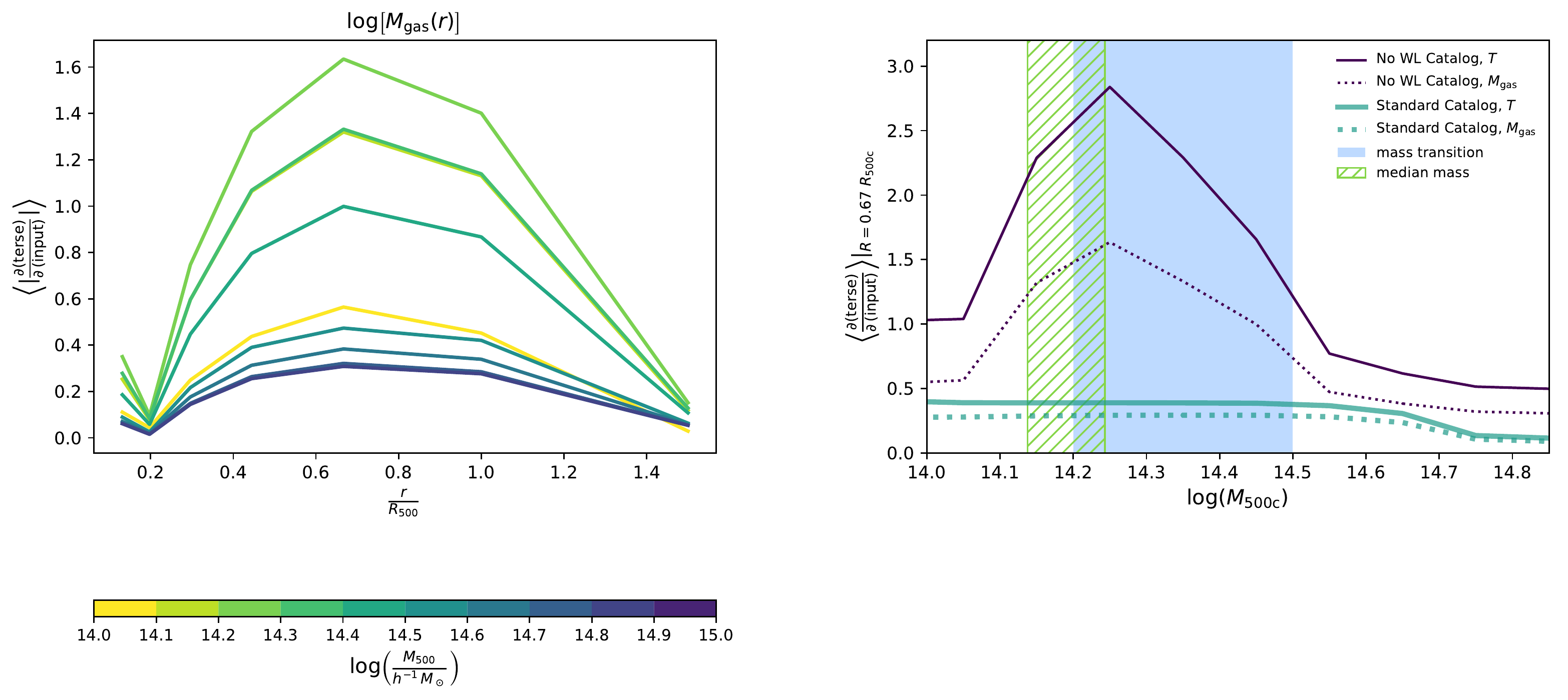}\\[3ex]
	 \end{centering}
	\caption[]{\new{Left:  Gas mass saliency for the No WL Catalog.  When weak lensing information is not included in training, a radial dependences similar to the Standard Catalog emerges (see the left panel of Figure \ref{fig:av_sal}).  As before, the average saliency of $\log\left[ M_\mathrm{gas}(r)\right]$ peaks near $R\sim0.6\,R_{500}$.  The more remarkable difference between the No WL and the Standard Catalog saliencies are as a function of mass (colorbar).  This trend is shown more clearly in the right panel. }
	Right: Average gas mass saliency at $R=0.67R_\mathrm{500c}$ for the Standard (teal) and No Weak Lensing (purple) Catalogs.  The largest absolute saliency values are for clusters with cluster masses between $10^{14.2}\,\msolarh$ and $10^{14.4}\,\msolarh$.  \new{Intriguingly, this peak is located near the mass range where the catalogs transition from flux- to volume-limited (pale blue rectangle) and the median mass range (hatched green rectangle).  The mass transition and median mass ranges shown extend from the minimum to the maximum values found in the 4,968 simulated cluster samples.}  When weak lensing is not provided to the model for mass calibration, the model learns sensitivity to this middle mass range, with less sensitivity to the lowest and highest masses, suggesting that the model has learned to calibrate the cluster masses by identifying the peak in mass function.}
       	\label{fig:altcat_saliency}	
\end{figure*}

One explanation is that the network has gained a sensitivity to the mid-mass range in order to calibrate the mass-observable relationships.  Figure \ref{fig:selfunc} shows that the number density and mass scale of the mid-mass clusters is important for this calibration.  The sensitivity to mid-mass clusters shown in Figure \ref{fig:av_sal} suggests that the network is using these clusters for calibration.

\subsubsection{A New Self-Calibration
  Mode for Flux- \& Volume-Limited Cluster Surveys}
\label{sec:nowlresults}

The above saliency analysis resulted in a discovery of a new
self-calibration mode for cluster surveys, an interesting result
that was possible only because the machine learning model was
interrogated rather than being accepted as a black box.  The analysis
was prompted by the results from a final alternative inputs test case
described in \S~\ref{sec:nonstdresults} where we remove weak lensing
masses from the supplementary inputs. 

\new{Omitting weak lensing masses effectively removes all
information about the absolute cluster mass scale from the ML inputs 
(except for calibration that could be derived from loose priors on the scaling
relation).  The scaling relation priors are equivalent to $\pm 30\%$ constraints on the
cluster mass for a given combination of observables (see
\S~\ref{sec:cosmomocks}). Without this mass calibration information, one naively
expects the same constraining power as from mapping only the total
number of clusters to \Sig. However, we find a significantly better
\Sig\ uncertainties compared to the $N_{\mathrm{cluster}}$-only case:
$\Delta\Sig=0.030$ ($\pm0.001$) for the No WL case compared to $0.065$ ($\pm0.003$) for the $N_{\mathrm{cluster}}$-only case (see Table
\ref{table:results} and accompanying sections).}

Note that the \Sig\ constraints in the $N_{\mathrm{cluster}}$-Only
case are indeed consistent with a 30\% a priori knowledge of the
cluster mass scale \cite[cf.~][with a 9\% systematic uncertainty on
masses translating to $\pm 0.023$ uncertainty in
\Sig]{2009ApJ...692.1060V}. A $\pm 0.03$ uncertainty in $\Sig$ should
be equivalent to a $\pm 12\%$ knowledge of the cluster masses. We do
not provide such information directly, therefore the CNN must
have inferred cluster masses internally. 

The ability to obtain
fundamental cluster properties together with the cosmological
parameters is referred to as ``self-calibration''
\citep{2002ApJ...577..569L,2004ApJ...613...41M,2003PhRvD..67h1304H}. 
Previously suggested self-calibration modes use either the cluster
clustering information together with the number counts
\cite{2003PhRvD..67h1304H,2004ApJ...613...41M}, or postulate a known
redshift evolution in, e.g., the mass-observable relations \citep[as
in][]{2002ApJ...577..569L}. We, however, explicitly do not provide
either the clustering information or redshift evolution as an
input. Therefore, the machine learning model must be using a
previously unknown self-calibration mode. This realization prompted a
further investigation.

\begin{figure}[]
	\includegraphics[width=0.5\textwidth]{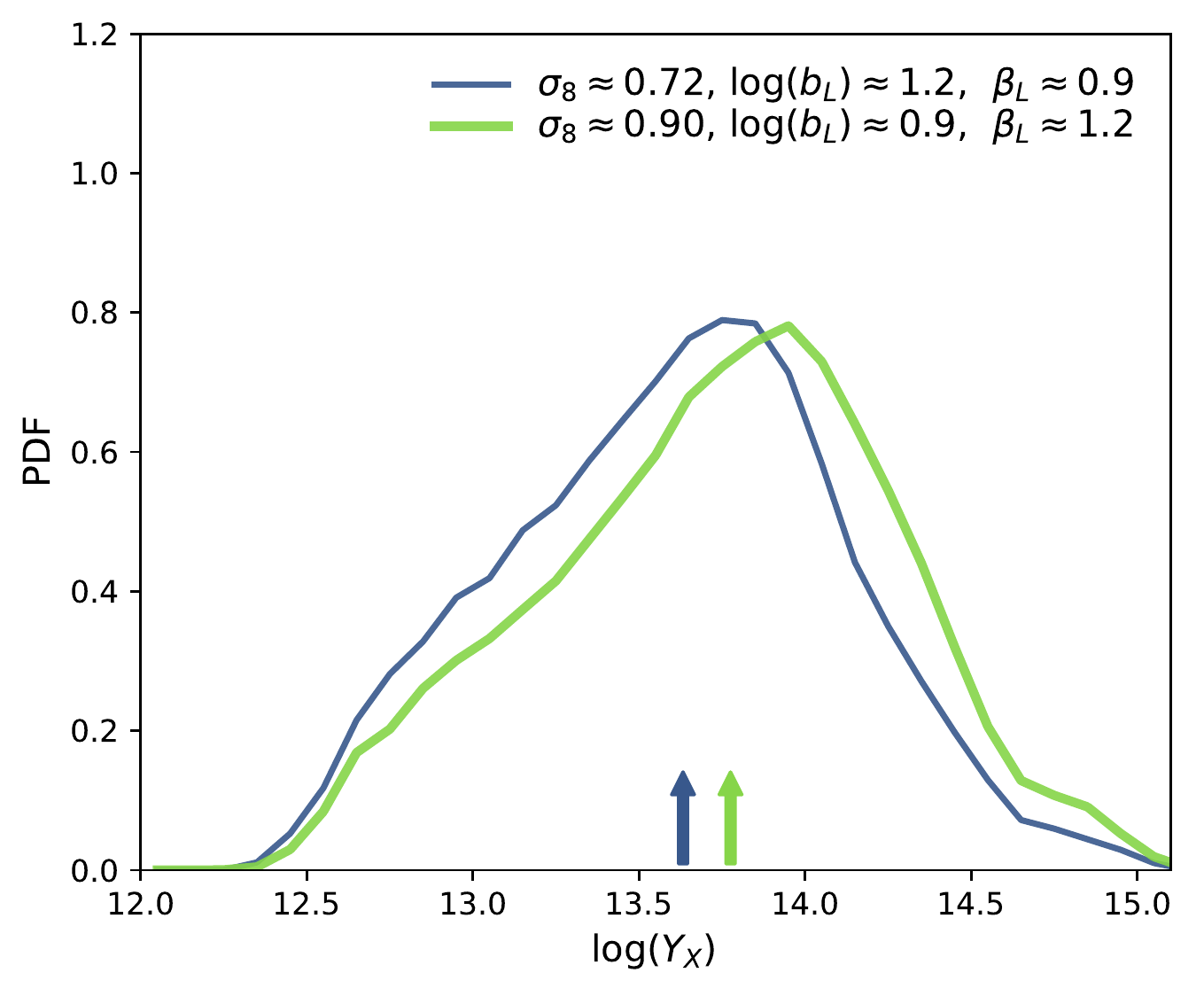} 
	\includegraphics[width=0.5\textwidth]{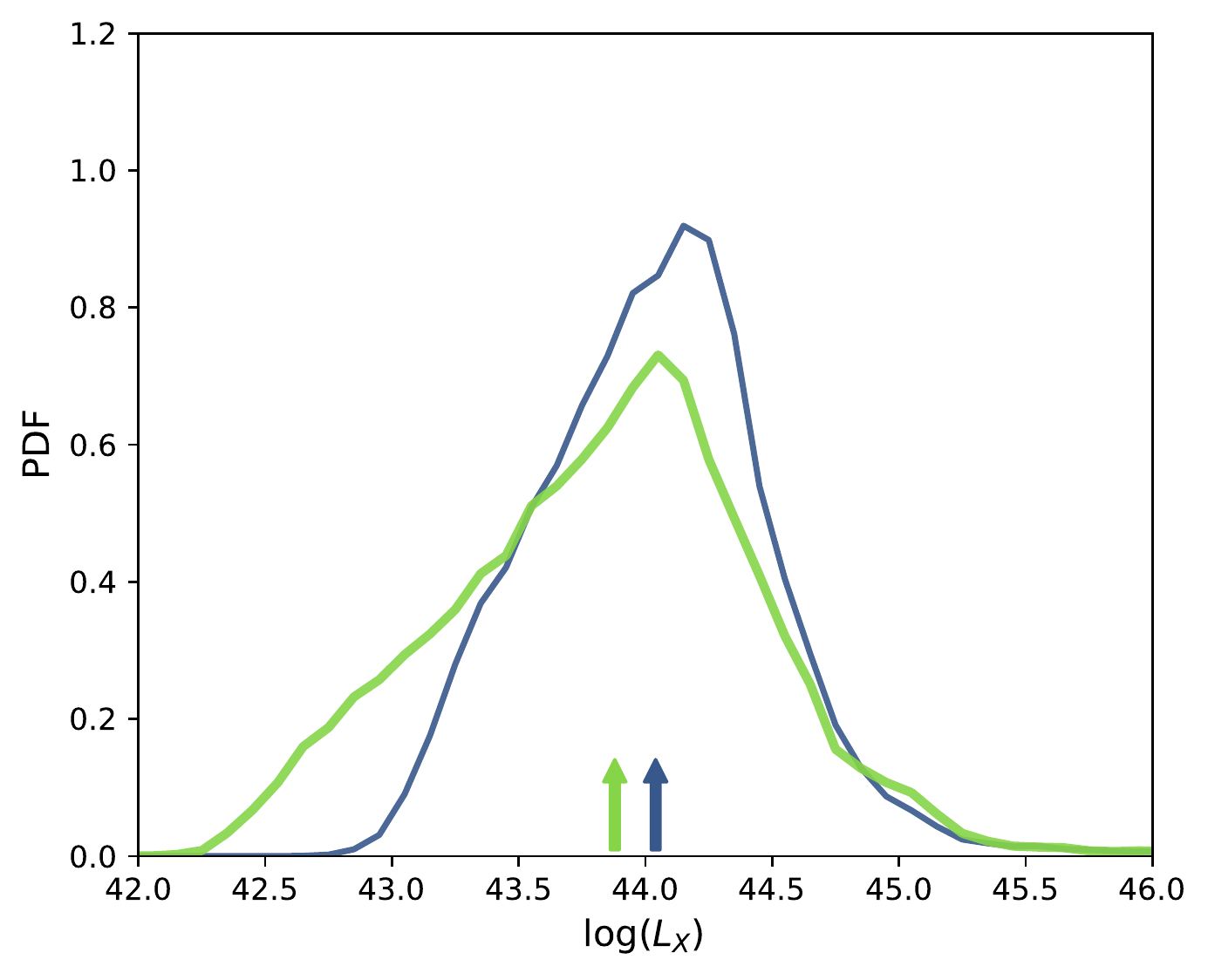} 
	\caption[]{\new{The averaged PDFs of $Y_X$ (top) and $L_X$ (bottom) for two sets of $\sim600$ simulated catalogs.  The median of these distributions (indicated by the arrows) are a good proxy for the mass scale at which the catalog transitions from flux- to volume-limited.}}
       	\label{fig:selfcal}	
\end{figure}

Using the average encoder saliency approach, we investigated which
mass range provides the most constraining power in this case. The
saliency at $r=2/3\, R_{500}$ as a function of cluster mass was used
for this analysis, by comparing the saliency for both the Standard and
No Weak Lensing Catalog results. Indeed, the saliency as a function of
mass shows a drastic rearrangement when the WL information is removed
(Figure \ref{fig:altcat_saliency}). Instead of fairly flat
distribution, both the gas mass and temperature saliencies show
maximum sensitivities to clusters close to the mass range where the
$V(M)$ function switches from the volume-limited mode to the
flux-limited mode.  This mode transition corresponds to a peak in the
PDF distribution of clusters (see Figure \ref{fig:selfunc}).  We,
therefore, are led to believe that the CNN identifies and uses the
peak in the distribution of selected clusters as a function of the
mass proxy. 

In our survey scenario (\S\ref{sec:selfunc}), the peak in
the cluster PDF corresponds to the break point in the selection
function $V(M)$, where the cluster sample transitions from a volume-
to an X-ray-flux-limited regime.  Figure \ref{fig:selfcal} shows that,
indeed, this break point is sufficiently well approximated by the \new{median
$L_X$ in each simulated cluster sample, providing a convenient and observable proxy for the transition point.  Furthermore, the median $L_X$ is something that the ML machinery
can easily infer from the data.}
  
\new{Next, we will explore whether there is sufficient information contained in 
two parameters \textemdash{} the total number of clusters and the 
location of mass PDF peak \textemdash{} to explain the surprising No WL results.
Consider the amplitude ($\hat{N}_{\mathrm{peak}}$) 
and location ($M_{\mathrm{peak}}$)  of the peak in the cluster PDF.
The information contained in these as a function of a low-scatter mass proxy or luminosity can be
expressed as the following systems of equations:}
\begin{equation}
  \begin{split}
    \hat{N}_{\mathrm{peak}} & = n(\Sig, M_{\mathrm{peak}}) \\
    M_{\mathrm{peak}} & = m_{Y}(\hat{Y}_{X, \mathrm{peak}}).
  \end{split}
\end{equation}
\new{where hat indicates observable quantities and we use
  $Y_{X}$ as a representative low-scatter mass proxy. The function
$n(\Sig,M_{\emph{peak}})$ follows from the cosmological cluster mass
function model, which we assume is precisely known a priori. 
The function $m_{Y}(Y_{X})$ is simply a power law with a tightly constrained slope, small scatter, and free
normalization and, therefore, is also tightly constrained a priori.
We now have a system of
two equations with two unknown (\Sig and $M_{\mathrm{peak}}$), which
can be solved for \Sig.}

\new{This analysis is not limited to the observable $Y_X$, and a similar argument can be made for the system of equations that considers the observable $L_X$:}
\begin{equation}
  \begin{split}
    \hat{N}_{\mathrm{peak}} & = n(\Sig, M_{peak}) \\
    M_{\mathrm{peak}} & = m_{L}(\hat{L}_{X, \mathrm{peak}}).
  \end{split}
\end{equation}
\new{Compared to $m_Y$, $m_{L}(\hat{L}_{X, \mathrm{peak}})$ is more loosely constrained (a power law,
with loosely constrained slope and scatter, and with free
normalization).  Because of this, using $Y_X$ to find the peak mass should produce
better results because its correlation with mass is more tightly
constrained.}

While we cannot establish that this is precisely how the CNN is
calibrating masses, we have clearly established a theory that can
explain the CNN's results: the information in the location and the
peak value of the cluster PDF distribution contains enough information
to constrain cosmology. The saliency analysis (\S \ref{sec:sal} and
Figures \ref{fig:sample_sal}-\ref{fig:altcat_saliency}) also clearly
indicates that the CNN is primarily using the data near the peak of
the PDF in this case. 
\new{Furthermore, we used the leave-one-out approach to confirm that the removal of a single cluster does not significantly change the output.  Independent of the mass of the omitted cluster, there is no strong systematic change to the estimated \sig, suggesting that the model has not learned strong dependencies on the shape of the halo mass function.} 
We can conclude that at the minimum, the CNN
performance and the analysis of saliencies in the no-WL case has led
to an identification of a previously unknown mode of self-calibration
for cluster surveys.

We can also verify that the constraining power from using the location
and amplitude of the cluster PDF peak approximately corresponds to the
\Sig\ uncertainty in the no-WL case.  Figure \ref{fig:ngalcorr} shows
this correlation between $N_{\mathrm{cluster}}$ and $Y_X$ more
plainly.  The correlation in the two highlighted sample regions is
$\sim 0.03$ and falls toward the edges of the allowed
$N_{\mathrm{cluster}}-Y_X$ space, where our priors on the allowed
models tighten the distributions of \sig.  This is roughly comparable
to the $0.0297\pm0.0013 $ constraints on \sig in the No WL
case. Therefore, we believe that a self-calibration mode similar to
that described above was in fact ``found'' and implemented by the CNN.

The discovery of this new self-calibration mode was facilitated by the process of attempting to interpret the machine learning model.  Deep learning models can be more flexible and more adept at finding complex patterns compared to more standard statistical techniques (such as Fischer matrix analyses).  Deep machine learning's flexibility, coupled with the interpretation schemes, aided the discovery of this previously unknown method for mass calibration.  
This self-calibration mode can be useful for large-scale X-ray surveys such as the upcoming eROSITA cluster sample, where the break point between the flux- and volume-limited samples can provide a handle on the absolute mass calibration.  \new{A further exploration of this self-calibration mode and its dependence on survey parameters is presented in Appendix \ref{sec:selfcal_appendix}.}

\begin{figure}[b]
	\includegraphics[width=0.5\textwidth]{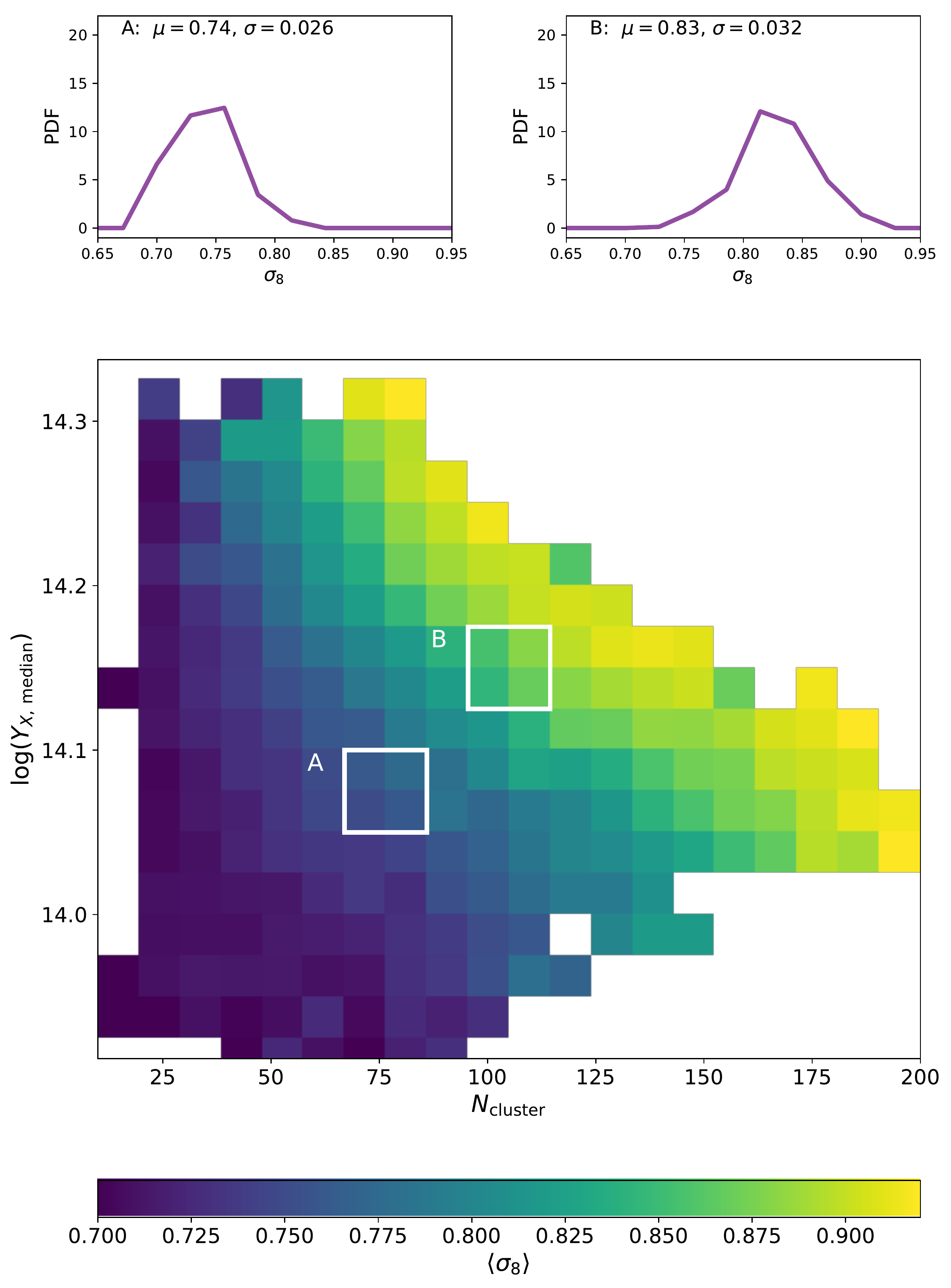} 
	\caption[]{Binned mean \Sig as a function of two parameters of the simulated catalog:  the number of clusters and the median $Y_X$ value (in $\msolarh\,\mathrm{keV}$).  In this two-dimensional space, there are clear correlations between the observables and the underlying \sig.  The constraining power of these two parameters is shown for two highlighted sample regions (labeled A and B, and shown in the popout figures, top).  For a given measurement of $N_\mathrm{cluster}$ and median $Y_X$, the underlying \sig is constrained to $\sim0.03$.  This self-calibration mode can be utilized in large X-ray surveys that have a break point between flux- and volume-limited samples; the upcoming eROSITA cluster sample is one such survey.}
       	\label{fig:ngalcorr}	
\end{figure}

\subsection{Correlations in the Terse Layer}
\label{sec:tersecorr}

\begin{figure*}[]
	\begin{tabular}{c c}
		 \includegraphics[width=0.43\textwidth]{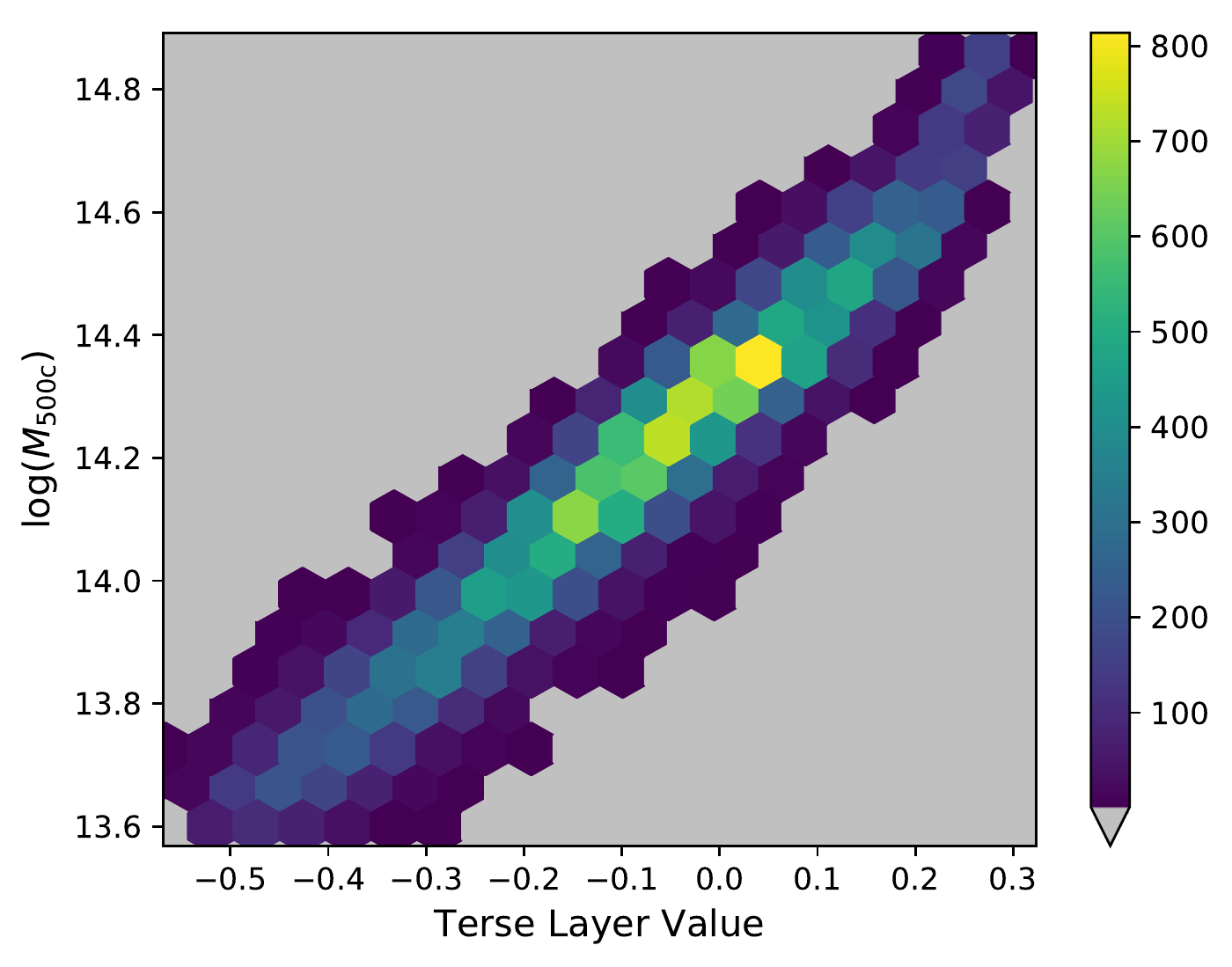} & \includegraphics[width=0.43\textwidth]{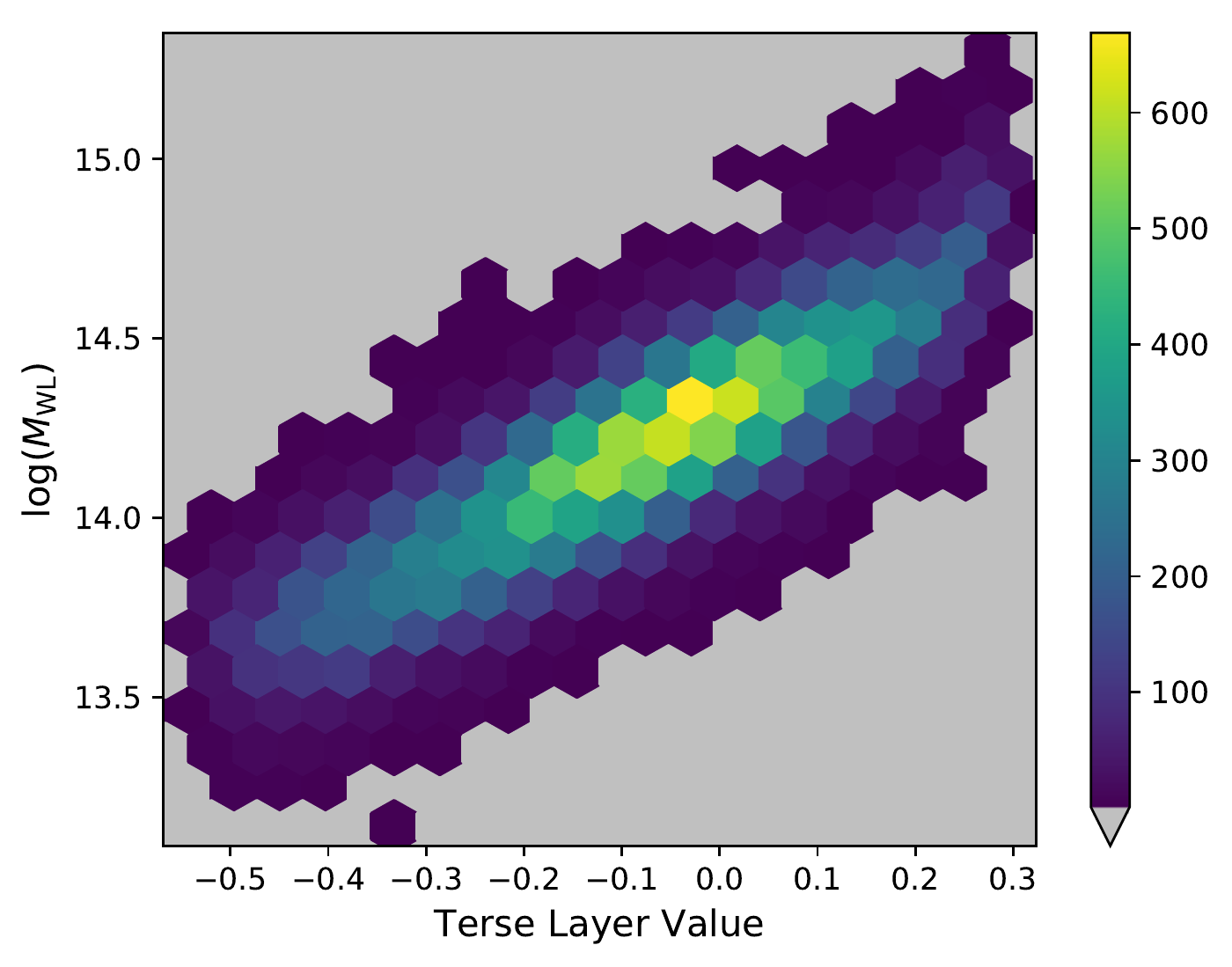}\\
		  \includegraphics[width=0.43\textwidth]{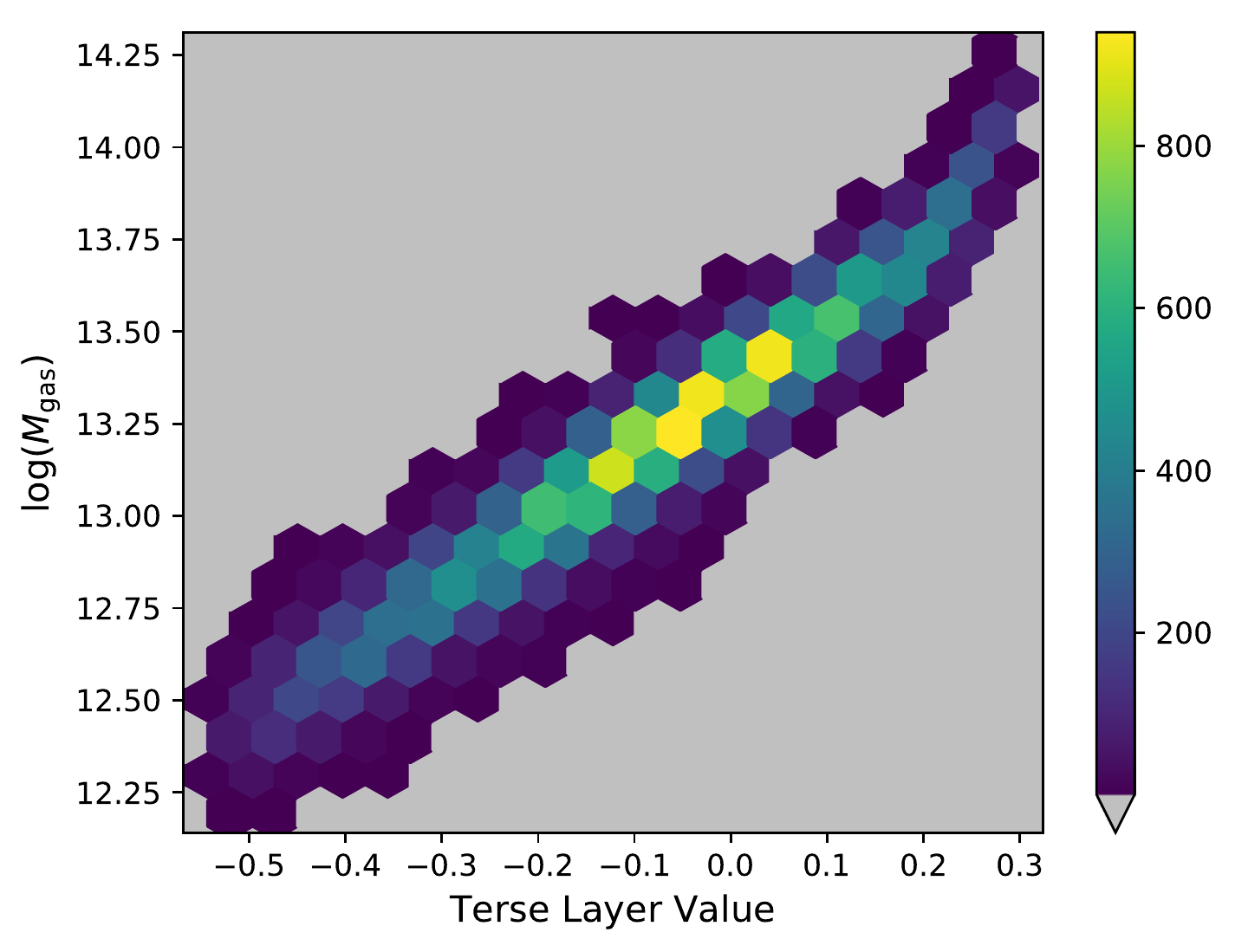} & \includegraphics[width=0.43\textwidth]{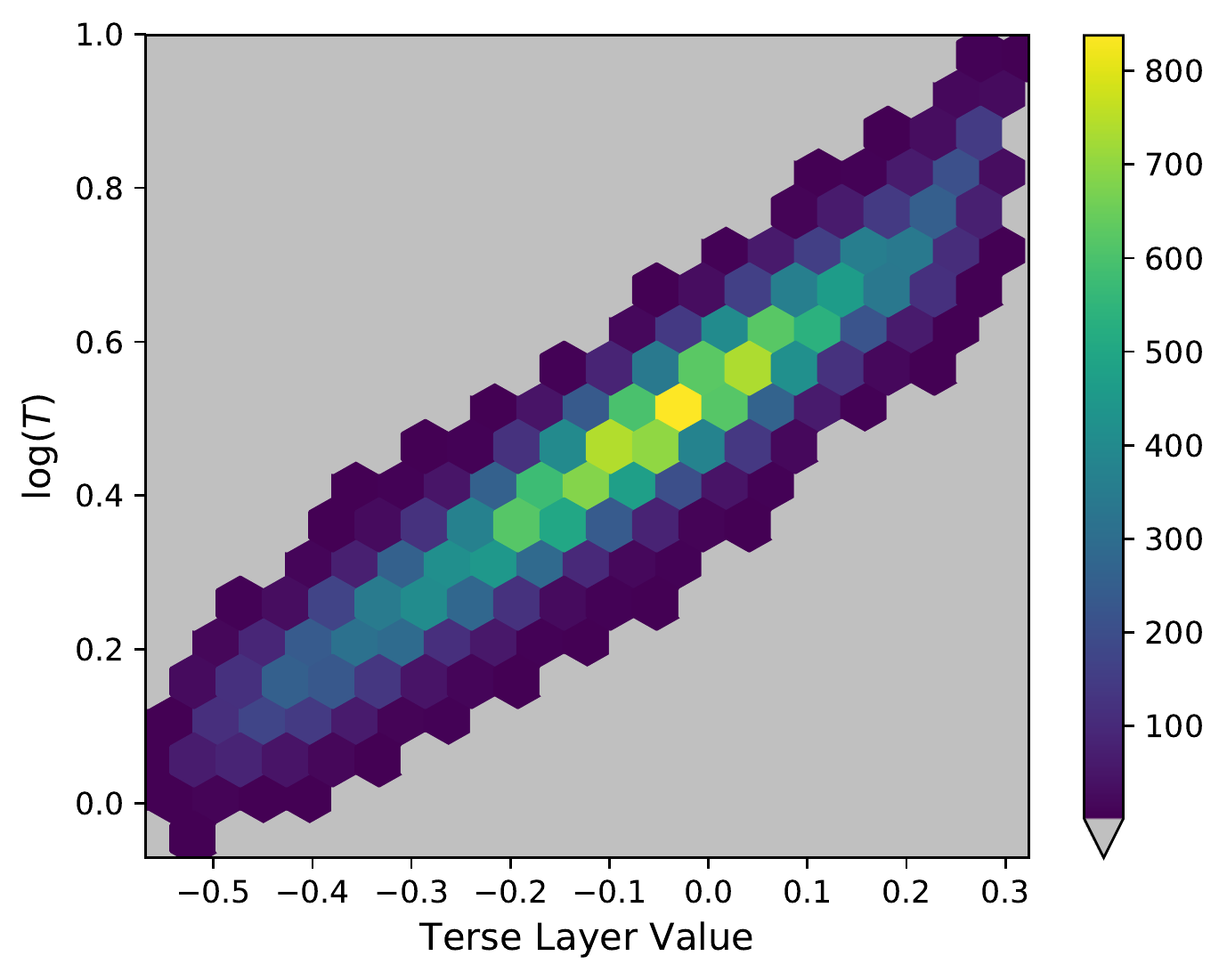}\\
		 \includegraphics[width=0.43\textwidth]{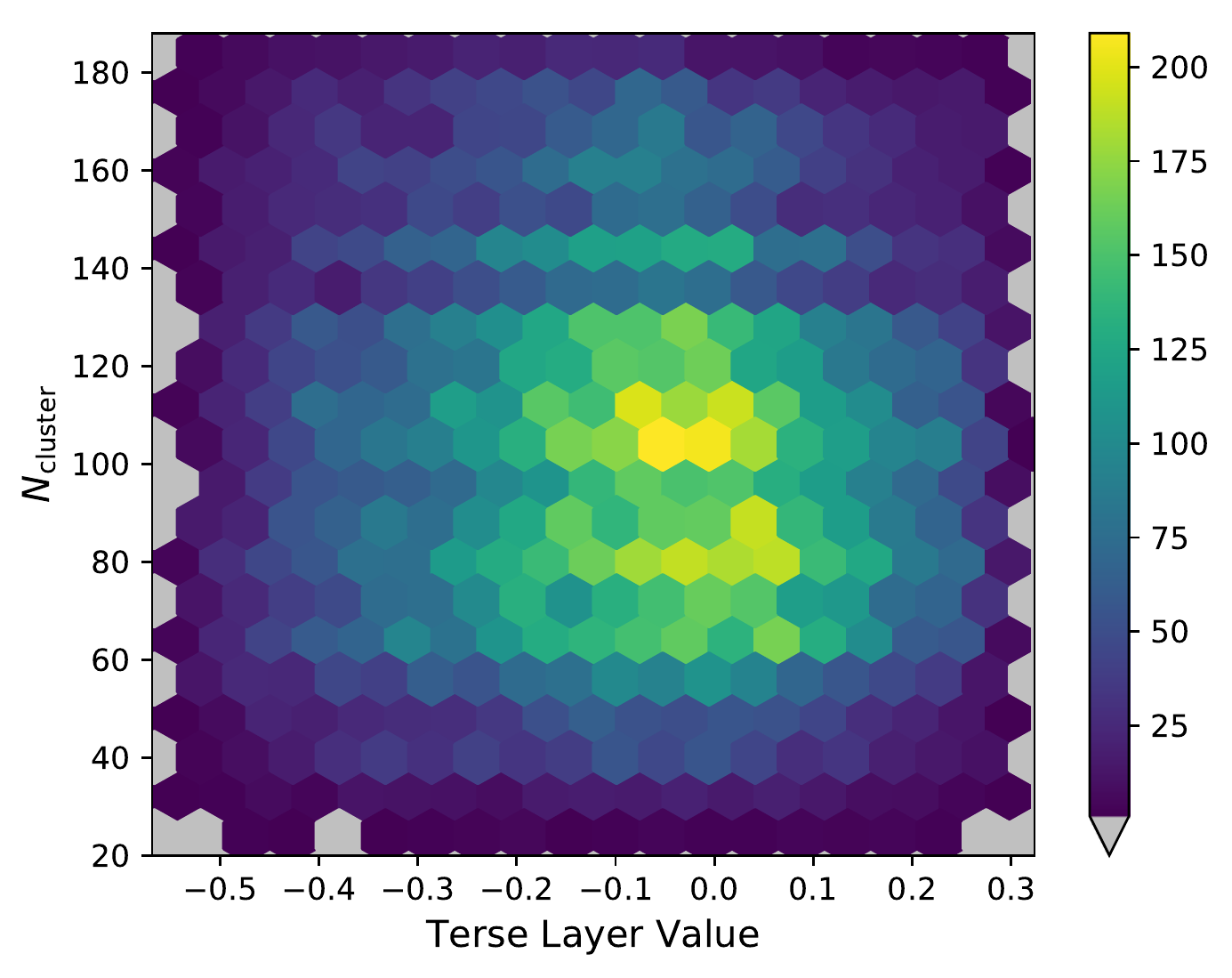} & \includegraphics[width=0.43\textwidth]{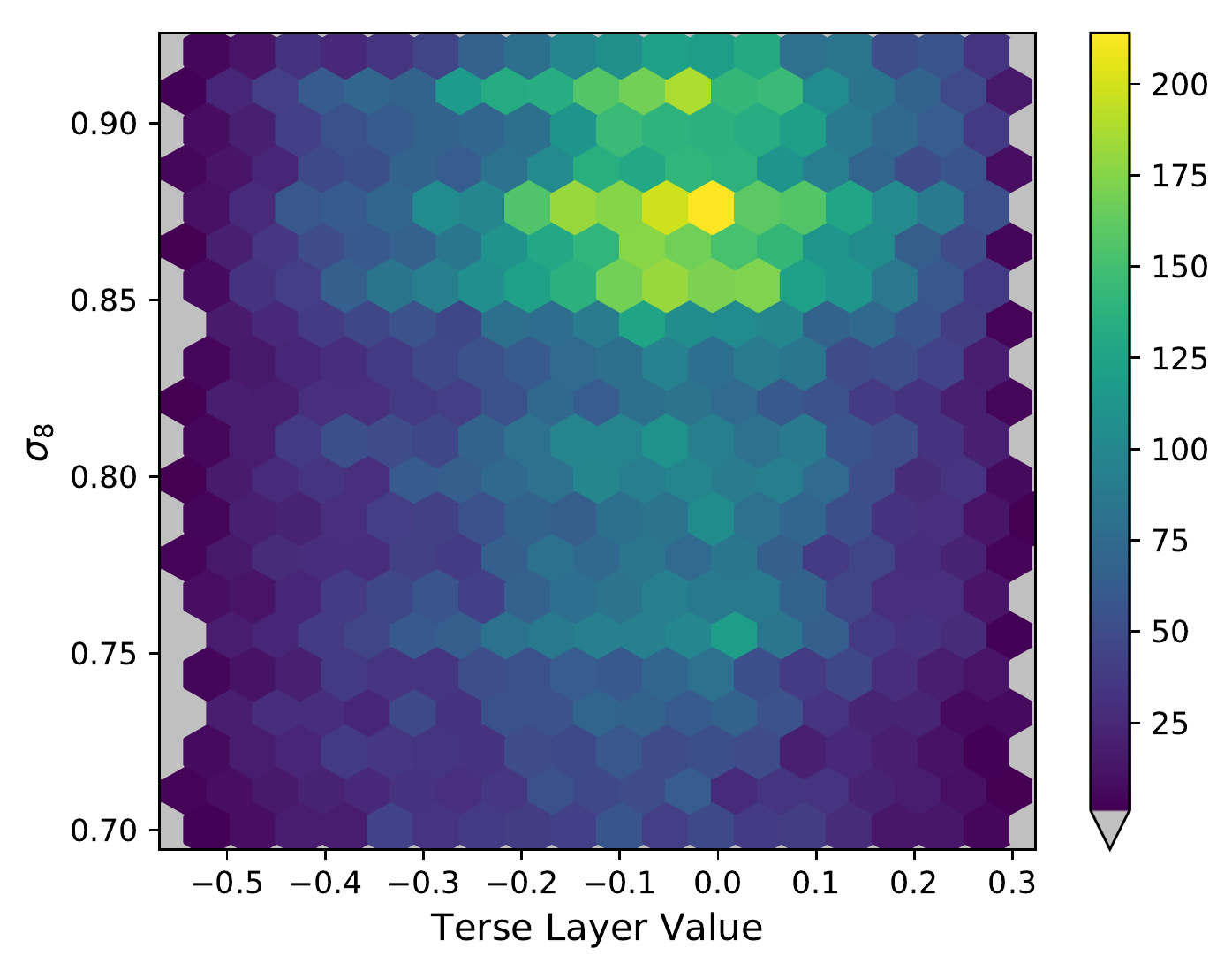}
	\end{tabular}
	\caption[]{2D Histograms of \new{all clusters in the testing set, showing} how the terse value correlates with physical parameters (the colorbar shows the number of clusters per hexagonal bin).  These correlations with the terse layer provide a valuable check of the physical underpinning driving the results of the machine learning model.  
	Top: the cluster mass (left) is correlated with the terse value.  If the terse value is some mass proxy, we expect its correlation to be more scattered with the noisy weak lensing proxy (right) and this is the case.  
	Middle:  Correlations of the terse value with input parameters can reveal whether the neural network is using one input exclusively or if it is building some nonlinear combination of the inputs.  Tight correlations with both gas mass (left) and also with temperature (right) suggest that the model has built a proxy from both of these parameters, in agreement with the saliency maps analysis.
	Bottom:  Because of the unique shape of the architecture, the terse value should encode information about a single cluster, but not about the cluster population.  \new{For these panels, the terse value for each cluster is labeled by the underlying global properties of the catalog.}  Any correlations of the terse value with the number of clusters (left) or with \Sig (right) would be worrisome, indicating that the model had accessed information about the underlying cosmology from a single simulated galaxy cluster.  Fortunately, these cosmological parameters are uncorrelated with the terse value. Assessing correlations in the terse layer is one way for a researcher to understand if and where a model can be applied safely.  }
       	\label{fig:terse}	
\end{figure*}

In the previous section, we explored saliency, the gradient of the terse value relative to changes in the input.  Here, we will explore how the terse values directly correlate with cluster parameters.

Figure \ref{fig:terse} shows a series of 2D histograms that show correlations between cluster parameters and the terse value.  The terse value correlates with cluster mass, suggesting that the encoder does, in fact, learn that cluster mass is a very useful parameter for extracting cosmological information.  Near the edges of the terse value range, the terse value loses sensitivity to mass, albeit subtly, where the 2D histogram twists vertically.  There are similar correlations between the terse value and the total gas mass $M_\mathrm{gas}$ as well as between the terse value and calculated $Y_X$.  Both $M_\mathrm{gas}$ and $Y_X$ are calculable from the \new{data contained in the 2D input}, but are not explicitly given.  

With this information about correlations in the terse value, we can return our attention to understanding the dark horizontal stripes in saliency from Figure \ref{fig:sample_sal}.  These do not represent a subpopulation of clusters (saliency does not correlate with, for example, the shape of the gas mass profile).  Instead, the dark stripes of low saliency are indicative of where a cluster falls in the terse-$M_{500}$ space.  These low-saliency clusters populate the low-sensitivity regime at the higest and lowest terse values.  Here, the terse value is stable even under larger-than-normal changes to the cluster parameters.  It is unsurprising, then, that saliency drops significantly for these clusters.

While correlations are important, the \textit{lack} of correlations is likewise informative.  Figure \ref{fig:terse} shows that there is no discernible correlation between the terse value and \Sig.  This is an important check \textemdash{} at the terse layer, the encoder has knowledge of a single galaxy cluster and should not be able to extract any cosmological information.  Correlations between the terse value and cosmological parameters would be concerning indications that the model was not learning as expected.  This is explored  in Appendix \ref{sec:cheat}.

A simple mass proxy can be constructed from the terse values in much the same way that a more traditional mass proxy would be fit to masses, by fitting the points in the terse-$\log(M_\mathrm{500})$ plane to a line.  The results of this fit are shown in Figure \ref{fig:massproxy}.  Recall that the scatter introduced in Section \ref{sec:YLTscatter} will introduce offsets to this linear fit.  However, we are really interested in the scatter on a single simulated cluster sample, not in the scatter marginalized over many cluster samples.  To roughly account for this, the mean error, $\bar{\delta}=\left<\log(M_\mathrm{predicted})-\log(M_\mathrm{true})\right>$, is calculated for each cluster ensemble and is subtracted to give an estimate of the typical scatter in this mass proxy.  An identical process produces the error PDFs for both gas mass and $Y_X$ as mass proxies.

As shown in Figure \ref{fig:massproxy}, this simple approach shows a tighter intrinsic scatter compared to gas mass, but which does not improve upon the state of the art, $Y_X$.  This result suggests that the model has learned about the anticorrelated errors of $M_\mathrm{gas}$ and $T$ as cluster mass proxies (see \cite{2006ApJ...650..128K} for more details), but has not fully taken advantage of the information in these correlated parameters.  Alternately, the simple linear fit and removal of offset may be inadequate to capture the role of the terse value as a mass proxy.

\new{We believe that we are working in a regime where the statistical uncertainties
(which are due to the small size of the cluster sample)
dominate those related to the intrinsic scatter in the cluster mass
estimates. This was explicitly verified by \cite{2009ApJ...692.1060V}
for the case of their selection function, which is similar to the one
we implement in this work. Specifically, they find that variations of
the scatter in the $M-Y_{X}$ relation by up to $\pm 50\%$ with respect
to the nominal values do not significantly change the inferred
cosmological parameters (see their \S8.4.1). Because the CNN is optimized 
to infer cosmological parameters, and because the largest source of error 
here is due to the small size of the cluster sample, 
there is little incentive for the model to identify the lowest-scatter mass proxy.}

\begin{figure}
	\includegraphics[width=0.45\textwidth]{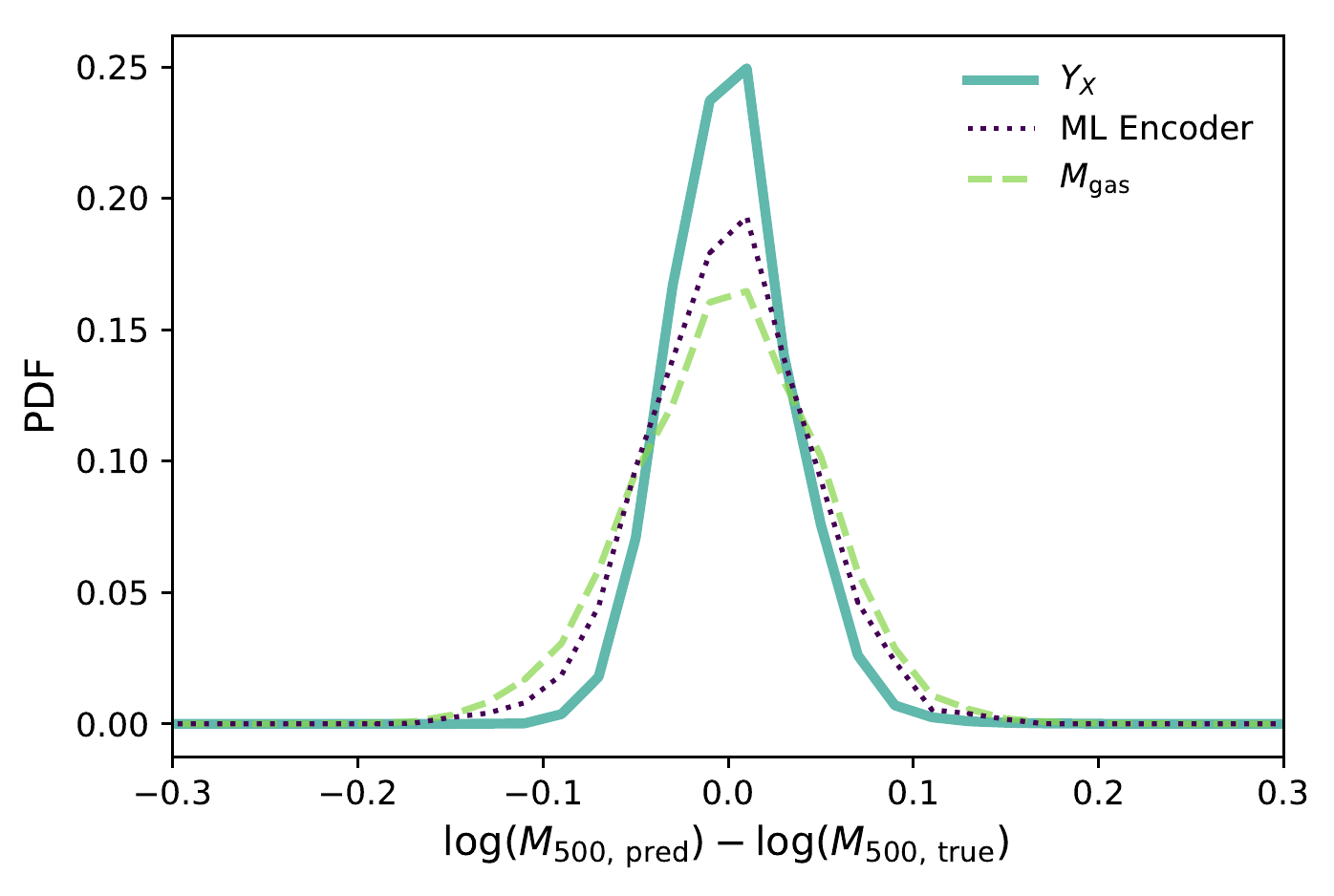}
	\caption[]{Gas mass (lime green dashed) is a low-scatter mass proxy; for our catalog, the proxy has intrinsic scatter $17.9\%$.  A second mass proxy, $Y_X$ (teal solid, calculated as the product of gas mass and spectral temperature) is a lower-scatter proxy, $7.3\%$.  When an ML-derived mass proxy is constructed from the terse values (purple dotted), the proxy has $16.1\%$ scatter, an improvement over using the cluster gas mass alone, but not as good as $Y_X$.  This may be because the machine learning model has failed to fully take advantage of correlated input data, or it may be because a simple linear fit to the terse-$\log(M_\mathrm{500})$ space is insufficient to capture the complex relationship between the terse value and the cluster mass.  Regardless of the origin, the exercise of checking the terse value as a mass proxy provides a simple test for understanding how the model derives \Sig from observables.}
	\label{fig:massproxy}
\end{figure}

\section{Discussion \& Conclusion}
\label{sec:conclusion}

Machine learning is often touted as a way to make order-of-magnitude improvements at the expense of understandability.  Yet, without interrogation and interpretation, the methods \textemdash{} and the results! \textemdash{} can be untrustworthy.  In this work, we have laid out a physically motivated, interpretable, supervised, multi-wavelength galaxy cluster encoder.  Our main conclusions are:
\begin{enumerate}
	\item Building an architecture carefully can make it inherently more interpretable.  Our supervised encoder architecture has a bottleneck that can be physically understood.  We have shown that, similar to the way that a human would approach the problem, the neural network constructs a cluster mass proxy for each cluster in the simulated sample.
	\item The model constrains \Sig to $0.0212\pm0.0009$, at approximately the theoretical limit of cosmological constraints for an observation of 100 clusters.
	\item A leave-one-out method is used to assess the information content in each cluster.  At low masses, the scatter in $\Delta\Sig$ sets the stability of the model at $\sim0.003$.  However, removing a cluster with $\log(M_\mathrm{500})\gtrsim14.2$ tends to lower the \Sig prediction.  This is expected, as removing a massive cluster from a simulated catalog will effectively shift the cluster mass function to mimic a lower-\Sig cosmology.
	\item Saliency maps, a visualization of $\partial(\mathrm{terse})/\partial (\mathrm{pixel})$, are used to assess the information content in input features.  We explore saliency averaged over many clusters to understand the terse value's sensitivity to the input features.  Consistent with previous literature, we find that gas masses in the $0.15$ to $1\:R_{500}$ range are the most informative, while the cluster cores carry less information. 
	\item We present a previously unknown self-calibration technique for cluster surveys.   This self-calibration mode was discovered by removing input data and investigating changes in saliency.   Through this investigation, we found that two catalog measurements \textemdash{} the total number of clusters in the sample and the median cluster $Y_X$ value \textemdash{} together contain sufficient information to calibrate masses and to constrain cosmology. Because this self-calibration model does not require additional cluster weak lensing observations, it can be useful for large-scale X-ray surveys such as the upcoming eROSITA all-sky survey, where the break point between the flux- and volume-limited samples can provide a handle on the absolute mass calibration.  
	\item We show that correlations in the terse layer can provide an essential check to determining whether a neural network has developed a reasonable and physically motivated model. \end{enumerate}

This work focuses on an often-ignored aspect of machine learning in astronomy \textemdash{} interpretability.  Machine learning techniques will be indispensable for the cosmological community to fully utilize the huge observational and simulated data sets on the horizon.  Though they offer enticing promise of improved results, machine learning methodologies cannot be applied blindly.  Interpretation schemes (such as leave-one-out, average saliency, and correlations in terse space) are an essential first step before confidently applying ML models to observational data.  \\

\acknowledgments{\new{We thank our anonymous reviewer for their thoughtful and helpful feedback.  In particular, for their suggestions that led to the results presented in Appendix \ref{sec:selfcal_appendix}.} Thanks to David Parkes for the useful discussions
  about overfitting and for suggesting the term ``overspecialized'' to
  describe a model that may not generalize. 
  We also thank Francie Cashman, Ana Maria Delgado, Klaus Dolag, Jesse Dunietz, Daniel Eisenstein, Gus Evrard, Doug Finkbeiner, Brianna Galgano, Sheridan
  Green, Matt Ho, Daisuke Nagai, Josh Peek, Melinda Soares-Furtado, and John Soltis for their helpful feedback on
  this project and manuscript.  MN was supported in part for this research by the
  Harvard Data Science Initiative.  The computations in this paper
  were run on the FASRC Cannon cluster supported by the FAS Division
  of Science Research Computing Group at Harvard University.\\}

\bibliographystyle{aasjournal}
\bibliography{references}

\appendix

\section{\new{Dependence of the Self-Calibration Mode on Survey Strategy}}
\label{sec:selfcal_appendix}

The Standard Catalog described in \S\ref{sec:input} included a simplifying assumption (the sharp transition between the flux- and volume-limited sample) and was explored for one particular volume (half of the simulation volume).  Forward modeling approaches, such as the new self-calibration mode presented in \S\ref{sec:nowlresults}, are dependent on properly modeling all relevant details of the observation and in this Appendix, we consider how our self-calibration results depend on choices in survey modeling.  

Recall that, while the discovery of the self calibration mode was ML-aided, the \textit{implementation} is purely statistical.  This appendix explores how the distributions of \sig{} for a given $N_\mathrm{cluster}$ and $Y_\mathrm{X,\,median}$ might differ when the catalog is modeled more realistically; it is a purely statistical approach.

\begin{figure}[]
\centering
			\includegraphics[width=0.45\textwidth]{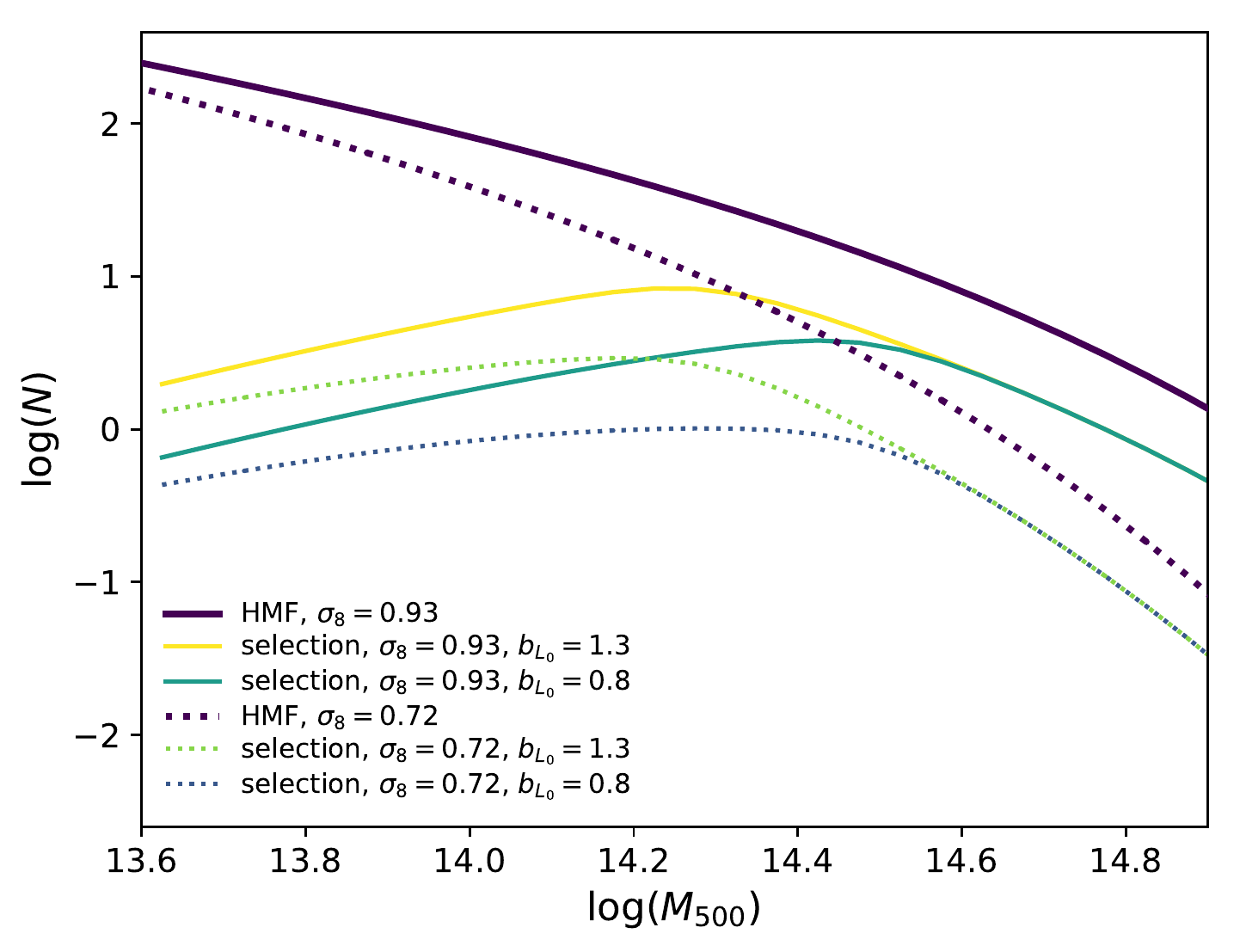} \\
			\includegraphics[width=0.45\textwidth]{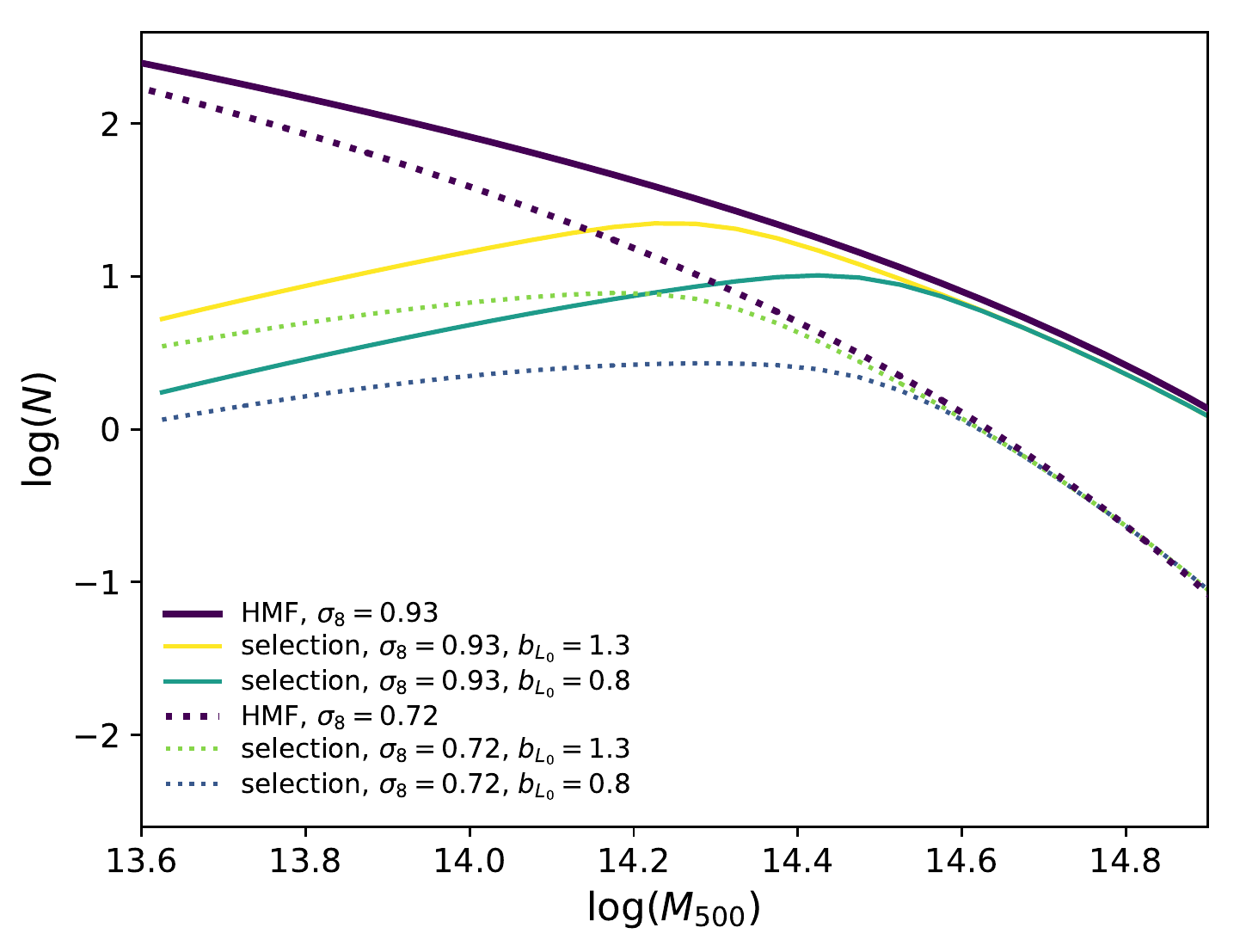} \\
    \caption[]{Survey cluster abundances that include two changes to the standard catalog: 1.~a smoothed transition from the flux-limited sample at low mass to the volume-limited sample at high mass and 2.~a change in survey volume.  The smaller survey volume (top) is 30\% of the simulation volume, while the larger survey volume (bottom) is 80\% of the simulation volume.  \\}
\label{fig:smoothed}
\end{figure}

\begin{figure*}[]
		\includegraphics[width=\textwidth]{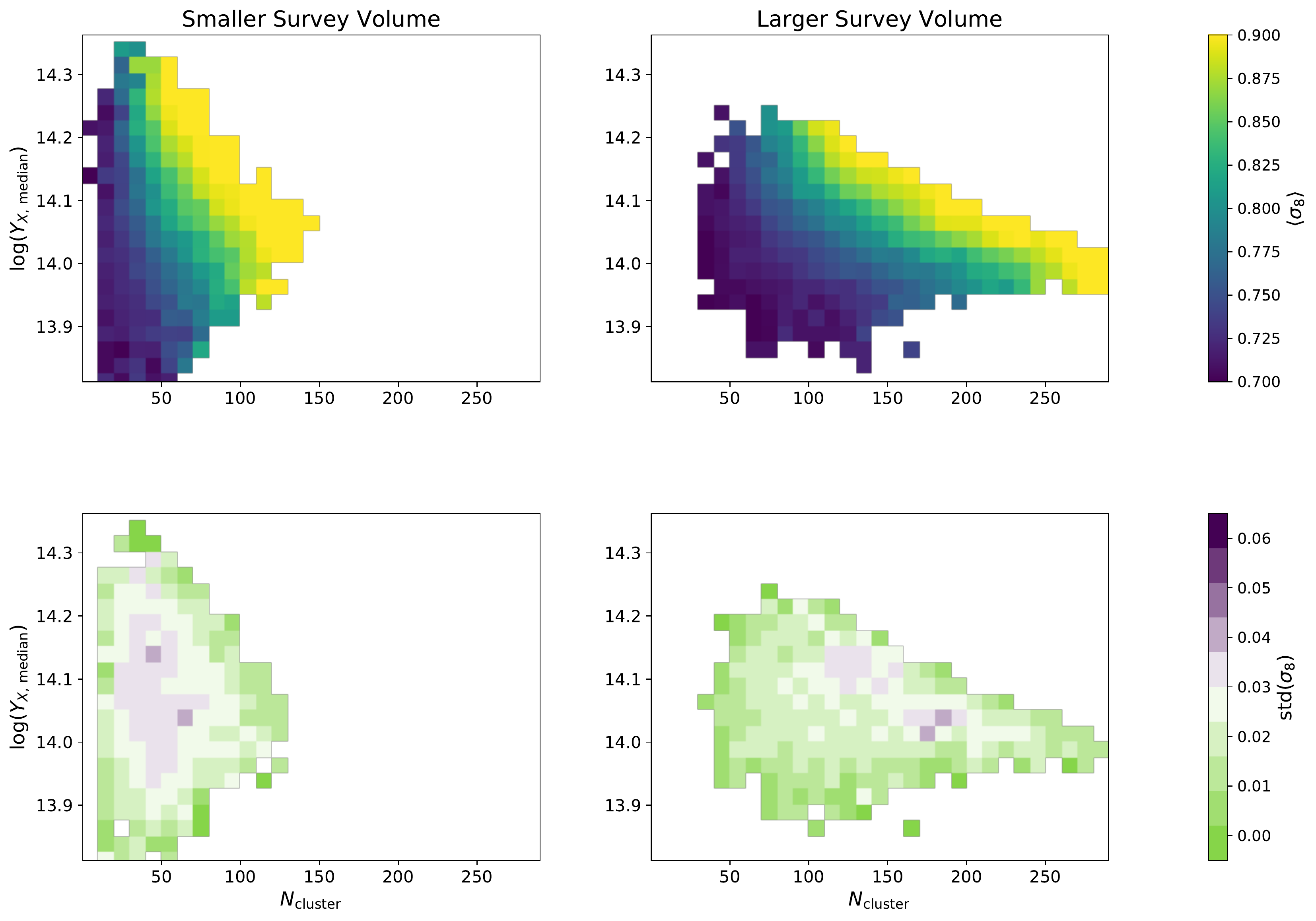} \\
    \caption[]{Top panels:  Binned mean \sig for a survey volume that is 30\% the size of the simulation (``Smaller Survey Volume,'' left panel) and 80\% the of simulation (``Larger Survey Volume,'' right panel).  The obvious differences between these two distributions highlight an important result:  average \sig as a function of $N_\mathrm{cluster}$ and $Y_{X,\,\mathrm{median}}$ depends strongly on the survey details.  Applying this self-calibration mode requires one to first forward-model the survey, including selecting appropriate priors on survey details. Bottom: the binned scatter in \sig values.  We saw in Figure \ref{fig:ngalcorr} that, for our Standard Catalog, \sig could be constrained to $\sim0.03$ with the new self-calibration mode.  For the Smaller (left) and Larger (right) Survey Volumes, we can place similar constraints on \sig.  In practice, an observed X-ray survey should be modeled as accurately as possible (e.g., survey volume and selection effects) before applying the self-calibration mode described in \S\ref{sec:nowlresults} because the model is dependent on survey details.}
\label{fig:fwdmodel}
\end{figure*}

To prepare the updated catalog, first, we smooth the transition between the flux- and volume-limited samples.  While the Standard Catalog cluster abundance has a sharp peak at this stitch point, the transition is much smoother in actual X-ray cluster surveys.  To model the smooth transition between the flux-and volume-limited samples, we apply a Gaussian smoothing to the original cluster abundance.  The scatter in $\log(M_\mathrm{500})$ for the $L_X-M$ scaling relation sets the smoothing width $\Delta \log(M) = 0.076$.  These cluster abundances are shown in Figure \ref{fig:smoothed}.  Compared to the Standard Catalog's sharply peaked cluster abundance (shown previously in Figure \ref{fig:selfunc}), these smoothed cluster abundances more realistically model the expected binned masses of an X-ray survey.

The second catalog improvement is to vary the survey volume.  The Standard Catalog assumes a volume of 50\% of the \textit{Magneticum} simulation box, but in this Appendix, we consider a smaller survey volume that is 30\% of the \textit{Magneticum} simulation box and a larger survey volume that is 80\%.  A sample of representative selections for the smaller and larger survey volumes is also shown in Figure \ref{fig:smoothed}.

Figure \ref{fig:fwdmodel} shows the binned mean \sig and binned \sig scatter for the smaller and larger volume surveys.  An interesting trend emerges in the \sig error figure: the error on \sig is smaller toward the edges of the region of interest.  This is to be expected because there are few parameter choices that result in $N_\mathrm{cluster}$ and $Y_{X,\,\mathrm{median}}$ values toward the edge of the regions of interest.  Toward the center of the region of interest, as before, we find errors on \sig of $\sim0.03$; the introduction of smoothing and the small changes in survey volume do not significantly change this result.

In Figure \ref{fig:fwdmodel}, the binned \sig values are derived directly from the simulated cluster samples.  This is a purely statistical, forward-modeling approach.  The obvious differences between left and right panels of this figure show the importance of properly forward modeling the details of the cluster survey, as varying the survey volume produces quite different $\left<\sig\right>$ for a given $N_\mathrm{cluster}$ and $Y_{X,\,\mathrm{median}}$.  

In practice, if one were to apply this self-calibration mode to an X-ray survey, the survey should be modeled as realistically as possible to include appropriate priors on \sig, simulation volume and redshift range, and details of the $L_X-M$ scaling relation.

\clearpage

\section{Overspecialization:\\How to Cheat with Deep Learning\\ (without really trying)}
\label{sec:cheat}

A single simulation is not precisely sufficient for describing a sample of galaxy clusters at an arbitrary cosmology, nor is it sufficient for capturing the range of allowed gas physics models.  In the context of cosmological constraints, galaxy cluster mass is often treated as the optimal compression of observations.  Though the cluster mass function is the primary cluster parameter that will be affected by the underlying cosmological model, it is not the only cluster observable that is affected.  X-ray luminosity, temperature, and gas mass, for example, also vary with cosmology.

Instead of introducing scatter as described in Section \ref{sec:YLTscatter}, one might also consider adopting a prescriptive scaling relation.  This scaling could be used to modify the simulated clusters to represent a sample from non-simulated cosmologies.  In this Appendix, we will proceed with this exercise of using a prescriptive scaling and assess the results.

\begin{figure}[b!]
	\includegraphics[width=0.45\textwidth]{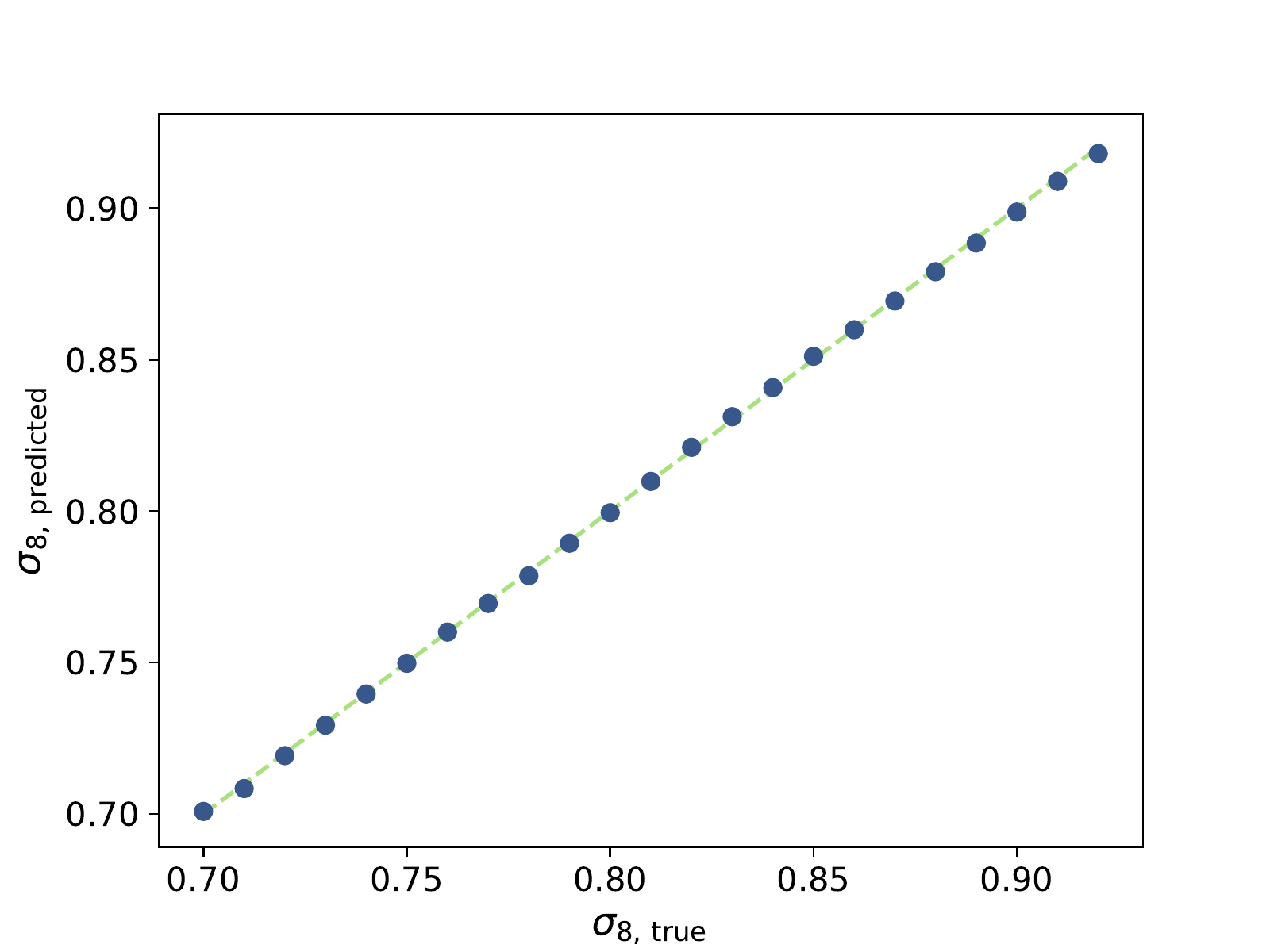}  
	\caption[]{Using a prescriptive scaling relation to modify cluster parameters (such as temperature and luminosity) as a function of \Sig gives dramatically improved cosmological constraints.  The predicted \Sig values (blue points) lie along the one-to-one line (green dashed) with no evidence of biasing toward the mean.  This is a remarkable, order-of-magnitude improvement in scatter ($\Delta \Sig = 0.0016$) compared to the results shown in Figure \ref{fig:results}.  Error bars on these predictions are too small to appear on the plot, but the ``improved'' results are not trustworthy.  As will be shown, the model has not learned to generalize and cannot be safely applied to true observations.  }
       	\label{fig:cheatresults}	
\end{figure}

For this exercise, the simulated catalogs differ from the method in Section \ref{sec:cosmomocks} in several key ways.  While the cluster abundance is identical to the one presented in Section \ref{sec:selfunc}, the supplementary data are not used and the 2D input matrices contain the full multi-wavelength information.  The 2D input contain the following features for each cluster:  the X-ray luminosity ($L_X$), the cluster global temperature ($T$),  the gas mass within $R_\mathrm{500}$ (\Mgas), $Y_\mathrm{500}$, and the weak lensing mass.  The prescriptive scaling relation for $L_X$, $T$, $M_\mathrm{gas}$, and $Y_{500}$ from \cite{2019arXiv191105751S} is adopted to scale these cluster properties to new cosmologies.

The architecture is identical to the one presented in Section \ref{sec:architecture}, with one exception: all instances of mean pooling layers are replaced with max pooling layers.  Training proceeds according to the method described  in Section \ref{sec:train} with the learning rate increased to 0.01.   

Figure \ref{fig:cheatresults} shows the resulting cosmological constraints.  These are a factor of $\sim10$ improvement over those presented in Figure \ref{fig:results}, but why?  To understand this, we must first define one term (\textit{overfit}) and introduce a second (\textit{overspecialized}).

A machine learning model is overfit when it has learned details of the training data set that do not generalize to the testing set.  Aggressive dropout \textemdash{} the practice of randomly turning off neurons during training so that the model cannot depend heavily on a few neurons \textemdash{} is one common way to combat overfitting \citep{srivastava2014dropout}.  Overfit results are illustrated in Figure \ref{fig:train}, at very high epochs where the MSE of the training set continues to drop while the validation and testing set MSE rise with epoch.  Overfitting is not seen seen in Figure \ref{fig:cheatmse}; the training and validation set MSE have not diverged during training.

When a model is \textit{overspecialized}, it has learned details of the training simulation that are not true of reality.  In contrast with overfitting, overspecialization is a problem with the simulated data, not with the training scheme.  Said another way, an overspecialized model may generalize \textit{to the data it's been given to work with}, but there is something inherently wrong with the data.  Simulations with subtle deviations from reality are one example of a flawed data set that may result in overspecialization.

\begin{figure}[t!]
	\centering
		\includegraphics[width=0.45\textwidth]{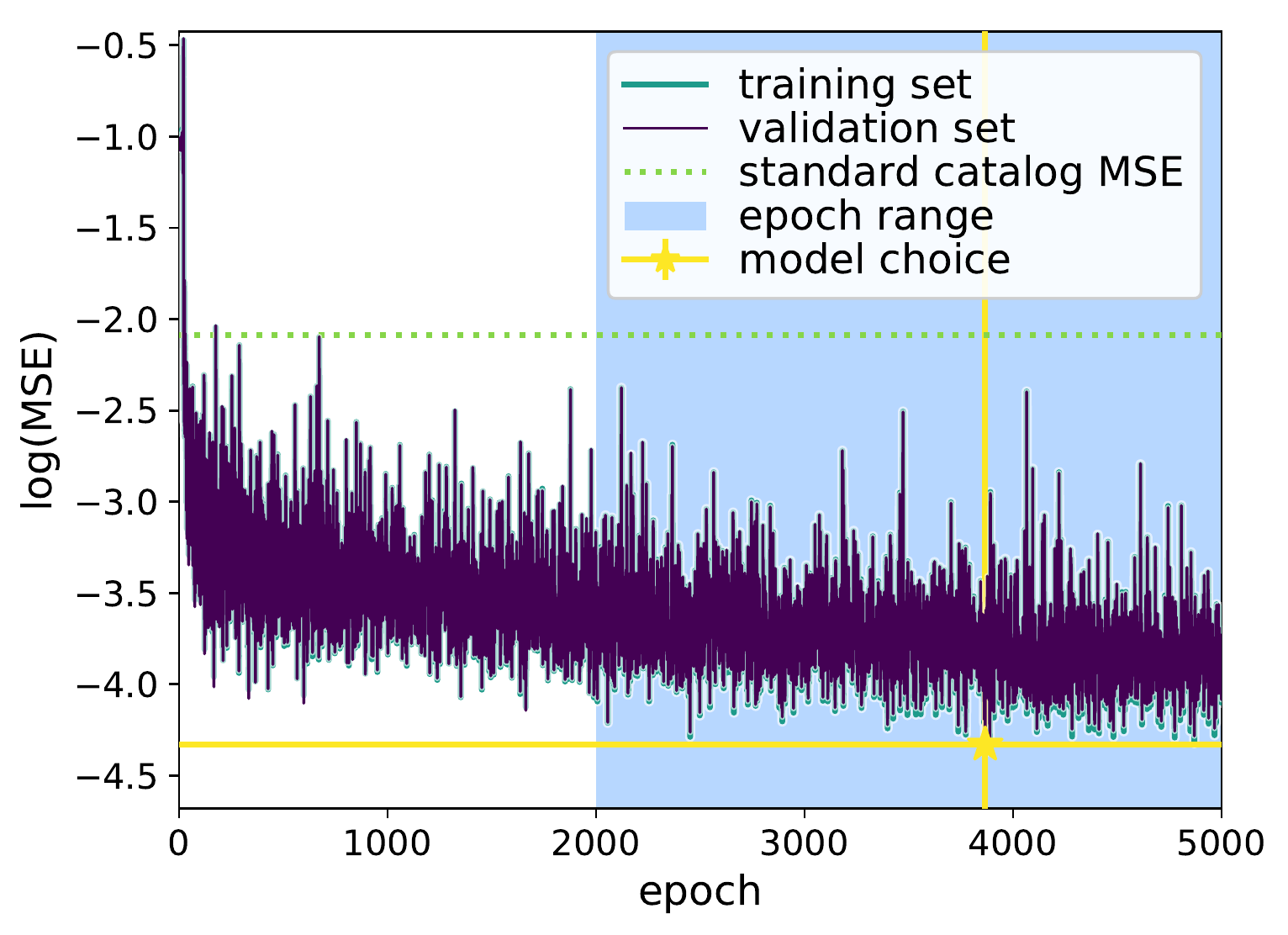}
	\caption[]{Mean squared error (MSE) as a function of training epoch for the training set (teal) and validation set (dark purple).  The best fit of the Standard Catalog and architecture (lime green dotted) is included for reference.  Note that this model is not overfit; the $\log(\mathrm{MSE})$ of the training and validation MSE track together so well that one is nearly imperceptible behind the other.  The cause of the ``remarkable'' tight constraints on \Sig cannot be determined from examining this figure, which shows no evidence of overfitting. }
       	\label{fig:cheatmse}	
\end{figure}

Overspecialization is difficult, though not impossible, to detect.  Interrogating a trained model by investigating correlations in the terse layer is one way to find evidence of overspecialization.  Figure \ref{fig:aliendragon2} shows concerning correlations between the terse layer values and cluster parameters for the toy model.  Cluster mass should encode the bulk of a cluster's cosmological information, yet the terse layer is virtually uncorrelated with cluster mass.  Furthermore, the terse layer should be completely uncorrelated with \Sig; this cosmological value requires a population of galaxy clusters.  But Figure \ref{fig:aliendragon2} shows a strong correlation between the terse value of a single cluster and the expected \Sig.  By investigating these often-unseen middle layers, overspecialization can be detected.

What could be causing the overspecialization?  In applying the prescriptive scaling relations to adjust $L_X$, $T$, $Y_X$, and $M_\mathrm{gas}$ as a function of \Sig, we have inadvertently encoded cosmological information in a way that the ML model can exploit.  The model can extract \Sig by reverse-engineering the prescriptive scaling relation.  (It is worth noting here that there is typically nothing wrong with prescriptive scaling relations for traditional statistical analyses, but a researcher must be careful when designing a training catalog for machine learning methods!)

As ML techniques become more prevalent in the astronomical community, overspecialization is one serious problem that will need to be addressed.  Understanding the ways that simulations and observations may disagree, coupled with a thorough interrogation of deep networks, is one way to find and manage overspecialization.

\begin{figure}[]
	\includegraphics[width=0.45\textwidth]{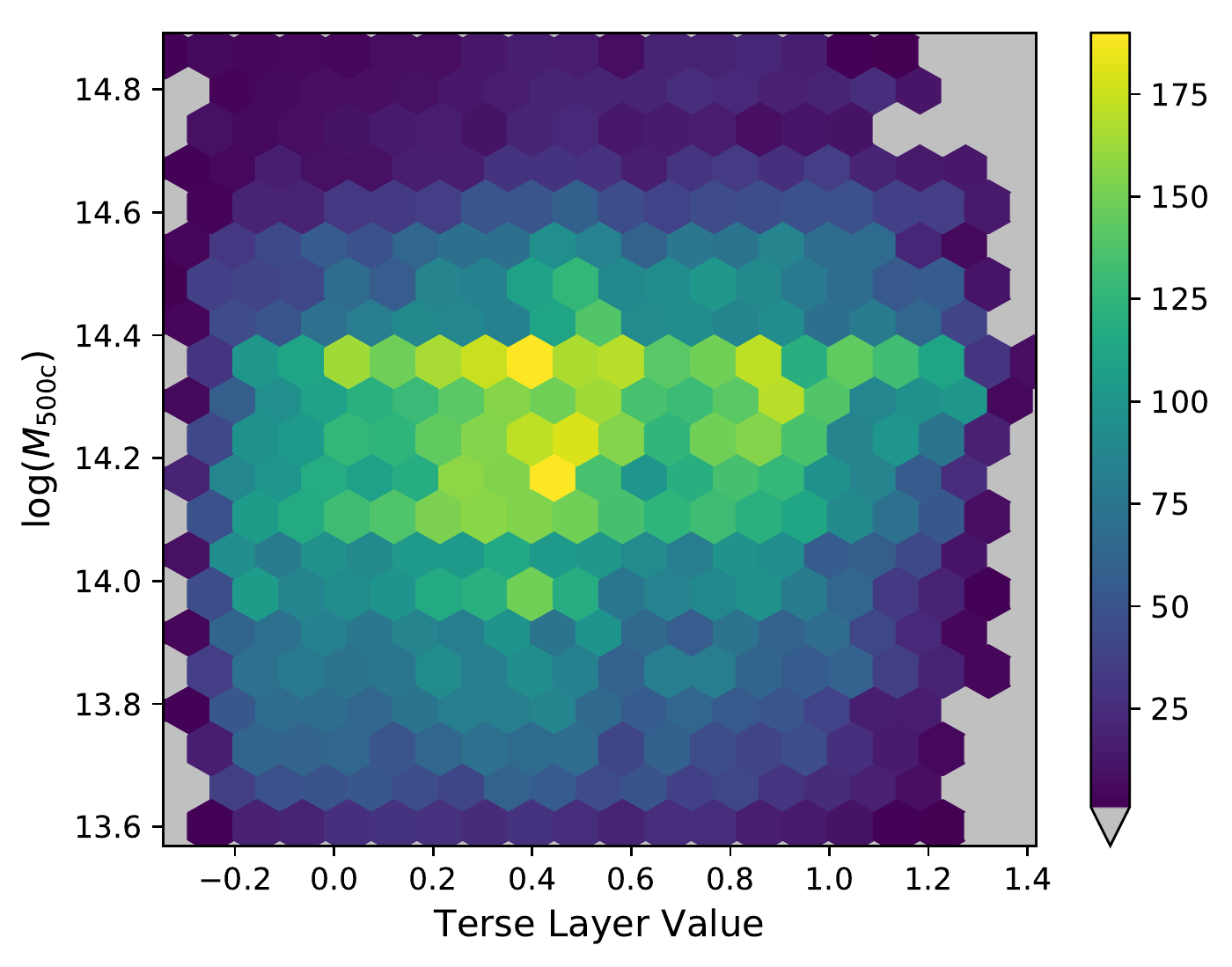}  
	 \includegraphics[width=0.45\textwidth]{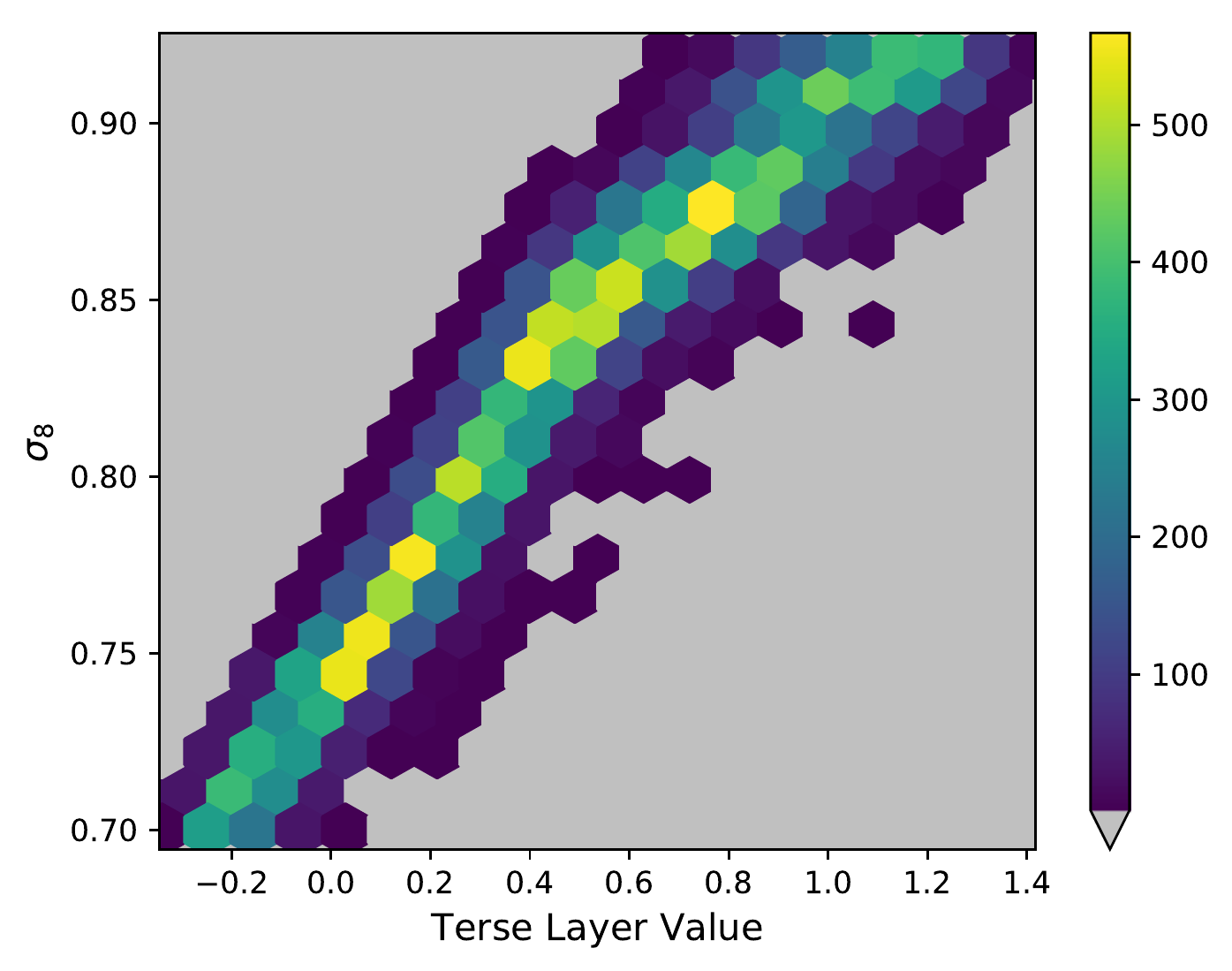}
	\caption[]{Top: Correlations between the terse value and cluster mass show a surprising result: the encoded value does not correlate at all with this important cluster parameter for cosmological analysis.  The tight cosmological constraints shown in Figure \ref{fig:cheatresults} would suggest that the ML model has developed a very low-scatter mass proxy, but this theory is not supported by the terse value.  Bottom:  The terse value is tightly correlated \Sig, a value that should not be encoded in a single cluster's data.  This result points to \textit{overspecialization}; the model has learned prescriptive details of the simulation and cannot safely be applied to observational data.}
       	\label{fig:aliendragon2}	
\end{figure}

\end{document}